\documentclass[11pt]{article}
\pdfoutput=1

\usepackage{graphicx}
\pdfoutput=1
\usepackage[utf8]{inputenc}
\usepackage{graphicx}
\usepackage{float}
\usepackage[numbers]{natbib}
\usepackage{textcomp}
\usepackage{caption}
\usepackage{subcaption}
\usepackage{textcomp}
\usepackage{amssymb,amsmath}
\usepackage[dvipsnames]{xcolor}
\usepackage{graphbox,graphicx}

\title{A Characteristic Dynamic Mode Decomposition}
\author{Jörn Sesterhenn$^{1}$ and Amir Shahirpour$^{2}$\\
		$^{1}$ Technical University of Berlin\\
		$^{2}$ Brandenburg University of Technology (BTU)}
\date{}
\begin{document}

\maketitle
\begin{abstract}
   Temporal or spatial structures are readily extracted from complex
  data by modal decompositions like Proper Orthogonal Decomposition 
  (POD) or Dynamic Mode Decomposition (DMD). Subspaces of such
  decompositions serve as reduced order models and define either spatial
  structures in time or temporal structures in space. On the contrary, 
  convecting phenomena pose a major problem to those decompositions. A structure
  traveling with a certain group velocity will be perceived as a
  plethora of modes in time or space respectively. This manifests
  itself for example in poorly decaying singular values when using a
  POD. The poor decay is counter-intuitive, since a single structure is expected
  to be represented by a few modes. The intuition
  proves to be correct and we show that in a properly chosen reference
  frame along the characteristics defined by the group velocity, a POD
  or DMD reduces moving structures to a few modes, as expected. Beyond
  serving as a reduced model, the resulting entity can be used to
  define a constant or minimally changing structure in turbulent
  flows. This can be interpreted as an empirical counterpart to exact
  coherent structures. We present the method and its application to a head vortex of a
  compressible starting jet.
  

\end{abstract}

\section{Introduction}
\label{intro}

The proposed method in this study aims at approximating large-scale coherent structures as dynamic modes
in space and time. Three topics come together: the so called coherent
structures, modal decompositions, and model reduction. The constituent
three parts shall be discussed briefly.\medskip

Study of coherent structures in turbulent flows has received increasing attention from 
scientists during the recent decades. It has become a common practice to try to understand
the complex and multi-scaled nature of turbulent flows by observing the instantaneous flow 
fields and inspecting organized motions which possess spatial and temporal coherence. The latter 
implies that such motions appear at some point in time while evolving in space and remain 
recognizable in space in a certain time span.\medskip

Relying basically on hot wire measurements and simple flow visualization measurements,
it was extremely difficult to come up with an idea of what structures are behind the 
observed quantities in turbulent flows. It was not until the advent of Particle Image Velocimetry (PIV) and Numerical
Simulations, that the idea of a coherent structure replaced the vague eddy in literature.\medskip

Some of the first observations of coherent structures were carried out by Theodorsen in 1952 
which gave rise to the notion of horseshoe eddies or hairpin vortices. His findings were
later supported by many studies including the experiments of \citet{Adrian2000} and the
simulations of \citet{WuMoin2009}. Such structures are observed to be originated from the wall
and form Large Scale Motions (LSM) when moving in groups at the same convective velocity 
\cite{Adrian2007}. Similar studies have reported existence of even larger structures which scale
on outer variables. They are commonly denoted in internal flows as Very Large Scale Motions
and as Super Structures in external flows \cite{BalakumarAdrian2007,HutchinsMarusic2007}.
In spite of the large range of studies carried out, many of the fundamental questions are still
unanswered regarding the origin, nature and evolution of such structures.\medskip

Besides differing views on the nature and origin of turbulent structures, the suitable approach to 
the analysis of such structures is also still under debate. The footprints of such motions can be 
followed by observing the premultiplied velocity spectra which represent the energy distribution in
the wave number space \cite {Rosenberg2013,Vallikivi2015}. The two peaks observed at relatively high Reynolds numbers in the outer 
region of the flow in the premultiplied velocity spectrum, are known to be associated with VLSM and LSM \cite {Rosenberg2013}.
Following the spectral peaks, which can be regarded as the signature of such structures, helps 
to determine their length scales and energy content at different wall-normal positions, but can not
provide any visualized insight into the evolution and interactions of the structures.\medskip

The availability of the strain rate tensor $S_{ij}=\frac{1}{2}\left(\frac{\partial u_i}{\partial
    x_j}+\frac{\partial u_j}{\partial x_i}\right)$ from numerical simulations has made the analysis
and perception of such structures accessible \cite{Chong1990} and has enabled their visualization
using for instance $Q$ criterion proposed by \citet{Hunt1988} and $\lambda_{2}$ criterion by \citet{JeongHussain1995}. 
They are categorized as Galilean-invariant but they fail to be invariant under more general changes 
such as rotation or accelerating reference frames \cite{Haller2005}.\medskip

Via a different method, \citet{Waleffe1998} looks at the coherent structures as fixed point solutions traveling 
in the flow. This route has been successfully followed such that today exact coherent structures can be 
computed for relatively high Reynolds numbers \cite{FaisstEckhadt2003,AvilaMellibovsky2013}.\medskip

On the other hand, data driven methods were adapted from other fields of science to extract structures
from turbulent flows. Proper Orthogonal Decomposition (POD) was for the first time introduced to fluid dynamics
by \citet{lumley1967,lumley1981}. POD serves as one of the methods to decompose a flow field into spatial or temporal
modes which are also regarded as characteristic features of the system. Being applied to flow fields, these
modes will then represent large scale energy containing structures of the flow. \citet{BackewellLumley1967} were
the first to apply the classical POD to experiments conducted in a turbulent pipe flow to find the dominant
large scale structure of the flow in the wall region. \citet{Glauser1987} also applied the method to turbulent
jet mixing layer and could show the existence of a large scale structure in the mixing layer containing 40$\%$ 
of the turbulent energy while providing proof that almost all the energy was contained in only the first three modes.\medskip

The Snapshot POD was later suggested by \citet{Sirovich1987} which was based on discretization of POD in the temporal domain
and was preferable for handling time resolved CFD data. During the recent years in studies by \citet{Hellstrom2011} 
and \citet{HellsromSmits2004}, Snapshot POD has been applied to cross-sectional PIV measurements to 
visualize the structure of LSM and VLSM in turbulent pipe flow.\medskip

Although POD has proven to be a powerful tool to study turbulent coherent structures and to extract the energetic modes, 
but the resulted patterns lack dynamics and have problems with convective flows. The first drawback
hampers the construction of reduced models of the flow, and in consequence the success of the method is limited.
This problem was addressed by the introduction of Dynamical Modes by \citet{SchmidSesterhenn} and was later followed 
by studies by \citet{Rowley2009} and \citet{Schmid2010} leading to vast applications of DMD afterwards. 
The study by \citet{Mezic2013} provides a detailed review on DMD and its correspondence to similar approaches.\medskip 

To overcome the difficulties of describing convective phenomena, several studies have focused on removing the discreet translational symmetries.
One of the first solutions was introduced by \citet{Rowley2001} via applying a POD in a shifted frame of reference, with the traveling
speed determined as a function of time {$c(t)$} using template fitting and a reconstruction equation. Later in a more general framework, 
the study by \citet{Rowley2003}, looked into self similar solutions by implementing both translation and scaling in space and time. 
More recently, Shifted POD was proposed by \citet{Reiss2015} aiming at model reduction by applying a shift in space to 
treat flows with multiple convective velocities. Regardless of the employed method to detect the group velocities, all similar works 
apply a spatial transformation on the dataset determined by the shift velocity. \medskip

In the present study we follow a new approach
inspired by gas dynamics and the theory of characteristics. We
propose to perform a modal decomposition in space and time along the
characteristics. This requires a spatiotemporal transformation rather than a spatial one, and the name ``Characteristic DMD'' is 
chosen to highlight this difference.
In the revision of this work, an archival version of this script was
cited by \citet{Mezic2016}, where the modes resulted from a
spatiotemporal transformation are interpreted as invariant solutions
of the Navier-Stokes equations.\medskip

The principal aim of this exercise is to extract low-dimensional subspaces of highly complex turbulent flows
along the characteristics having the slope of the group velocities of the structures. The subspaces serve as tangential linear 
approximations of the nonlinear events and will accommodate the large-scale scale coherent structures in the flow.
Several modes, traveling and interacting along the
characteristics, shall be defined as an
empirical coherent structure. \medskip

To come up with a mathematical
description term, we may start off from a simple 1D problem as for
example $\partial_t q+ A \partial_x q=0$ and chose a group velocity
which is of particular interest to us. (In the present case it could
be an eigenvalue of $A$, if the system is hyperbolic, but in general
it is a group velocity $u_{g}$, defined differently). Next we introduce a
rotated coordinate system, which points along the $\tau$ direction.
\begin{eqnarray}
  \label{eq:rot}
  \xi= c x + s t,\\
  \tau=- s x + c t,
\end{eqnarray}
where $c=cos(\theta)$ and $s=sin(\theta)$. The angle $\theta$ corresponds to the group velocity
and is defined as $\theta = u_{g}\,dt/dx$, with $dt$ and $dx$ being respectively the timestep between the snapshots and 
the spatial distance between the points along $x$.
Introducing the new variables, we look for
the principal eigenfunctions of the system
\begin{equation}
  \label{eq:Along_tau}
\partial_\tau q+(I c + As)^{-1}(-Is+Ac)\partial_\xi q=0.  
\end{equation}
Several eigenfunctions with a small decay rate along $\tau$ and
probably interacting (since the system is not symmetric) shall be
investigated as candidates for empirical coherent structures in future
work.\medskip

In what follows the method is first applied to the solutions of KDVB equation to address the problem for a one dimensional 
structure. In chapter \ref{sec:validation}, in order to validate the method, it has been applied to a two dimensional Lamb-Oseen vortex propagating
in space and time with a known frequency and decay rate. In chapter \ref{sec:startingJet}, the Characteristic DMD has been implemented to detect the 
vortex head of a starting jet. Finally in chapter \ref{sec:StVsSp}, a comparative analysis is carried out, to highlight the differences between a decomposition
in the spatiotemporal and in a shifted frame of reference. Application of the presented method to fully turbulent structures in wall bounded 
flows will be the subject of a future study.

\section{The Problem}
\label{sec:problem}
From an empirical point of view a structure can be defined as an entity
in space which appears somehow recognizable elsewhere at a later
time. A clear example for this definition would be a solution $u(x,t)= u(x-\lambda t)$ to
the convection equation
\begin{equation}
\label{eq:convect}
\partial_t u + \lambda \partial_x u=0.
\end{equation}
Even when non-linearly distorted, damped and dispersed, e.g. for the
Korteweg -- de Vries -- Burgers (KDVB) equation 
\begin{equation}
\label{eq:KdVB}
\partial_t u + u \partial_x u -\nu \partial_x^2 u + \delta \partial_x^3 u =0,
\end{equation}
it is possible to find an analytic solution fitting the above
definition in form of a soliton. A solution of KDVB equation developing from a given
initial condition will serve as an introductory example below. Even 
in more complex situations, for example a boundary layer, the
flow exhibits structures which lack an analytic solution but clearly
fit the above definition. They might be found as fixed point solutions
of the Navier-Stokes equations. In what follows, we concentrate on the
general case where we do not have descriptive equations yet. \medskip

The main example presented in this study will be the vortex ring of a starting supersonic jet.  
A long and a short high-pressure pulse being released from an orifice will be studied separately.
A short pulse forms a laminar vortex ring and a long one will lead to a vortex ring
followed by a jet. Both the vortex head and the jet will become turbulent provided 
that the Reynolds number is sufficiently high. POD, DMD or other model 
reduction techniques, might then be expected to easily reduce
the flow and to result in a principal mode representing the structure, followed by 
the higher modes modifying the main mode to some degree. Unfortunately, this is not the case.\medskip

\begin{figure}
    \centering
    \includegraphics[scale=0.4]{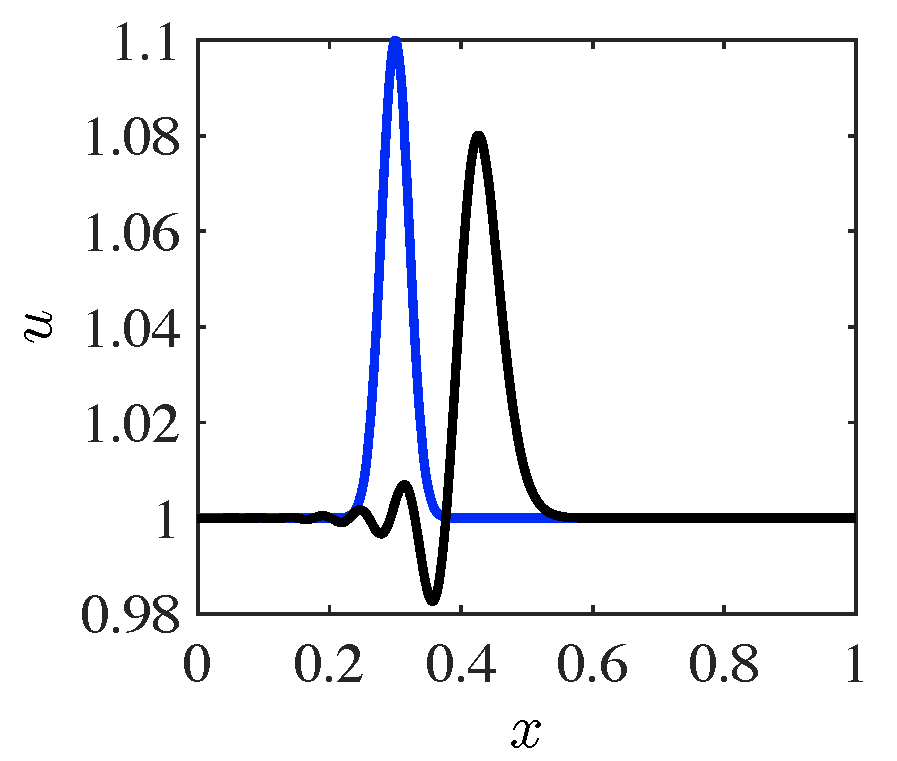}
    \caption{Solution (black) to equation \ref{eq:KdVB} at $t=0.1$ for a Gaussian
     initial condition (blue).}
  \label{fig:KdVB-Solution}
\end{figure}

\begin{figure}
     \begin{subfigure}[b]{0.5\linewidth}
          \centering
          \includegraphics[scale=0.4]{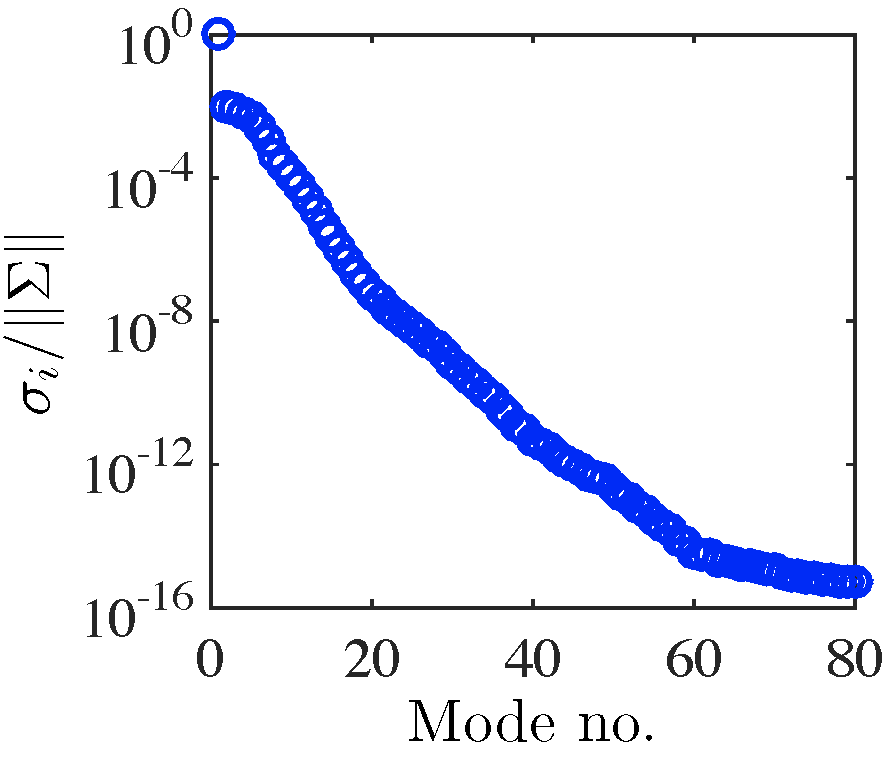}
          \caption {}
          \label{fig:KDVB_SingVals}
     \end{subfigure} 
     \begin{subfigure}[b]{0.5\linewidth}
          \centering
          \includegraphics[scale=0.4]{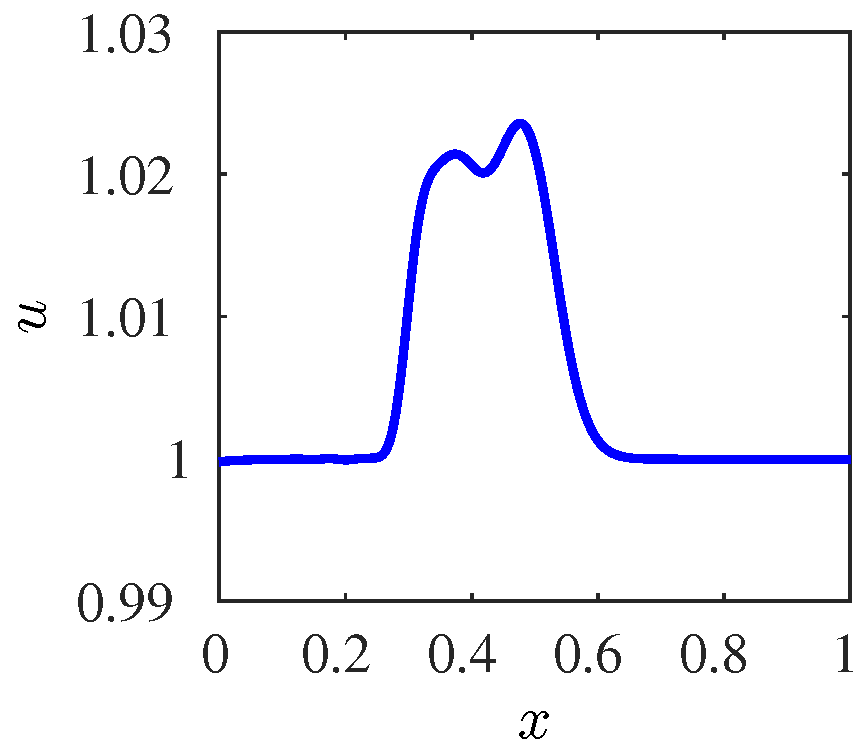}
          \caption{}
          \label{fig:KDVB_SVD_Mode_01}
     \end{subfigure}%
   \vspace{0.2 cm}
     \begin{subfigure}[b]{0.5\linewidth}
          \centering
          \includegraphics[scale=0.4]{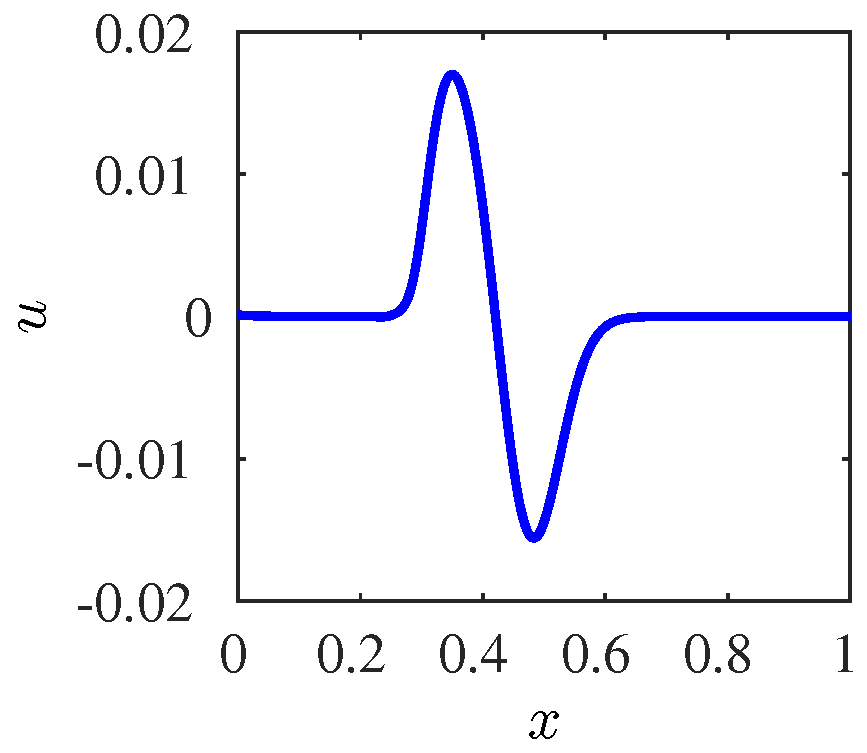}
          \caption{}
          \label{fig:KDVB_SVD_Mode_02}
     \end{subfigure}%
     \begin{subfigure}[b]{0.5\linewidth}
          \centering
          \includegraphics[scale=0.4]{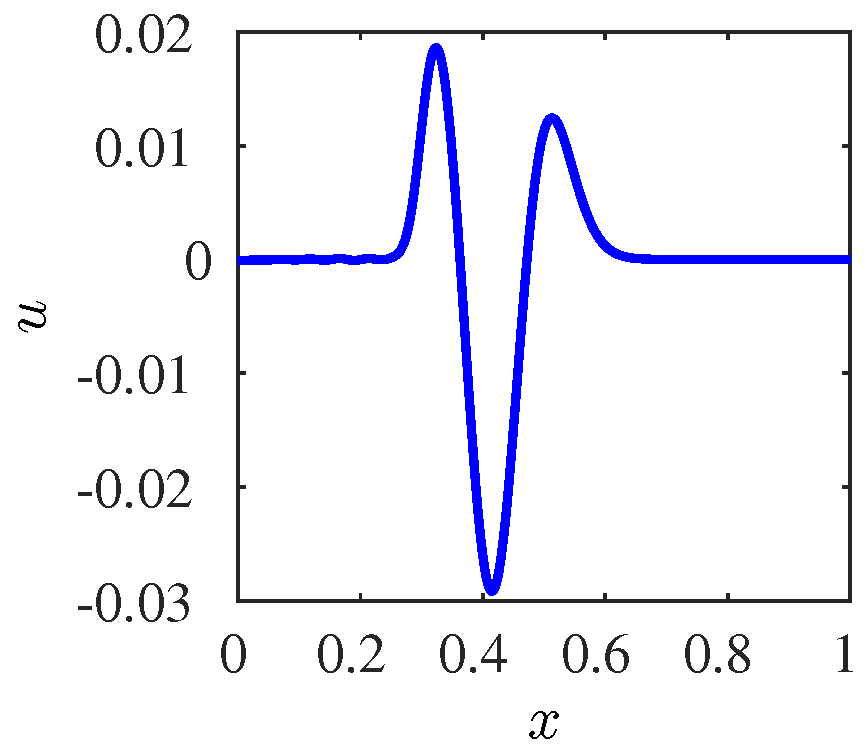}
          \caption{}
          \label{fig:KDVB_SVD_Mode_03}
     \end{subfigure}
     
     \caption{Singular values of snapshots of solutions to equation (\ref{eq:snaps}) normalized by the norm of singular values vector $\|\Sigma\|$ (a) and
     the first three POD modes of the solutions of the same equation (b,c,d).}
     \label{fig:svd-t}
\end{figure}

The failure can be demonstrated
already for a solution of the KdVB equation (\ref{eq:KdVB}). The chosen parameters are
$\nu=5\times10^{-4}$, $\delta=4\times10^{-5}$ with initial condition of
$u(x,0)=1+\alpha e^{-(x-x_o)/\beta^2}$ where $\alpha=0.1$ and
$\beta=0.03$. The solution at $t=0.1$ is given in figure \ref{fig:KdVB-Solution}, 
which is distorted, damped and dispersed, but has primarily
experienced a shift in $x$ direction and would still qualify as an evolving
structure in space and time. A POD of the data
\begin{equation}
  \label{eq:snaps}
  X=\left[u(x,t_0),u(x,t_1),...u(x,t_{n-1})\right]
\end{equation}
using an SVD
\begin{equation}
  \label{eq:pod}
  X=U\Sigma V^T
\end{equation}
yields the poorly reducing singular values, depicted in figure \ref{fig:KDVB_SingVals}. 
The POD-modes, being an average over all shifted solutions and
necessary distortions, therefor sum up to the desired solution. 
This is illustrated in figure \ref{fig:KDVB_SVD_Mode_01}. The dominant mode
appears to be the swallowed elephant. Any real
instance in time is made up by subtracting a large number of modes
like the ones depicted in figures \ref{fig:KDVB_SVD_Mode_02} and \ref{fig:KDVB_SVD_Mode_03}. 
The resulting linear combination will not yield zero in many spatial locations and give rise
to substantial spurious structure, where there should be none.
A DMD along $t$ suffers from the same problem.

\begin{figure}
  \centering
  \includegraphics[scale = 0.4]{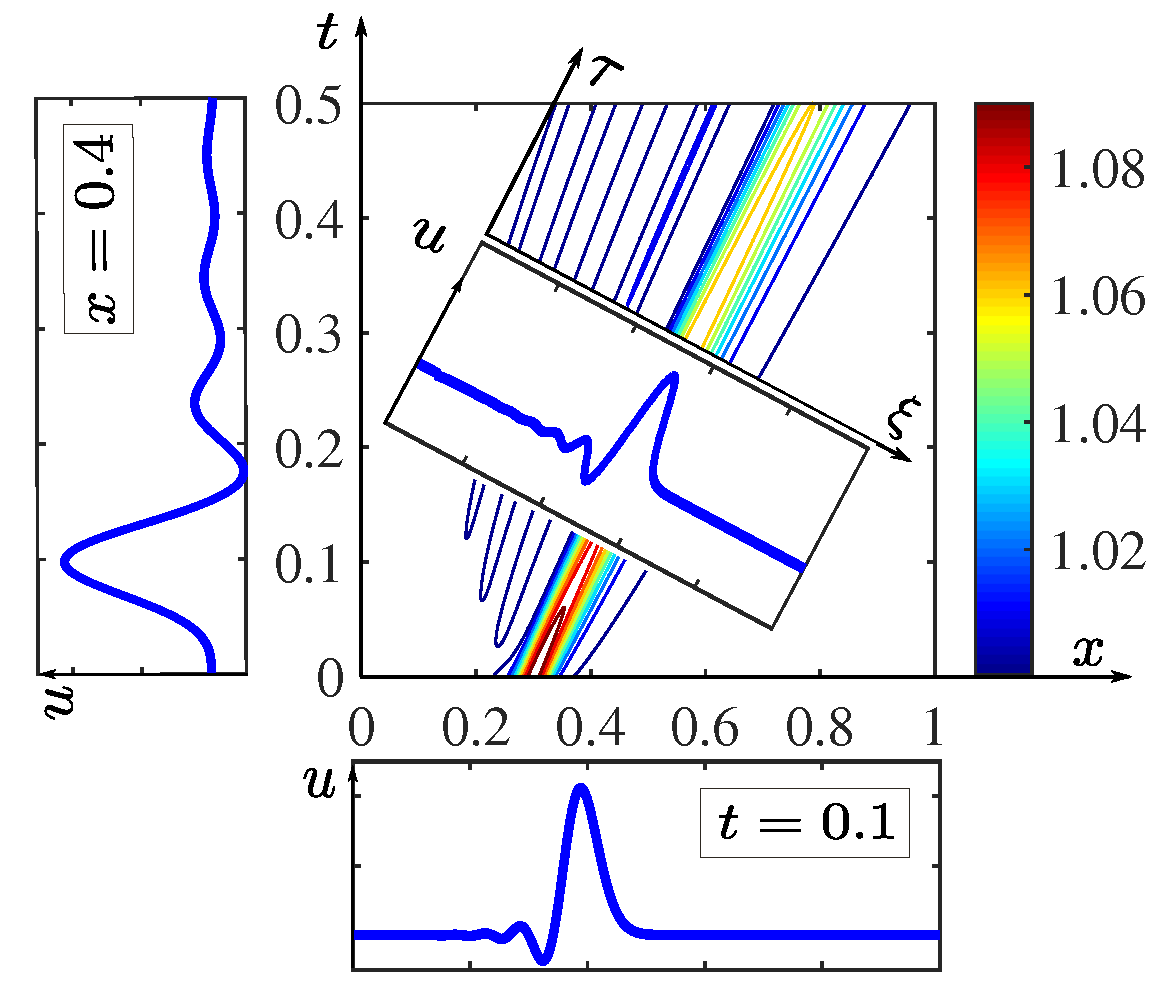}
  \caption{Spatial, temporal and spatiotemporal representation of the KDVB solutions.}
  \label{fig:CharacteristicDiagramm}
\end{figure}

\section{A  Remedy}
\label{remedy}

The main problem above, comes neither from the nonlinearity nor the
other factors, rather, the mere translation. The fact that the flow has a
relatively simple structure is easily inferred from the characteristic
diagram in figure \ref{fig:CharacteristicDiagramm}. It can be observed that the solutions travel
relatively unmolested in the direction of $\tau=x-\lambda t$.\medskip

To treat the mentioned problem for a structure traveling in time along the direction
given by the wave-vector $\kappa$, the modal decomposition is to be
sought in a plane normal to that direction in space-time
\begin{equation}
  \label{eq:Spacetime}
  x_i=\left\{t,x_1,x_2,x_3\right\}~~~~~~~i=0...3.
\end{equation}

\begin{figure}
  \centering
       \begin{subfigure}[b]{0.45\linewidth}
           \includegraphics[scale = 0.4]{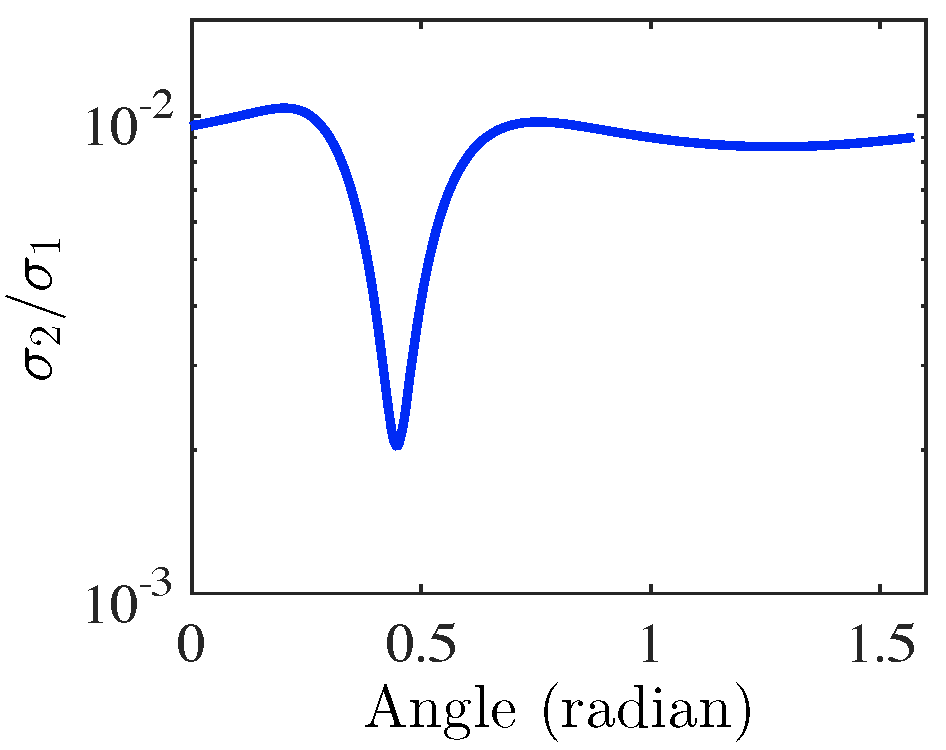}      
           \caption{}
           \label {fig:KDVB_SingValDrop_vs_angle}
       \end{subfigure}
       \begin{subfigure}[b]{0.45\linewidth}
            \includegraphics[scale = 0.4]{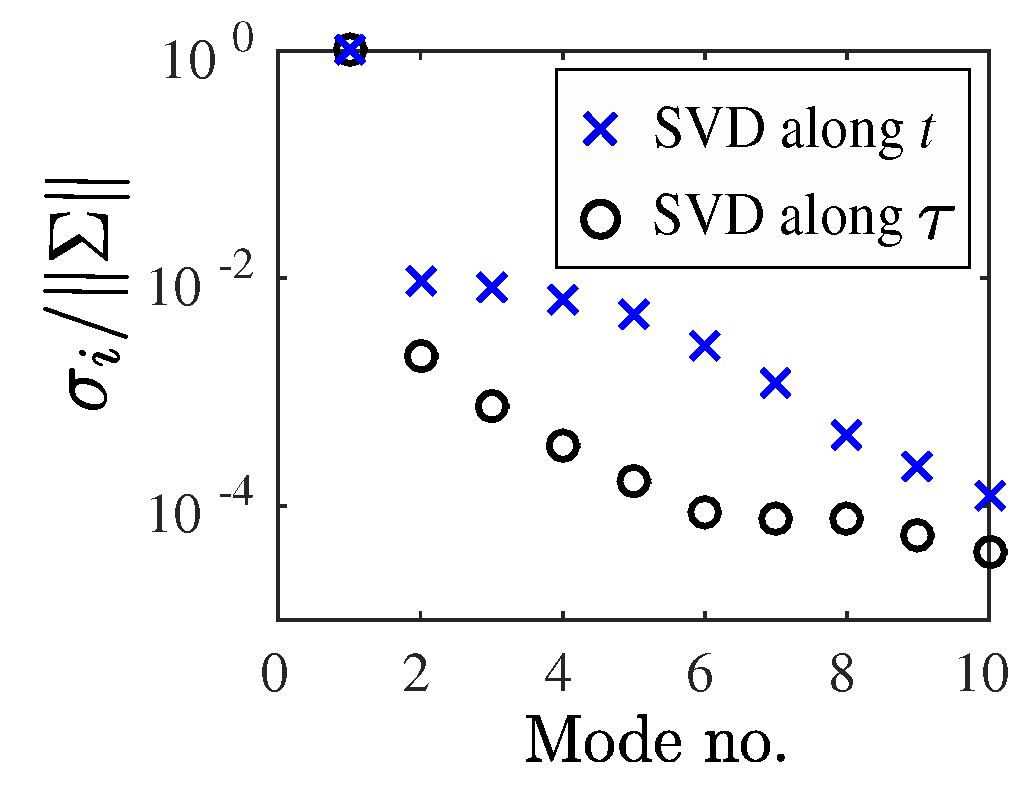}      
            \caption{}
           \label {fig:KDVB_SingValDrop}
       \end{subfigure}
  \caption{Ratio of singular values for SVD along different directions (a) and
    the first 10 normalized singular values along t and $\tau$  (b).}
  \label{fig:drop}
\end{figure}

Given the example above, the ratio of the second to the first singular
value of eq.(\ref{eq:snaps}) are plotted in figure \ref{fig:KDVB_SingValDrop_vs_angle} 
for decomposition in different directions. There
is a dramatic drop when the proper direction is chosen for the
snapshots, as also shown for the first 10 modes in figure \ref{fig:KDVB_SingValDrop}. The 
new frame of reference is in fact reached by transformation of snapshots matrix via a rotation in space and time,
to align the new time coordinate ($\tau$) with the direction which leads to the maximum drop of singular values.
The resulted snapshots matrix will thereby accommodate the structures in spatiotemporal space as:

\begin{equation}
    \label{eq:Rotated Snapshots}
    X_{0..n}=\{u(\xi,\tau_o),u(\xi,\tau_1)...u(\xi,\tau_n)\}.
\end{equation}

\emph{The essence of the method presented here is to
perform a modal decomposition along the direction which leads to the maximum drop of singular values
 and later transform the snapshots back into physical space.}\medskip

\begin{figure}
\centering
      \begin{subfigure}[b]{0.45\linewidth}
	      \includegraphics[scale = 0.4]{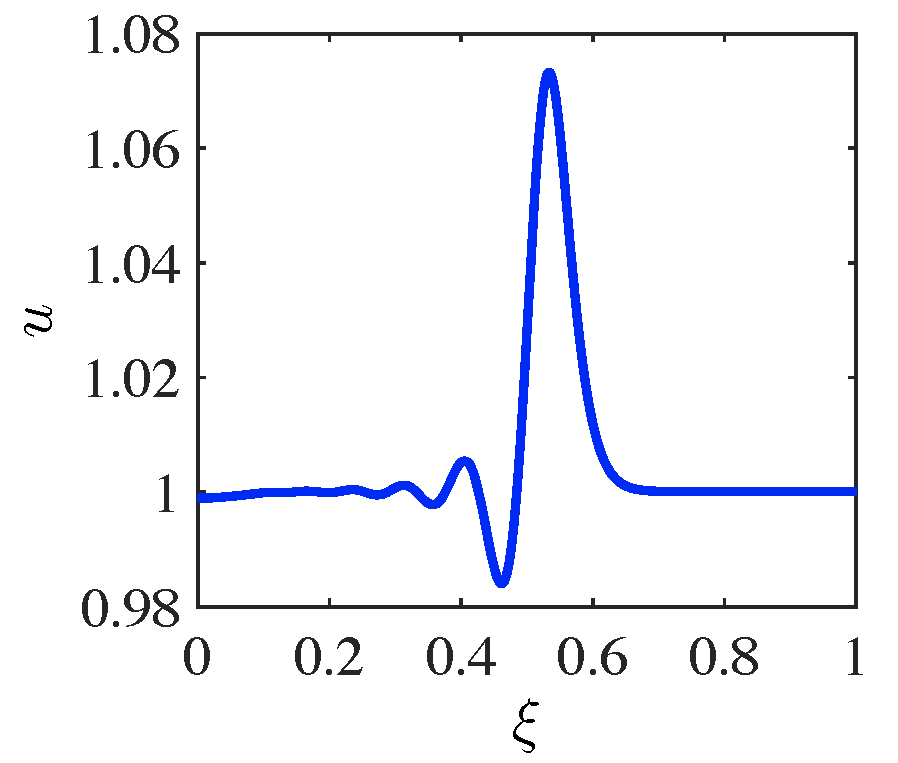}
	      \caption{}
               \label {fig:KDVB_LDMD_SpatTemp}
      \end{subfigure}
      \begin{subfigure}[b]{0.45\linewidth}
	      \includegraphics[scale = 0.4]{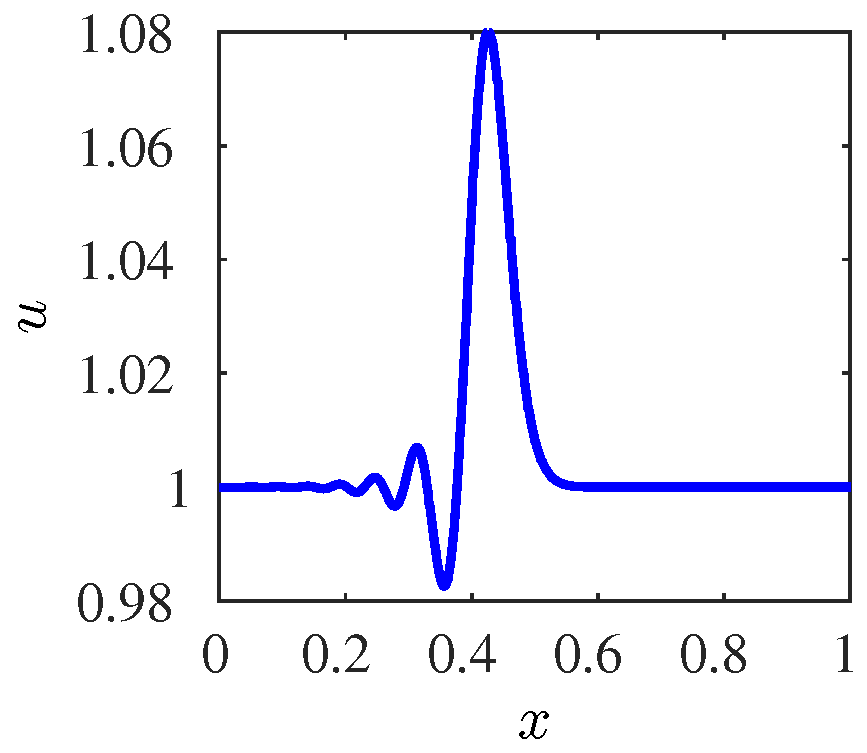}
	      \caption{}
              \label {fig:KDVB_LDMD_phys}
      \end{subfigure}
  \caption{KdVB-Structure in spatiotemporal space (a) and physical space (b) at $t=0.1$.}
  \label{fig:KdVB-Structure}
\end{figure}

After that operation, the first singular vector, shown in figure
\ref{fig:KdVB-Structure}, looks as expected for the structure of the
solution of the KdVB-equation (\ref{eq:KdVB}). Higher modes drop off
fast and minimally change the overall shape of the mode. It should be noted 
that figure \ref{fig:KDVB_LDMD_SpatTemp} represents the
spatiotemporal structure. A back transformation to physical space will be
necessary to either see the temporal evolution in a given space interval as depicted in 
figure \ref{fig:KDVB_LDMD_phys}, or conversely, the spatial changes in a time interval.\medskip

\section{Validation of the Method}
\label{sec:validation}

\subsection{Detection of a traveling Lamb-Oseen vortex}

In order to validate the method and to demonstrate how the decay rate and frequency of a structure can be correctly captured along the characteristics,
analytical solution of Lamb-Oseen vortex is used. The tangential velocity 
distribution along the radius is defined as:
\begin{equation}
    \label{eq:lamb-oseen}
    U(r)=U_{max} (1+\frac{2}{2\alpha}) (\frac{R}{r}) \left(1-exp\left(\frac{-\gamma r^{2}}{R^{2}} \right)\right),
\end{equation}
with the constant of $\gamma=1.256$, where $U_{max,i}$ and $R$ corresponding respectively to the initial maximum velocity and the vortex radius. While 
the vortex propagates in space, its maximum velocity is dictated to oscillate and decay in time 
as:
    \begin{align}
    &U_{max}(t) = U_{max,i}\,exp(2\pi t(-d^{*}+f^{*}i)),\\
    &d_{v}^{*}= \frac{2\,d\,R}{U_{max,i}} = 0.01, \;\;f_{v}^{*}= \frac{2\,f\,R}{U_{max,i}} = 0.08.
    \end{align}
with $d^{*}$ and $f^{*}$ being the dimensionless decay rate and frequency of the vortex respectively.\medskip

The main aim of this chapter will be to detect the vortex 
with only one mode with the eigenvalues presenting a frequency and a decay rate similar to those of the vortex. Therefore two 
decompositions will be carried out for the introduced set up, one along the characteristics direction defined by the group velocity of the vortex,
and the other along the time axis in physical space, in order to emphasize the differences between both sets of results.\medskip

\begin{figure}
\centering
      \begin{subfigure}[b]{0.45\linewidth}
         \includegraphics[scale = 0.4]{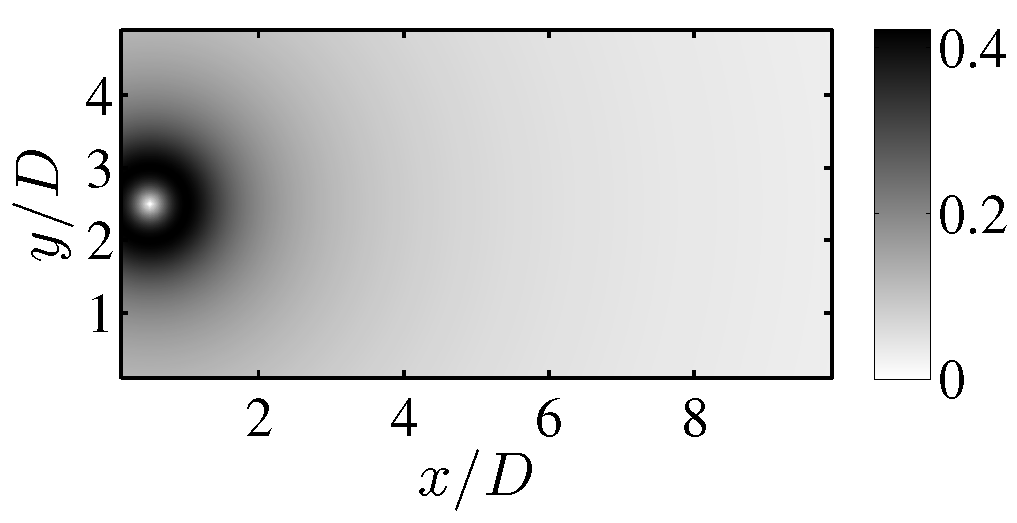}
         \caption{}
         \label {fig:lamb-Oseen_contour}
      \end{subfigure}
      \begin{subfigure}[b]{0.45\linewidth}
         \includegraphics[scale = 0.4]{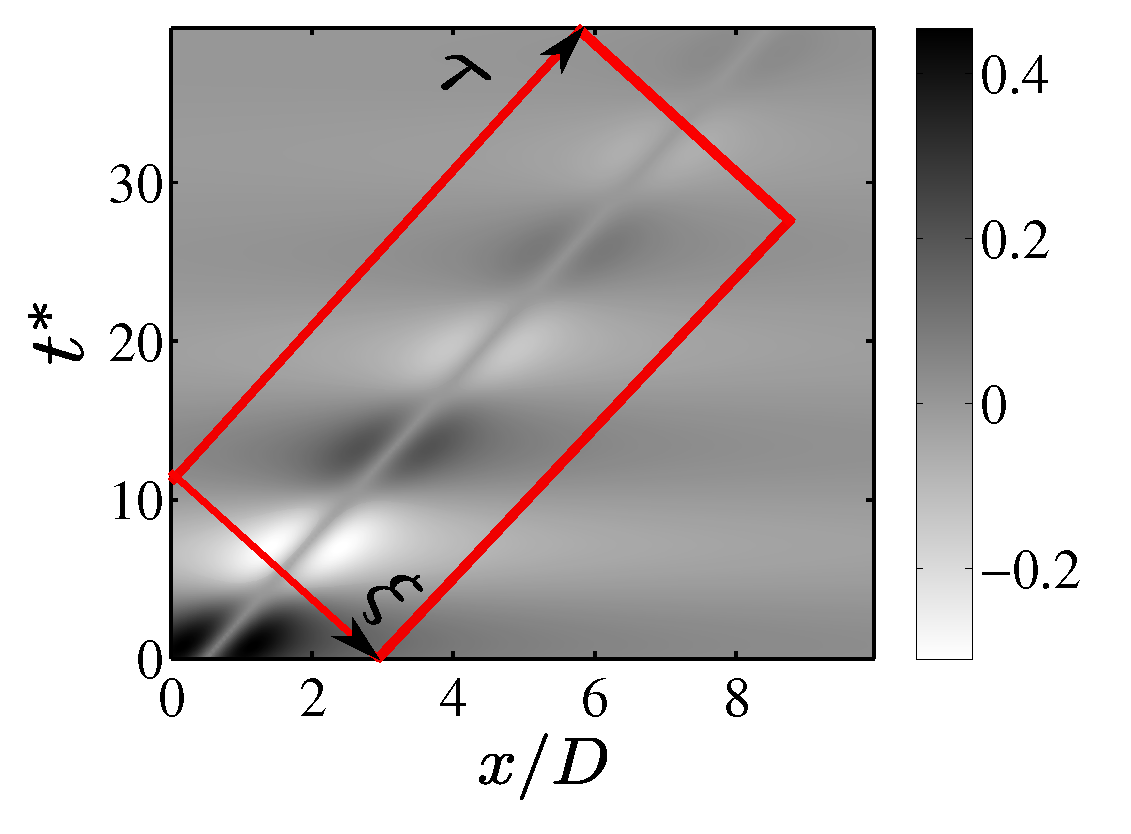}
         \caption{}
         \label {fig:lamb-Oseen_spaceTime}
         \end{subfigure} 
    \caption{Contours of tangential velocity normalized by $U_{max,i}$ used as the initial condition (a), and space time diagram(b).}
    \label{fig:lamb-Oseen}
\end{figure}

\begin{figure}
\centering
      \begin{subfigure}[b]{0.45\linewidth}
         \includegraphics[scale = 0.4]{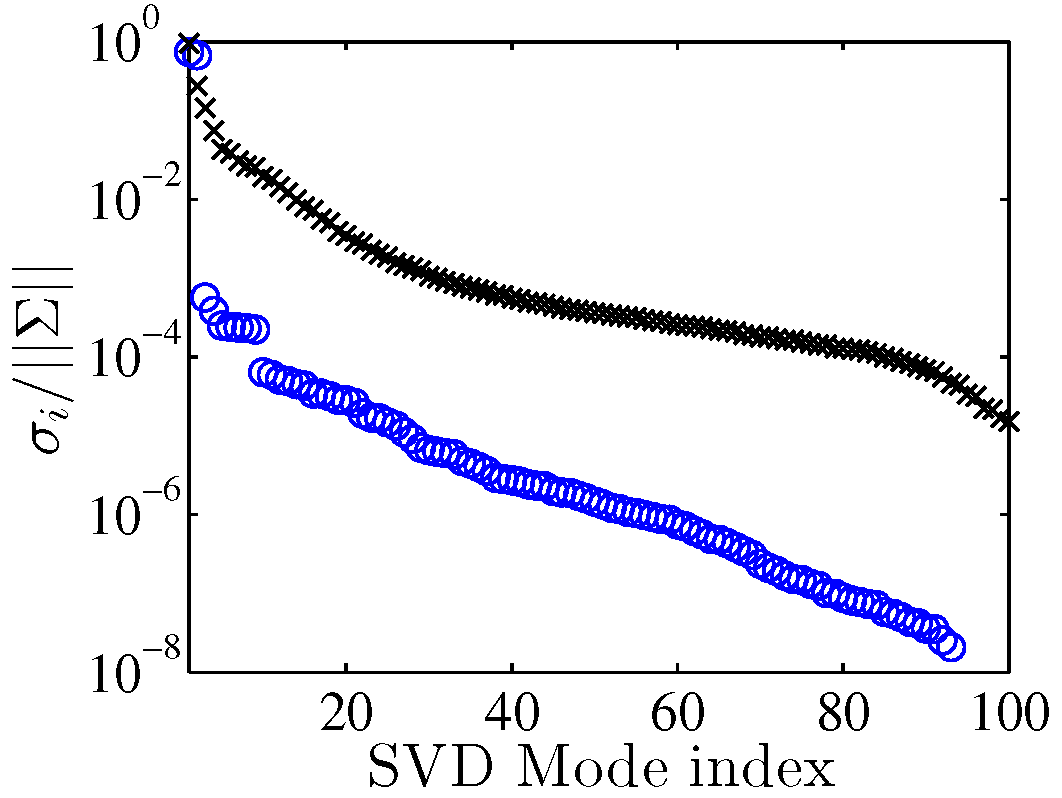}
         \caption{}
         \label {fig:lamb-Oseen_sing}
      \end{subfigure}
      \begin{subfigure}[b]{0.45\linewidth}
         \includegraphics[scale = 0.4]{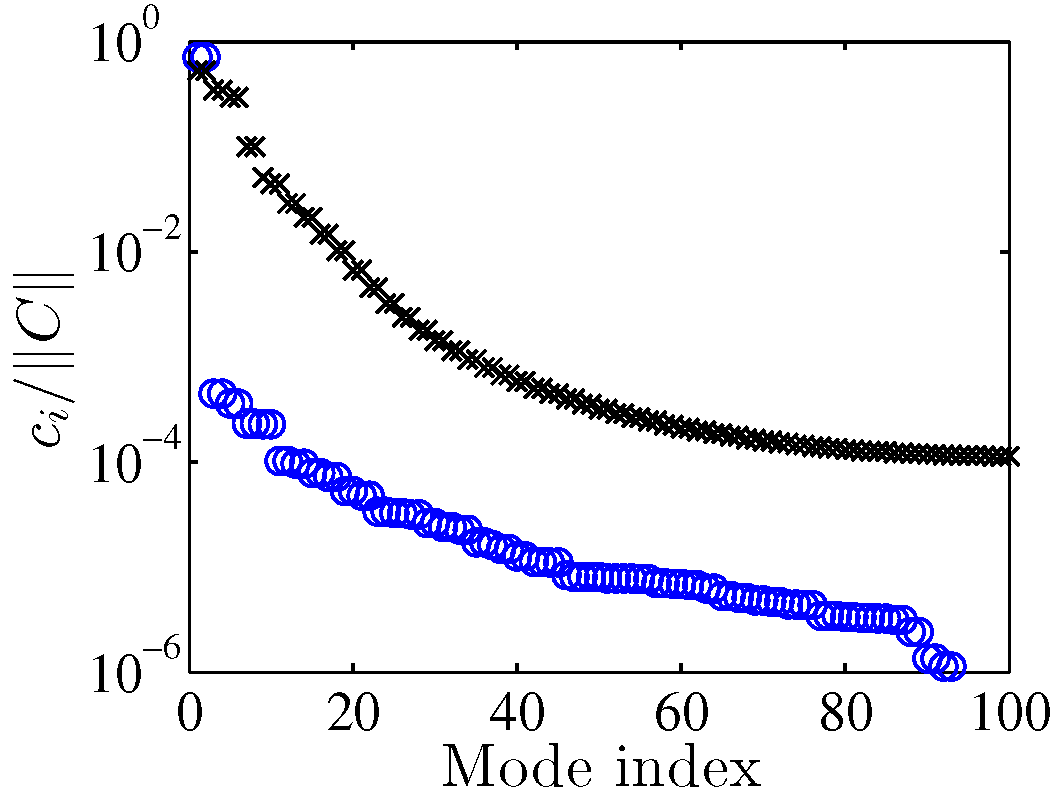}
         \caption{}
         \label {fig:lamb-Oseen_decay}
         \end{subfigure} 
    \caption{Singular values (a) and modal decay (b) of DMD modes along $\tau$ ($\circ$) and $t$ ($\times$).}
    \label{fig:lamb-Oseen}
\end{figure}

Starting from the initial condition shown in figure \ref{fig:lamb-Oseen_contour} as contours of dimensionless velocity ($u^{*}=u/U_{max,i}$), the vortex propagates 
in space and time with the group velocity of $u_{g}^{*}=0.2$, which is observable in the space time diagram in figure \ref{fig:lamb-Oseen_spaceTime}.
A rotation in space and time will provide the proper
frame of reference aligning the new coordinates with the direction yielding the best drop of singular values. As expected, 
it can be clearly seen in figure \ref{fig:lamb-Oseen_sing}, that a much faster drop is reached along $\tau$ (values normalized by the 
norm of singular values vector $\|\Sigma\|$).\medskip

At this step, since the development of the modes while traveling can be analyzed better
using a DMD, the snapshots which are now in spatiotemporal space will be decomposed using the following algorithm, known as
the standard DMD \cite{Schmid2010}.  For this purpose, a linear mapping is then assumed as:

\begin{equation}
   \label{eq:DMD}
   X'=AX
\end{equation}
with $X$ and $X'$ being the first and last $n$ snapshots in $X_{0..n}$. The transition 
matrix $A$ can be approximated using SVD of matrix $X$,

\begin{equation}
    X = U\Sigma V^T
\end{equation}
and using the projected matrix $\tilde{A}$,

\begin{equation}
   \tilde{A}= U^TAU = U^TX'V\Sigma^{-1}.
\end{equation}

By computing the eigenvalues and eigenvectors of $\tilde{A}$, 

\begin{equation}
   \tilde{A}w= \Lambda w
\end{equation}
the DMD modes will be given by

\begin{equation}
   \phi= Uw.
\end{equation}

What follows in this chapter, clarifies how a DMD in the rotated frame of reference along $\tau$ compares with a traditional DMD along $t$. The red frame in 
figure \ref{fig:lamb-Oseen_spaceTime} \footnote{Here we note that we do not make use of all
data using this approach, specially if the dataset is periodic. This is unlike the method of \citet{Rowley2001} 
which implements a pure shift. But as we are not willing to chose the
method on economic grounds, there is not much to do about this for the
moment.} shows the bounds of data which is decomposed in the new frame of reference. For a rotation in $x-t$ plane, the maximum number of snapshots that can be
acquired in the rotated frame ($n_{\tau (max\,)}$), is a function of number of snapshots along the time axis ($n_{t}$), the group velocity ($u_{g}^{*}$), 
timestep between the snapshots ($dt^{*}$) and the spatial distance between the grid points along $x$ ($dx^{*}=dx/2R$). Therefore the transformation
coefficient of $\alpha_{t}$ can be defined as:
\begin{equation}
   \alpha_{t}=\sqrt{1+ \left(u_{g}^{*} \,\frac{dt}{dx}\right)^{2}},
\end{equation}
for the ratio of:
\begin{equation} \label{eq:alpha}
   \frac{n_{\tau\,(max)}}{n_{t}}=\alpha_{t},
\end{equation}
stating that for a certain number of snapshots in physical space, the value of $n_{\tau \, (max)}$ will be essentially larger than $n_{t}$. 
Nevertheless, as it can be inferred from figure \ref{fig:lamb-Oseen_spaceTime}, the final value of $n_{t}$, is also defined by the choice of 
frame width along $\xi$. Furthermore, the latter coefficient also provides a measure for the spatial resolution of the desired structure in 
the spatiotemporal space $Res_{(st)}$ as:
\begin{equation}
   \frac{Res_{(phys)}}{Res_{(st)}}=\alpha_{t},
\end{equation}
with $Res_{(phys)}$ being the spatial resolution in physical space. In other words, if a structures travels in space $(x)$ and time $(t)$ with a constant 
length and resolution of $Res_{(phys)}$, it will be observed in the rotated frame of reference with resolution of  $Res_{(phys)}/\alpha_{t}$ along $\xi$.
Consequently, to maintain the spatial resolution of the structure in spatiotemporal space, temporal resolution of the snapshots
should be adjusted accordingly to keep the value of $\alpha_{t}$ as close as possible to unity. Therefore, as it holds true for any type of 
modal decomposition, a high temporal resolution would be crucial for this method as well.\medskip 

For this test case, transformation coefficient 
of $\alpha_{t}=1.4$ was chosen and the decomposition was carried out acquiring 96 snapshots along $\tau$ and 100 snapshots along the time axis
on the domain size of $10D\times 5D$ with the resolution of $200\times 100$ in $x$ and $y$ directions respectively.\medskip 

In the next step the snapshots at all timesteps are projected onto DMD eigenmodes and the modes are sorted by their projection coefficients. 
Time averaged mode amplitudes are plotted in figure \ref{fig:lamb-Oseen_decay} normalized by the norm of full mode for decomposition in both frames. It is clear that
the modes decay much faster along the characteristics direction implying that fewer modes will be needed to reconstruct the vortex. \medskip 

Having reconstructed the modes in spatiotemporal space, they will be transformed back to physical space.
The corresponding eigenvalues calculated in the spatiotemporal space $\lambda_{st}$, should be also transformed back to physical space 
using the rotation angle $\theta$ corresponding to the group velocity as:

\begin{equation}
   \label{eq:EV_transform}
   \lambda = exp\left(\frac{\log \lambda_{st}}{cos(\theta)}\right).
\end{equation}

\begin{figure}
\centering
      \begin{subfigure}[b]{0.45\linewidth}
         \includegraphics[scale = 0.4]{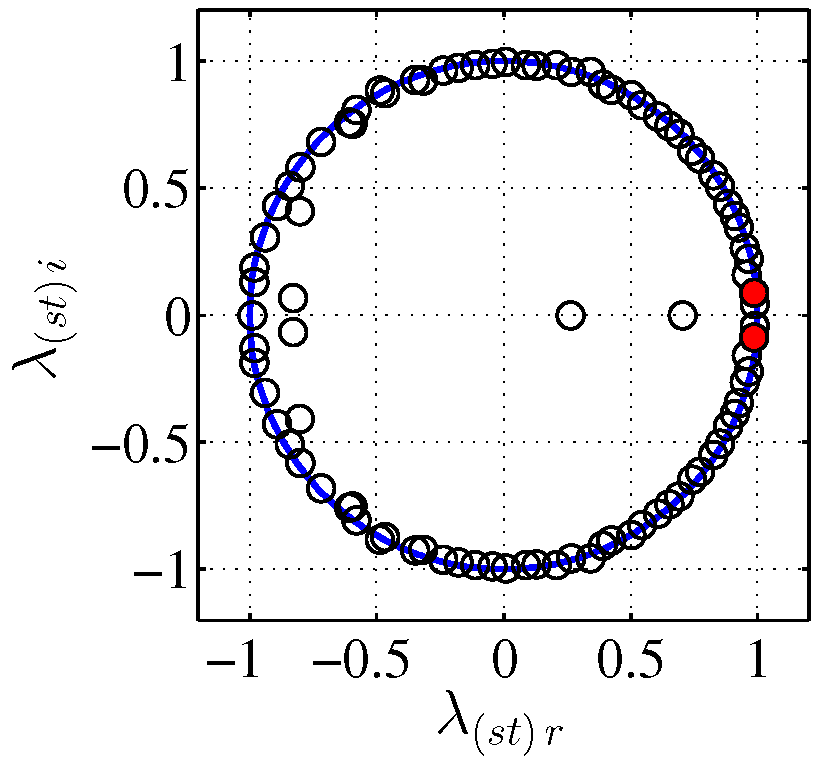}
         \caption{}
         \label {fig:LO_LDMD_EV_st_circle}
      \end{subfigure}
      \begin{subfigure}[b]{0.45\linewidth}
         \includegraphics[scale = 0.4]{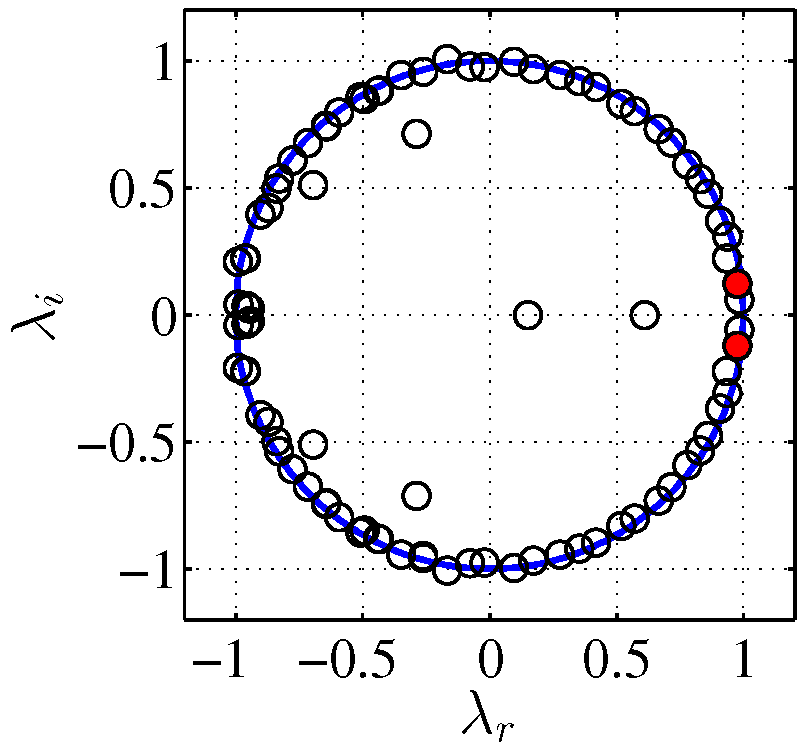}
         \caption{}
         \label {fig:LO_LDMD_EV_ph_circle}
      \end{subfigure} 
      \begin{subfigure}[b]{0.45\linewidth}
         \includegraphics[scale = 0.4]{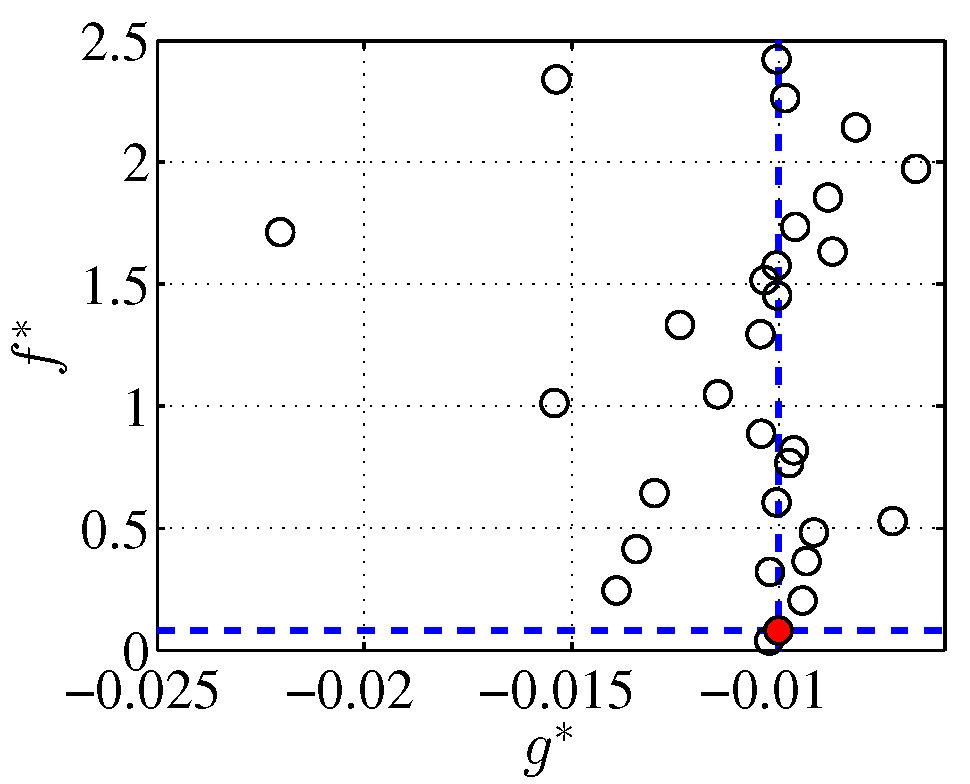}
         \caption{}
         \label {fig:LO_LDMD_EV_ph}
      \end{subfigure} 
      \begin{subfigure}[b]{0.45\linewidth}
         \includegraphics[scale = 0.4]{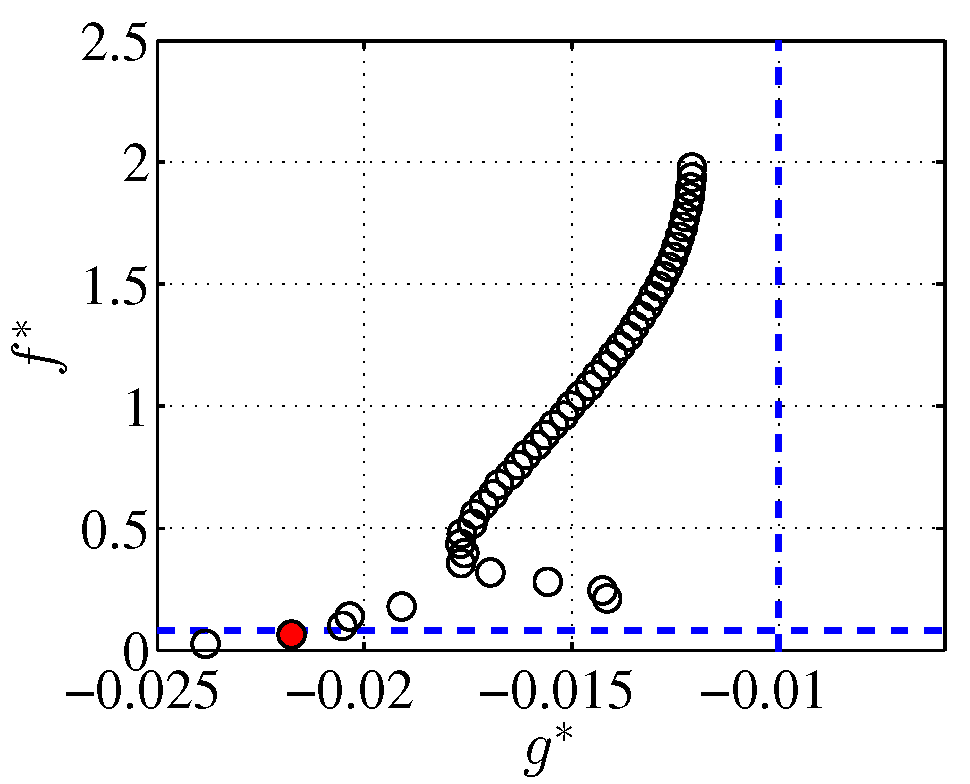}
         \caption{}
         \label {fig:LO_DMD_EV}
      \end{subfigure} 
    \caption{Eigenvalues of $\cal{C}$DMD modes in spatiotemporal space (a) and physical space (b),
    frequencies and growth rates ($g^{*}=-d^{*}$) of the $\cal{C}$DMD modes in physical space (c) and those of DMD modes (d).}
    \label{fig:lamb-Oseen_EV}
\end{figure}

The frequencies and decay rates of ($\cal{C}$DMD) modes, will be defined along the physical time, as a 
function of the timestep between the snapshots $dt$ and rotation angle $\theta$ using equations \ref{eq:freq} and \ref{eq:decay}. 
The eigenvalues in spatiotemporal space are compared against the transformed ones in figures \ref{fig:LO_LDMD_EV_st_circle} 
and \ref{fig:LO_LDMD_EV_ph_circle} with the filled markers 
showing the eigenvalue of the first mode. Dimensionless frequencies and decay rates of $\cal{C}$DMD modes are also presented in figure 
\ref{fig:LO_LDMD_EV_ph} in comparison with those of DMD modes in figure \ref{fig:LO_DMD_EV}. The blue lines in both figures, 
show the frequency and decay rate of the vortex, and the filled red markers correspond to the first modes in each reference frame.\medskip

\begin{equation}
   \label{eq:freq}
   f^{*}=\Im \left(  \frac{\log \lambda_{st}}{2\pi\,dt\,cos (\theta )} \right),
\end{equation}

\begin{equation}
   \label{eq:decay}
   d^{*}=\Re \left(  \frac{\log \lambda_{st}}{2\pi\,dt\,cos (\theta )} \right).
\end{equation}

It can be seen that the first $\cal{C}$DMD mode has captured the expected features of the vortex accurately. The DMD modes on the other hand,
have overestimated the decay rate. This is due to the fact that in the original frame of reference the structure moves downstream
and therefore this is understood by the DMD as a fast decay rate.\medskip

\begin{figure}
\centering
      \begin{subfigure}[b]{0.45\linewidth}
      \captionsetup{justification=centering}
         \includegraphics[scale = 0.4]{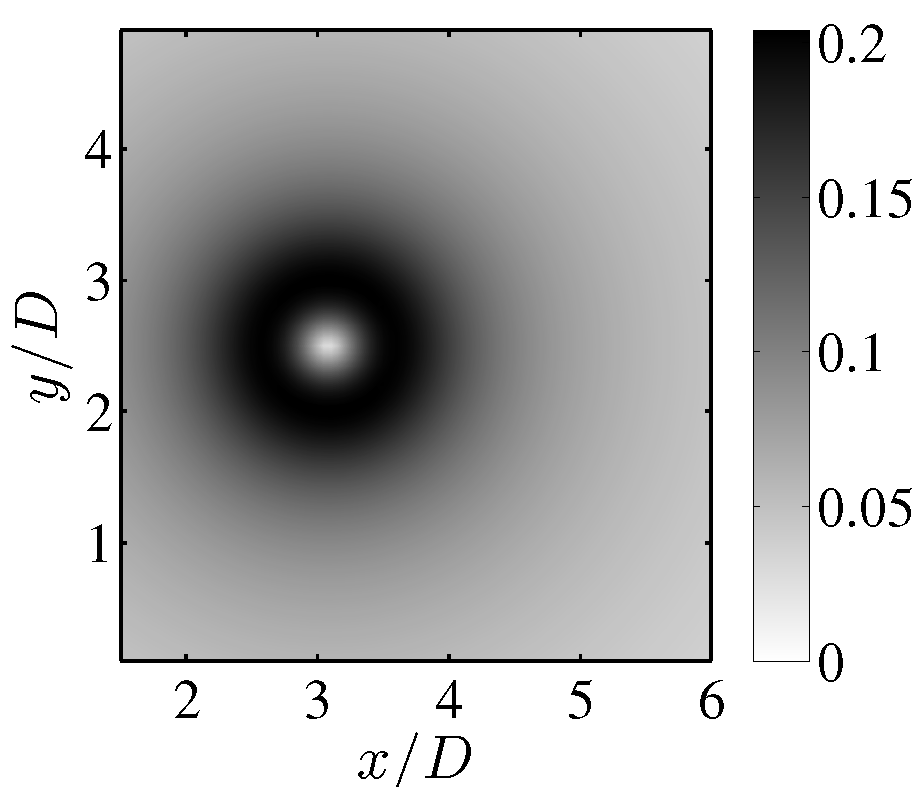}
         \caption{$\cal{C}$DMD, m=1\\
         $d^{*}$= 0.010,\; $f^{*}$= 0.080\\
         $\lambda_{1} = 0.97\pm 0.12i$}
         \label {fig:LO_LDMD_mode_01}
      \end{subfigure}
      \begin{subfigure}[b]{0.45\linewidth}
      \captionsetup{justification=centering}
         \includegraphics[scale = 0.4]{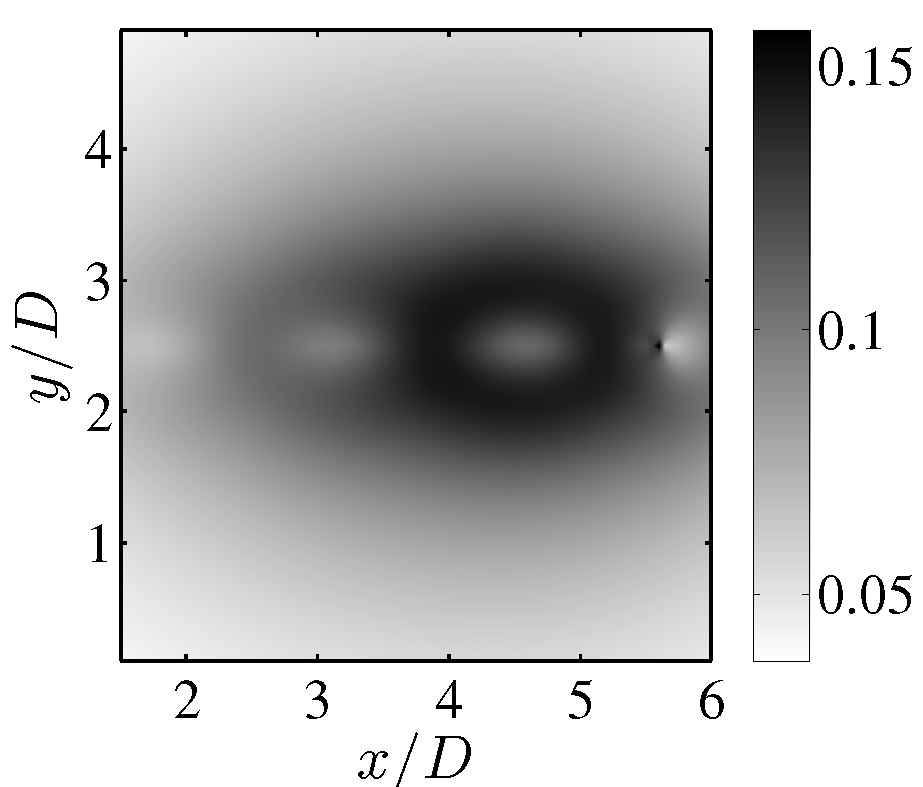}
         \caption{DMD, mode=1\\
         $d^{*}$= 0.022,\; $f^{*}$= 0.63\\
         $\Lambda_{1} = 0.96\pm 0.09i$}
         \label {fig:LO_DMD_Mode_01}
      \end{subfigure}
      \begin{subfigure}[b]{0.45\linewidth}
         \includegraphics[scale = 0.4]{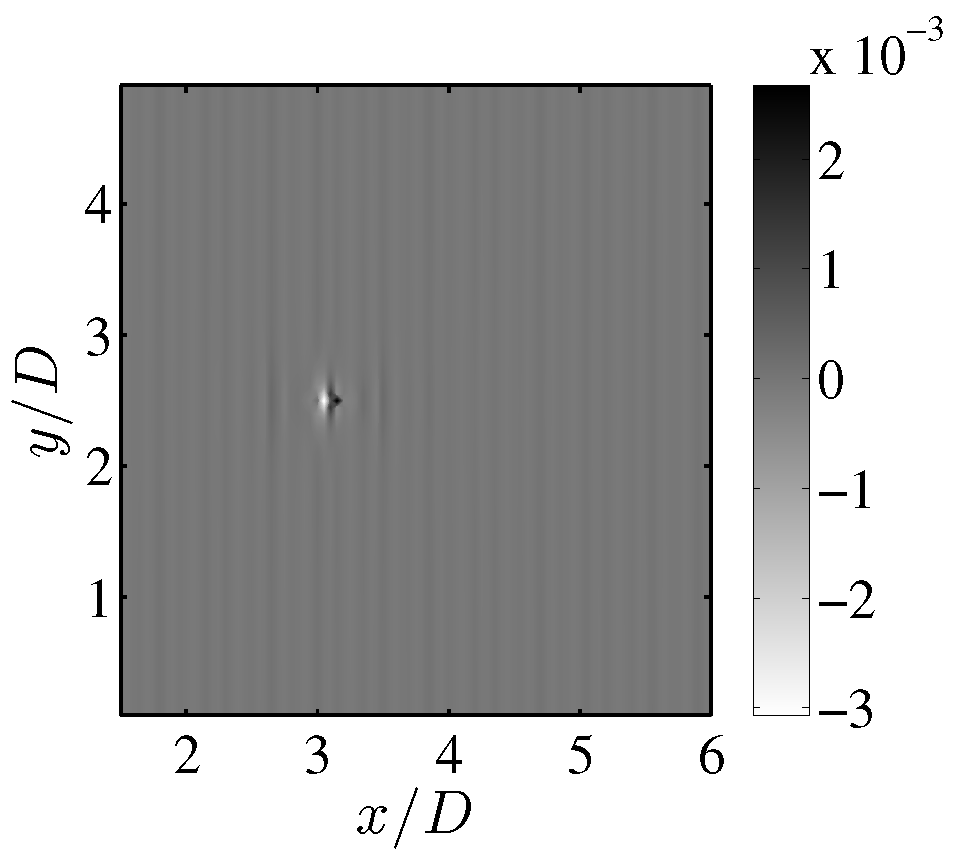}
         \caption{Relative Error for\\
         $\cal{C}$DMD, m=1}
         \label {fig:LO_LDMD_relError}
      \end{subfigure} 
      \begin{subfigure}[b]{0.45\linewidth}
         \includegraphics[scale = 0.4]{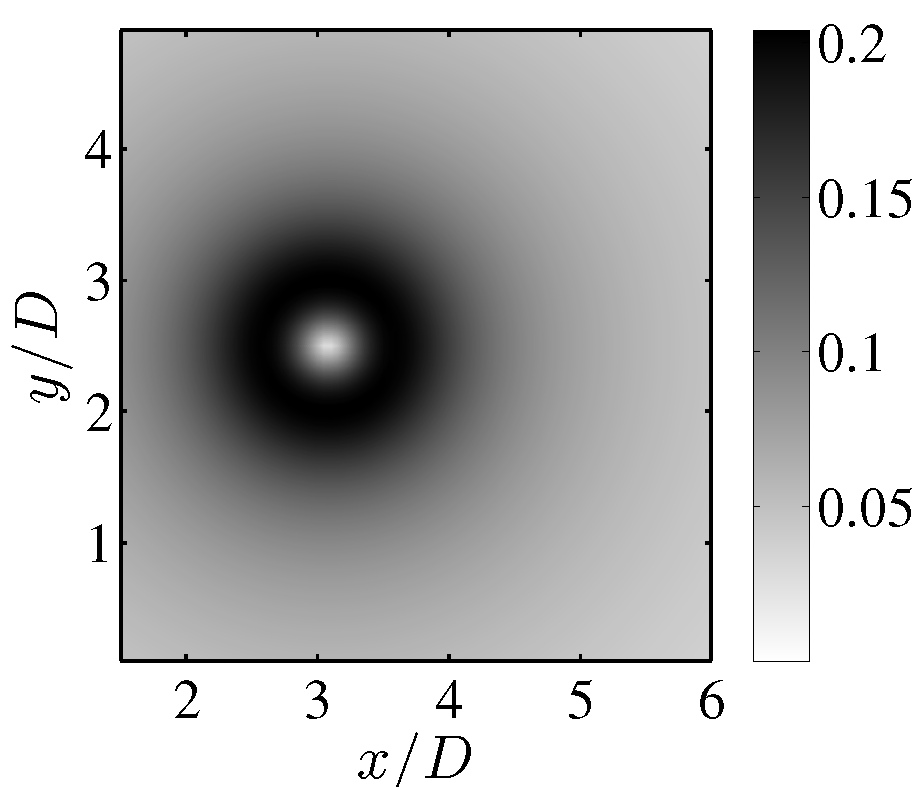}
         \caption{fullfield\\
         $d_{v}^{*}$= 0.01,\; $f_{v}^{*}$= 0.08}
         \label {fig:LO_fullfield}
      \end{subfigure}  
    \caption{First $\cal{C}$DMD mode reconstructed in physical space (a), fullfield vortex (b), relative error of the first $\cal{C}$DMD mode (c),
    and reconstruction of the first DMD mode (d) at time $t^*=12$. }
    \label{fig:lamb-Oseen_modes}
\end{figure}

Reconstruction of the first $\cal{C}$DMD and DMD modes at time $t^*=12$ show the drastic difference between decompositions along the two directions in comparison 
 with the full-filed (figures \ref{fig:LO_LDMD_mode_01}, \ref{fig:LO_fullfield} and \ref{fig:LO_DMD_Mode_01}). 
Only one mode along the characteristics suffices to capture the vortex with the mode eigenvalues having correctly detected the expected decay rate and frequency. 
This is while multitudes of DMD modes are needed along $t$ to reconstruct the vortex.
The relative error for the first $\cal{C}$DMD mode is depicted in figure \ref{fig:LO_LDMD_relError}.\medskip

\emph{Given the example above, the important step in applying the Characteristic DMD ($\cal{C}$DMD) is to
take the columns of  $X_{0..n}$ normal to the characteristics direction
and the rows along it}. The resulting structures are defined in planes normal to the
group velocity of the structure in space-time. That means they have no
immediate temporal or spatial interpretation. For a structure traveling from right to left for instance,
the values at the top of the snapshots correspond to a later time than those at the bottom. A backwards 
rotation in space and time will then result in the spatial representation of the structures.\medskip
 
 \subsection{Interpretation of periodic data in spatiotemporal space}
\begin{figure}
\centering
      \begin{subfigure}[b]{0.45\linewidth}
         \includegraphics[scale = 0.4]{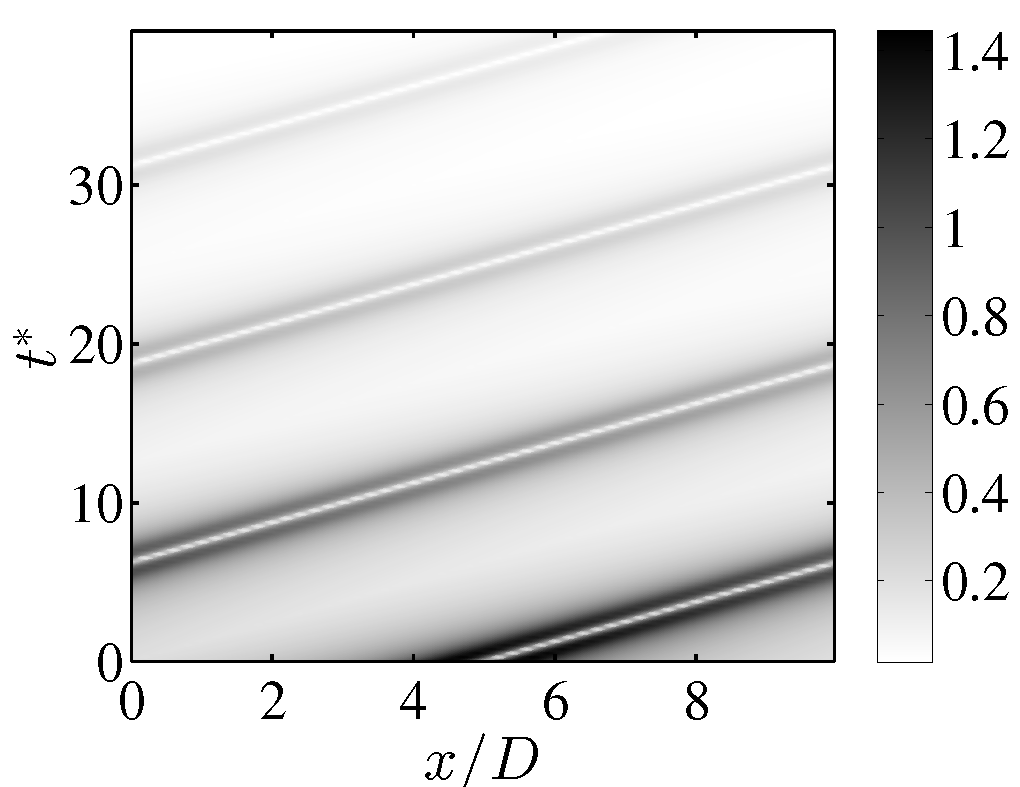}
         \caption{}
         \label {fig:lamb-Oseen_p_charDiag}
      \end{subfigure}
      \begin{subfigure}[b]{0.45\linewidth}
         \includegraphics[scale = 0.4]{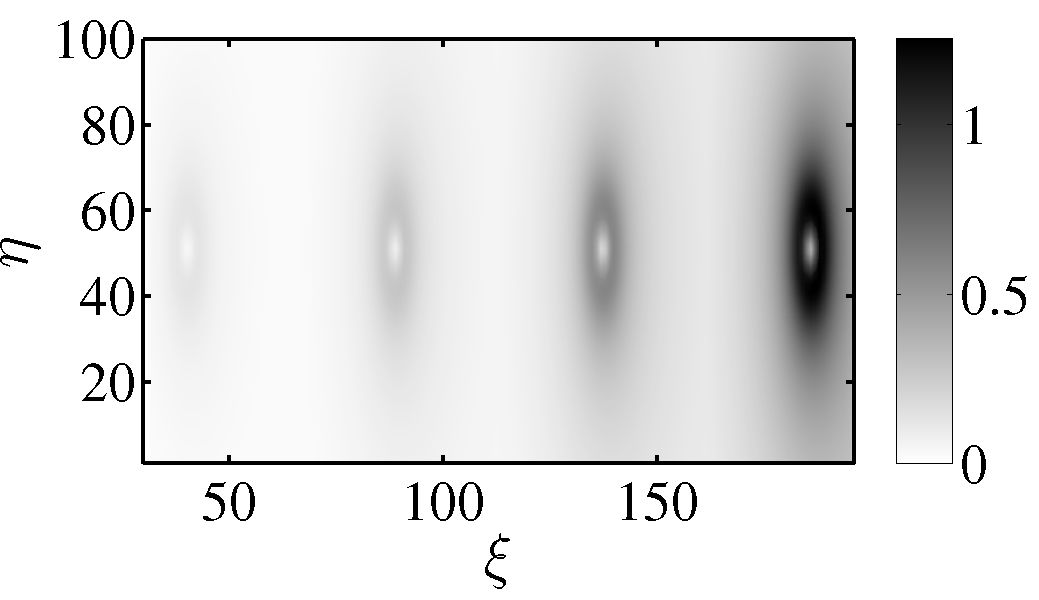}
         \caption{}
         \label {fig:lamb-Oseen_p_spTemp}
         \end{subfigure} 
    \caption{Characteristic diagram for a periodic Lamb-Ossen vortex head (a) and the spatiotemporal representation of the vortex (b).}
    \label{fig:lamb-Oseen_periodic}
\end{figure}

 In order to demonstrate how a set of data with periodicity in the translation direction, can be interpreted in spatiotemporal space, a periodic Lamb-Oseen vortex 
 is considered with the decay rate and frequency of $d^{*}=0.01$ and $f^{*}=0$ respectively. In this setup, the flow is considered to be
 periodic in the $x$ direction with the vortex traveling in space with the group velocity of $u^{*}_{g}=0.8$. Space time diagram is depicted in figure \ref{fig:lamb-Oseen_p_charDiag}
 where the gradual decay of the periodic vortex can be observed. \medskip
 
 \medskip

 By applying a rotation in space and time, with the rotation angle corresponding to the vortex group velocity, one can 
 look at the spatiotemporal representation of the dataset. Along with the transformation, the spatial periodicity is also 
 transformed to accommodate several instances of the same structure in each spatiotemporal set of data at at each point along the characteristics of the flow.
  Figure \ref{fig:lamb-Oseen_p_spTemp}, represents one of the spatiotemporal sets along $\tau$. \medskip
 
 As explained earlier, the structures here do not belong to a certain point in time or space.
 Rather, they carry spatial information from a range of physical timesteps. As it will be shown in chapter \ref{sec:StVsSp}, this can be regarded as one of the main reasons why a decomposition along the 
 characteristics will result in a faster drop of singular values, where the events can be described using fewer modes. This example serves only to show how a simple traveling structure can be interpreted in the spatiotemporal space, specially in presence of translational periodicity. 
 In chapter \ref{sec:StVsSp}, it will be analyzed how this approach compares to looking for the structures in a frame of reference which is shifted only in space with respect to the group velocity.\medskip

\section{Modal Analysis of a Starting Jet}
\label{sec:startingJet}

In this section the method explained above, is applied to existing three dimensional DNS of
a starting jet carried out by \citet{Fernandez2016}. The initial condition in the 
mentioned study is a tube-like shock, formed by a pressurized reservoir 
which discharges fluid through a nozzle into an open chamber with ambient pressure. The pressure
ratio of $p_1/p_2=3.4$ has been chosen to ensure that eventually a supersonic jet will
develop. The Reynolds number is approximately $Re=10^{4}$ based on the fully
expanded conditions. One crucial parameter besides the pressure ratio, is
the non-dimensional mass supply of the jet. It can be expressed as the
ratio of length to diameter of the pipe $L/D$. If this ratio is close to
unity, a vortex ring will form. In order to develop a trailing jet, the ratio of 
$L/D$ needs to be larger than $5$.\medskip

\begin{figure}
\centering
        \begin{subfigure}[b]{0.23\linewidth}
              \includegraphics[scale =0.5]{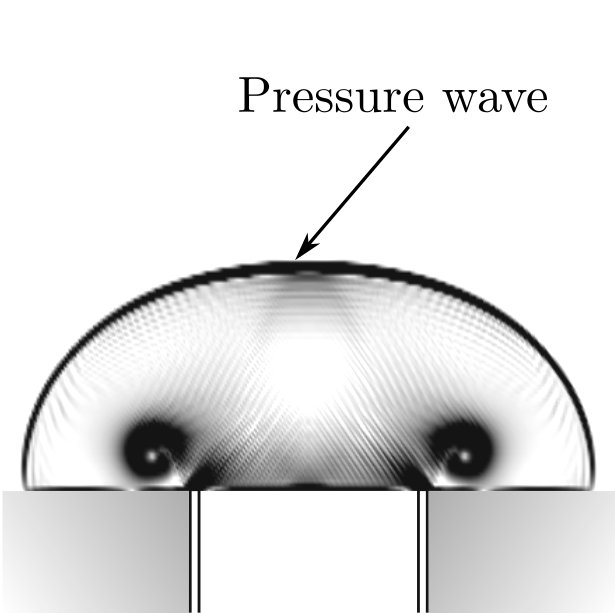}
              \caption{Pressure wave}
              \label {fig:JetShlieren_a}
        \end{subfigure}
        \begin{subfigure}[b]{0.23\linewidth}
              \includegraphics[scale =0.5]{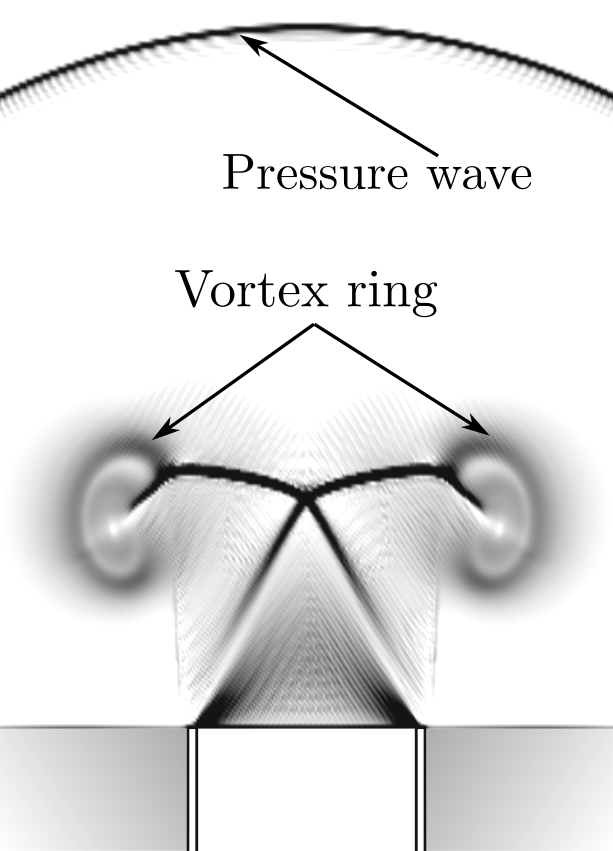}         
              \caption{Vortex ring}
              \label {fig:JetShlieren_b}
        \end{subfigure}
        \begin{subfigure}[b]{0.23\linewidth}
              \includegraphics[scale =0.5]{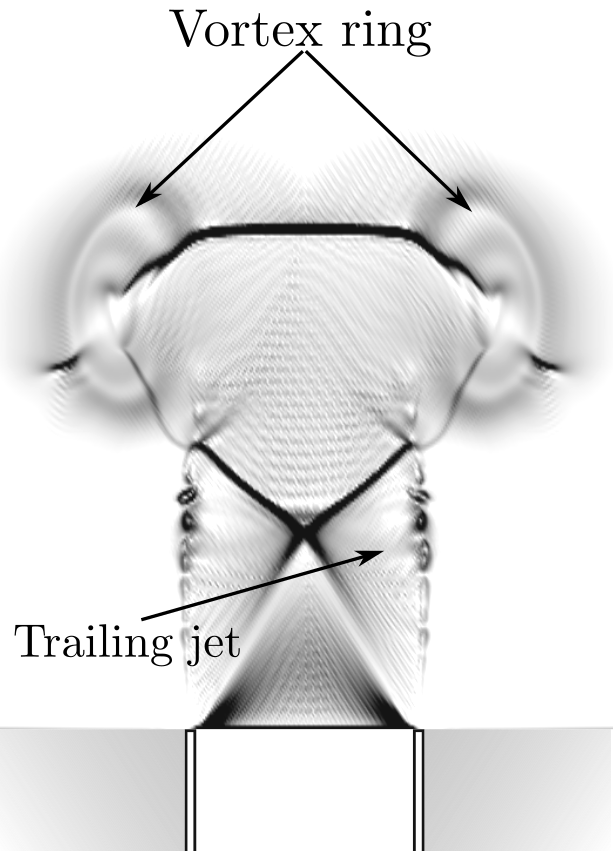}         
              \caption{Trainling jet}
              \label {fig:JetShlieren_c}
        \end{subfigure}
        \begin{subfigure}[b]{0.23\linewidth}
              \includegraphics[scale =0.5]{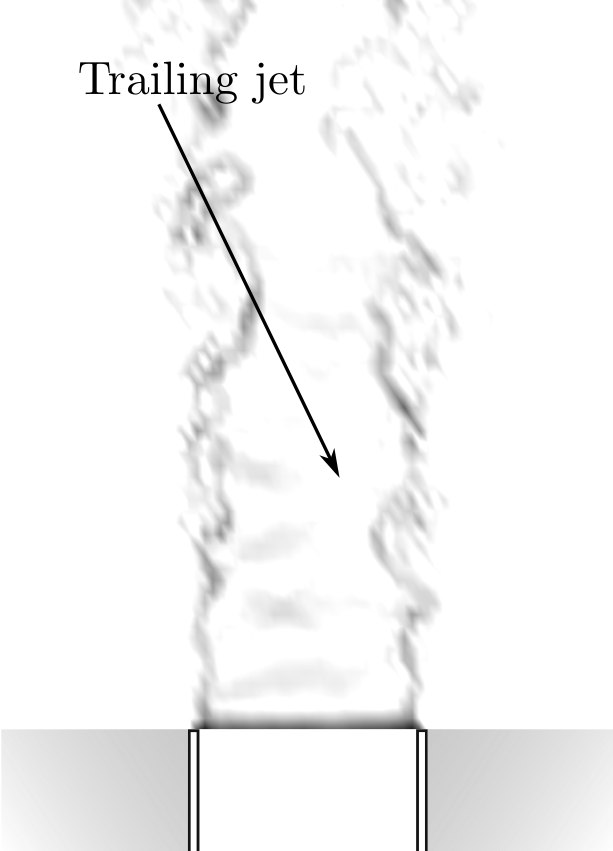} 
              \caption{Decay stage}
              \label {fig:JetShlieren_d}
        \end{subfigure}

    \caption{Pseudo-schlieren images of starting jet in time by \citet{Fernandez2016}.}
    \label{fig:StartingJetEvolution}
\end{figure} 

The temporal evolution is shown in figure
\ref{fig:StartingJetEvolution} as a pseudo-schlieren image in a two
dimensional cut through the jet and the vortex ring. Figure 
\ref{fig:JetShlieren_a} and \ref{fig:JetShlieren_b} show respectively the initial 
pressure wave and the developing vortex ring at the wall. If enough vorticity is generated, the
self--induced velocity of the vortex ring makes it accelerate and
travel in flow direction and slightly expand its diameter. Figure \ref{fig:JetShlieren_c} 
shows the vortex ring and the trailing jet, which is formed, if
enough mass is supplied. The last image on the right (figure \ref{fig:JetShlieren_d})
shows the full jet when the mass supply vanishes and the vortex ring has moved away. 
We wish to identify the flow for the case of a vortex ring with trailing jet.

\subsection {The Vortex Head with Trailing Edge} 
\label{sec:EarlyVortHead}

One of the dominant features of the starting jet is the vortex
ring. It is initially formed at the tube lip and detaches later to first move with constant velocity and finally travels with
a velocity decaying as square root of time. The aim in this chapter is to detect this
vortex and describe it with a few DMD modes. For this test case mass supply ratio of $L/D=10^{7}$ has been chosen so that formation of a vortex ring will be followed by trailing edge. For the results presented in this chapter, the decompositions were carried out on 2D cuts of the flow field. \medskip

\begin{figure}
    
    \begin{subfigure}[b]{0.5\linewidth}
        \centering
        \includegraphics[scale=0.4]{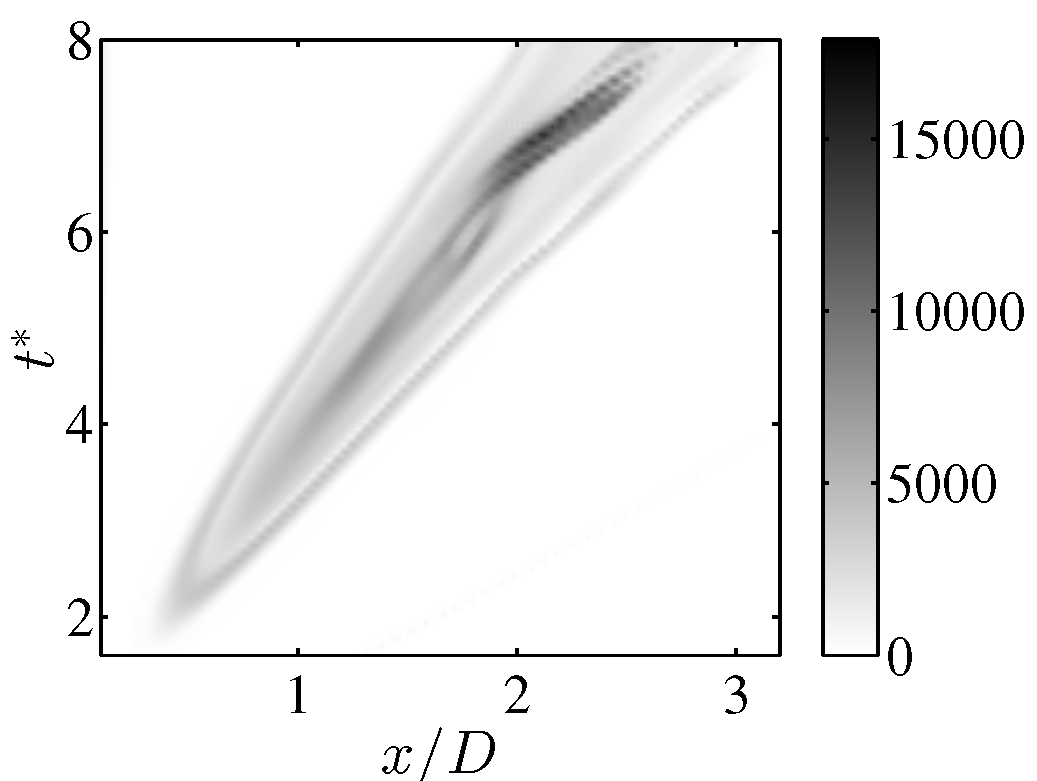}
        \caption{}
        \label{fig:EarlyVortex_CharDiag}
    \end{subfigure}
    \begin{subfigure}[b]{0.5\linewidth}
        \centering
        \includegraphics[scale=0.4]{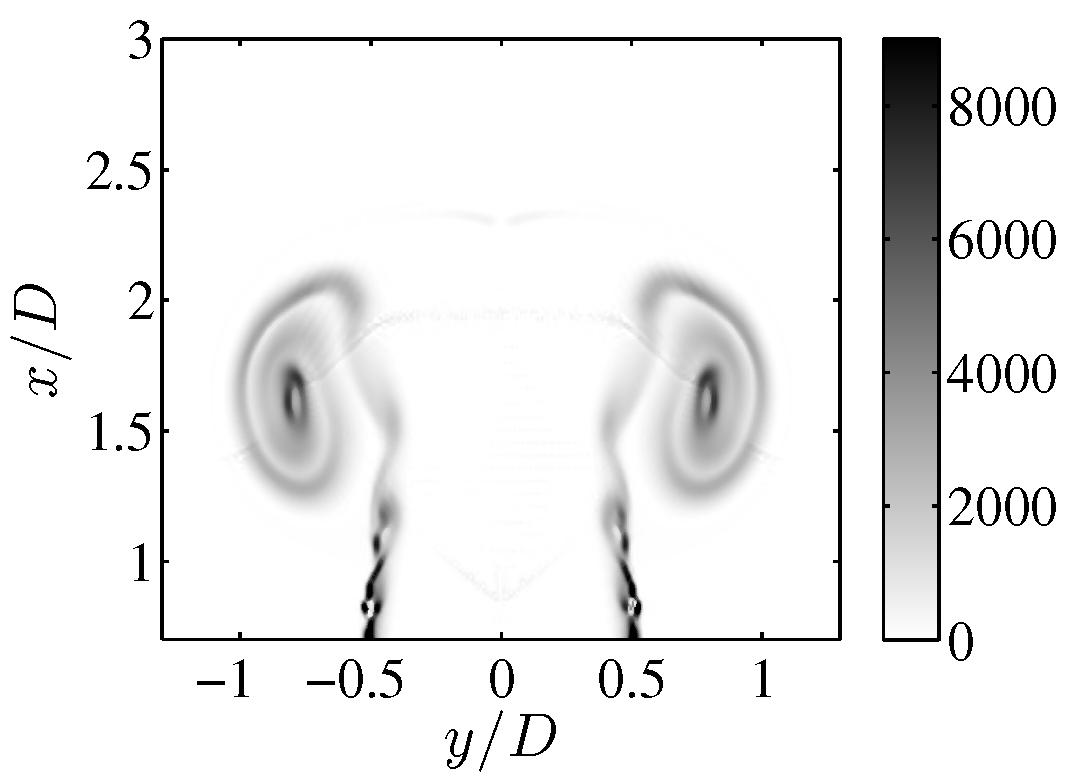}
        \caption{}
        \label{fig:Fullfield_first}
    \end{subfigure}      
       \caption{Characteristic diagram along the vortex centerline at $y/D=0.8$ (a) and the vortex head at time $t^{*} = 5.5$. Both figures show contour plots of vorticity magnitude.}
    \label{fig:Lam_vortHead}
\end{figure}

The characteristic diagram is given as space-time plot of the
vorticity magnitude along the center of the vortex head in figure \ref{fig:EarlyVortex_CharDiag}, showing the trace of the
vortex head traveling in space and time. In this figure $t^{*}=t/(D/U)$ is the dimensionless time while $U$ and $D$ being 
the characteristic velocity and the jet diameter respectively. The vortex head is shown in figure \ref{fig:Fullfield_first}  at time $t^{*}=5.5$.\medskip

\begin{figure}
    \begin{subfigure}[b]{0.5\linewidth}
        \centering
        \includegraphics[scale=0.4]{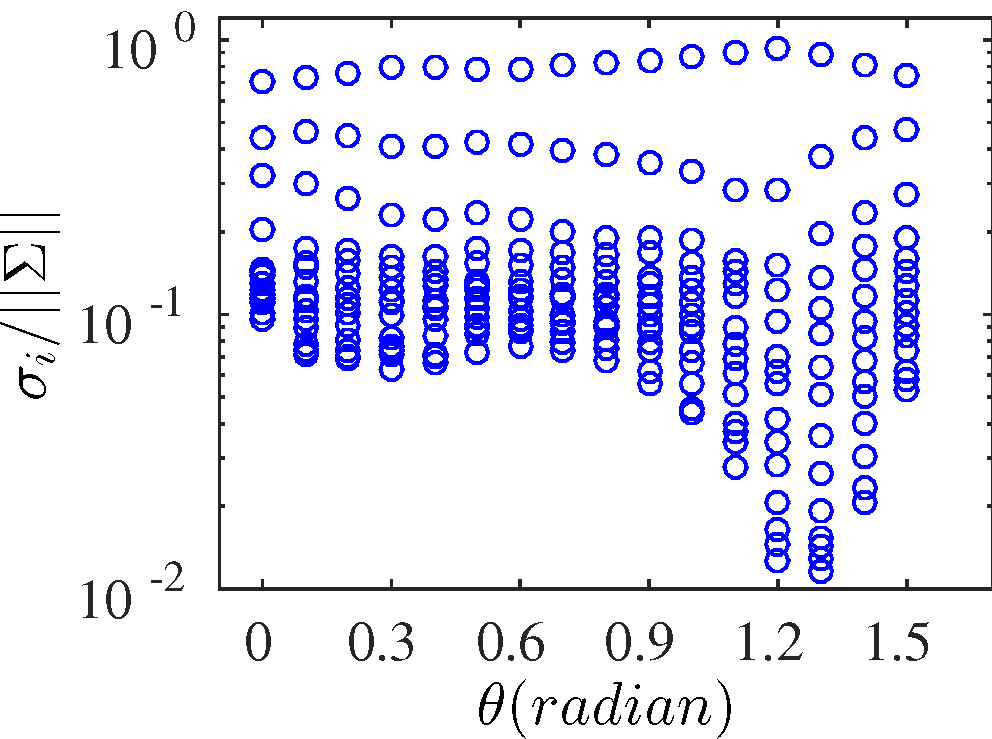}
        \caption{}
        \label{fig:dirSweep}
    \end{subfigure}
    \begin{subfigure}[b]{0.5\linewidth}
        \centering
        \includegraphics[scale=0.4]{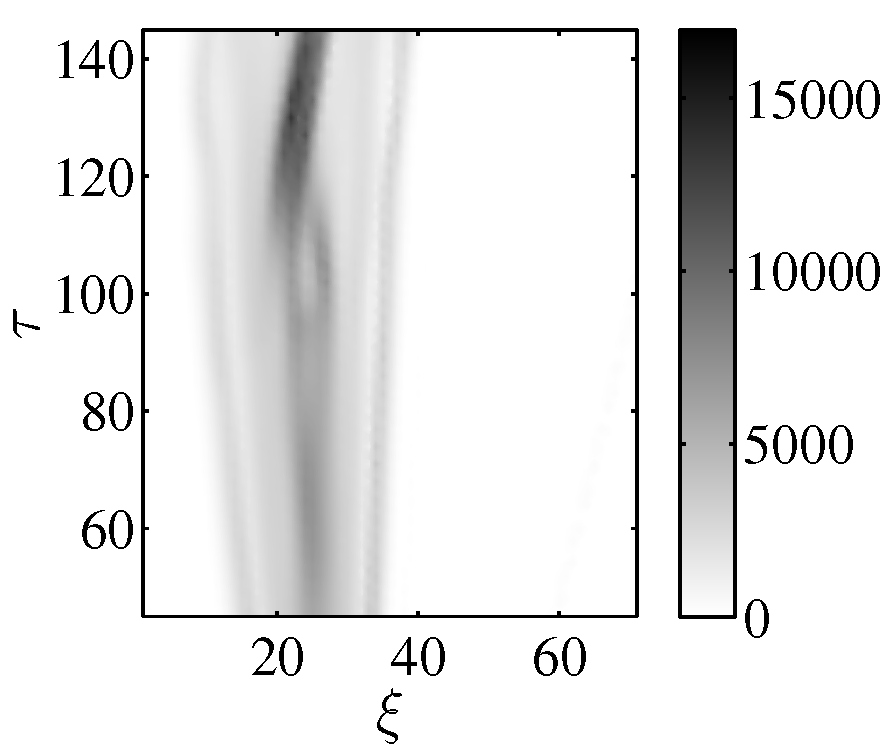}
        \caption{}
        \label{fig:EarlyVortex_ChaDiag_r}
    \end{subfigure} 
      
      \caption{Drop of the first 15 singular values for a range of rotation angles (a) and the characteristic diagram rotated with the optimal angle $\theta = 1.2 \; radian$, 
      corresponding to the group velocity of $u^{*}_{g} = 0.37$ (b).}
    \label{fig:dirSweep_CharDiag_r}
\end{figure}

To detect the optimal direction which also highlights the largest group velocity in the flow, 
a singular value decomposition is carried out for a range of rotation angles in $x-t$ space and the first 15 singular values 
are plotted against the rotation angle in figure \ref{fig:dirSweep}. The first SVD for $\theta = 0$ is performed on the unrotated snapshots matrix.
After that, the snapshots matrix is rotated counterclockwise, with the rotation angle being increased with increments of $\Delta \theta = 0.1 \;radian$.\medskip

It is clear in this figure that there is a faster drop for rotation angle $\theta = 1.2 \; radian$ which corresponds to the
dimensionless velocity of $u^{*}=u/U=0.37$, and can be 
understood as the most dominant group velocity $u^{*}_{g}$ in the space-time diagram.\medskip

In the next step, a coordinate transformation of the $xyzt$-space into the detected direction has to be performed.  
One can observe in figure \ref{fig:EarlyVortex_CharDiag} that the group velocity changes with time. Thus, 
to describe the vortex head with the minimum number of modes, also a temporal transformation
$t'=\frac{at}{b\sqrt{t}+c}$ with suitable coefficients should be employed. 
This complication is left for later\footnote{Transformation in time
can be achieved by choosing the snapshots equidistantly in $t'$ as
defined above. Since abundant time-steps are available from the existing numerical
simulation, this can be easily done by choosing the right snapshots. On the other hand, since the vortex head is expanding with time, 
via a more general approach proposed by \citet{Rowley2003}, also a scaling can be
employed. We refer the reader to that article for further information.} and for now the $xyzt$-cube of 
data is transformed to the spatiotemporal space treating the data as if the group velocity is constant in time.\medskip

The transformation procedure is performed as a decomposition of the rotation matrix into three
shears $q'=S_1S_2S_1q$ which is a fast and accurate algorithm \cite {Paeth1990}.
The result is a new $\xi\eta\zeta\tau$-cube in which a satisfyingly straight
part on the characteristic is chosen for modal analysis (figure \ref{fig:EarlyVortex_ChaDiag_r}).\medskip

In general, any rotation in four
dimensional space can be represented by two rotations in two suitably
chosen planes. In our example, where the main flow is in
$xt$-direction, a single rotation in the $xt$-plane suffices.\medskip


Having transformed the data to the spatiotemporal space (with $\alpha_{t}=2.64$), a DMD is carried out along $\tau$ 
to capture the spatiotemporal modes. To compare the results with those of a traditional DMD, another decomposition is 
performed along $t$ on the stationary frame of reference.
For this purpose, 65 snapshots were taken along $t$ on a domain size
of $4D$\texttimes$5D$ with the resolution of $326$\texttimes$600$
(in $x, y$ directions) and 100 snapshots were emplyed along $\tau$ with resolution of $70$ along $\xi$.\medskip

\begin{figure}[H]
     \begin{subfigure}[b]{0.5\linewidth}
         \centering
         \includegraphics[scale=0.4]{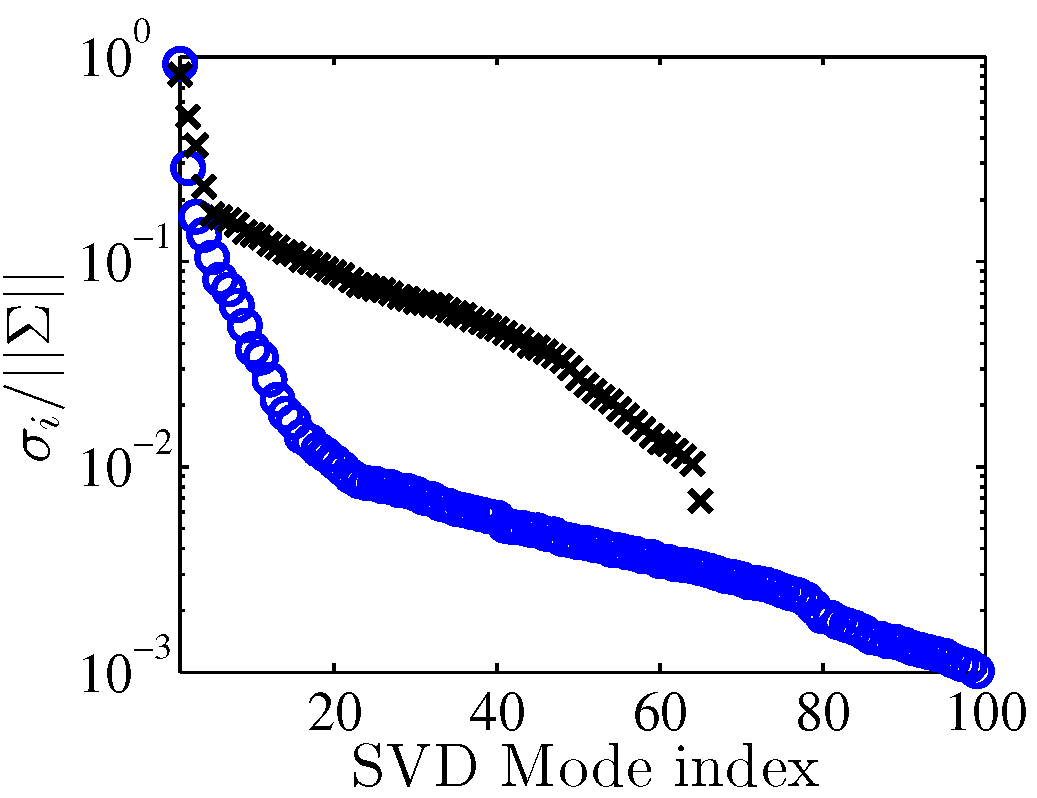}  
         \caption{}
         \label{fig:EarlyVort_SingVal}         
    \end{subfigure} 
    \begin{subfigure}[b]{0.5\linewidth}
         \centering
         \includegraphics[scale=0.4]{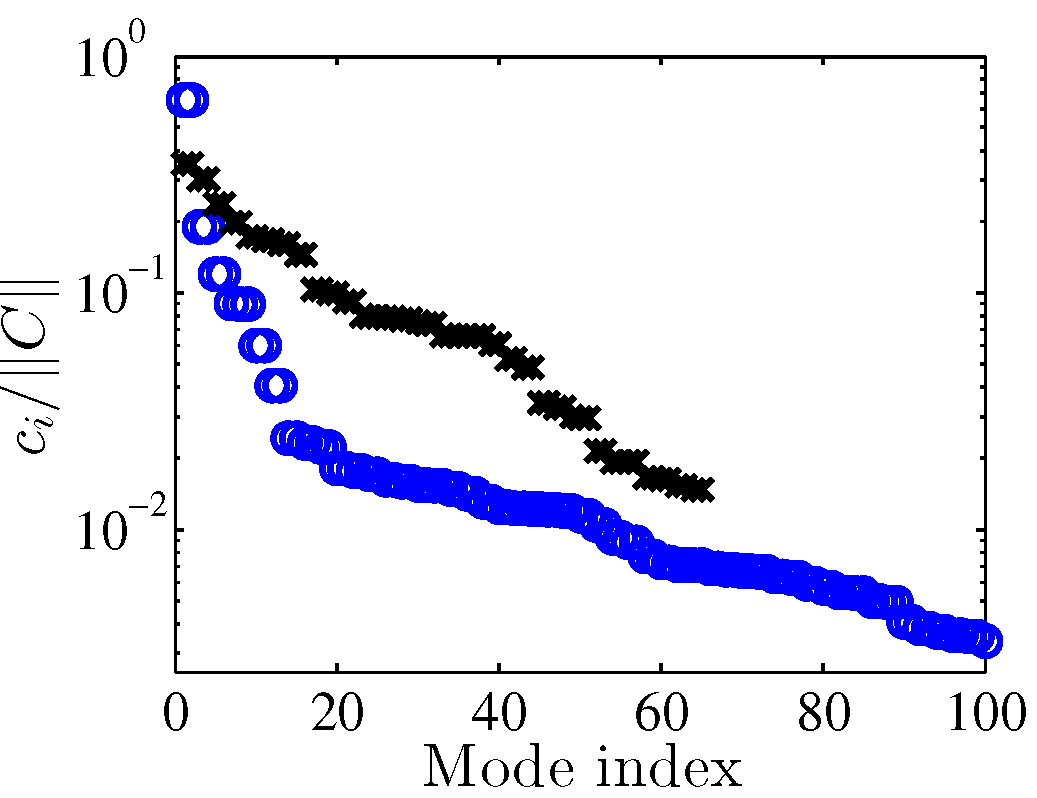}
         \caption{}
         \label{fig:EarlyVort_ModalDecay}         
    \end{subfigure}   
   \caption{Normalized singular values (a) and modal decay (b) for $\cal{C}$DMD ($\circ$) vs. DMD ($\times$).}
   \label{fig:EarlyVortHead_SV_MD}
 \end{figure}

The yielded modes in both frames are then sorted by their averaged amplitudes. The singular values and Averaged mode amplitudes are 
compared for both decompositions in figures \ref {fig:EarlyVort_SingVal} and \ref {fig:EarlyVort_ModalDecay}
, demonstrating a steeper decay and a faster drop in the rotated frame. It is observable that the first $4$ $\cal {C}$DMD modes represent the snapshots up to a relative remainder of 
less than $10^{-1}$.\medskip 
 
\begin{figure}
     \begin{subfigure}[b]{0.5\linewidth}
         \centering
         \includegraphics[scale=0.4]{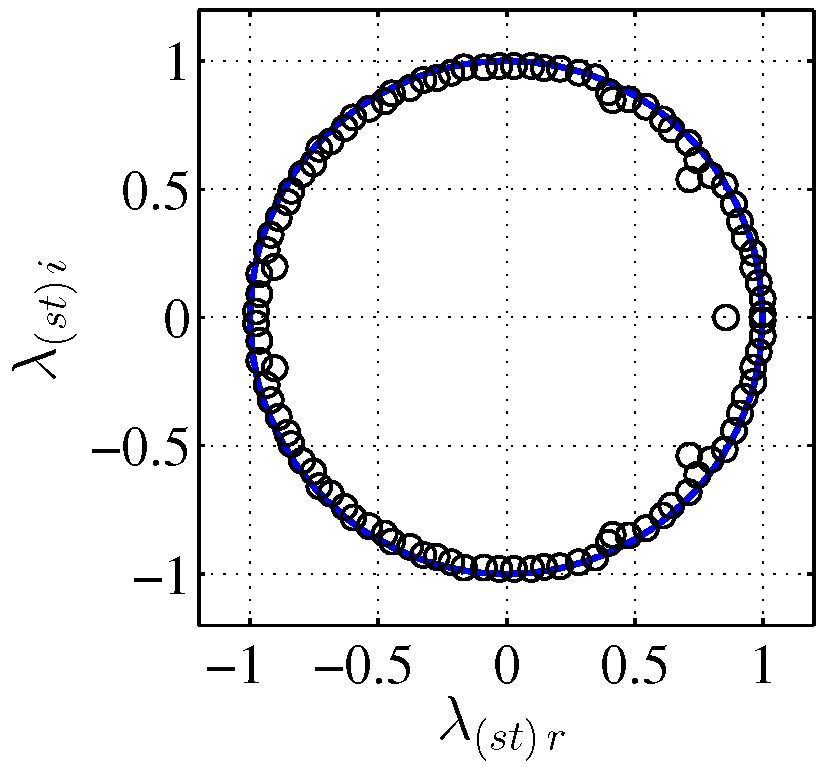} 
         \captionsetup{justification=centering}
         \caption{$\cal{C}$DMD spectrum \\
	  in spatiotemporal space}
         \label{fig:EarlyVort_EV_ST}         
    \end{subfigure} 
    \begin{subfigure}[b]{0.5\linewidth}
         \centering
         \includegraphics[scale=0.4]{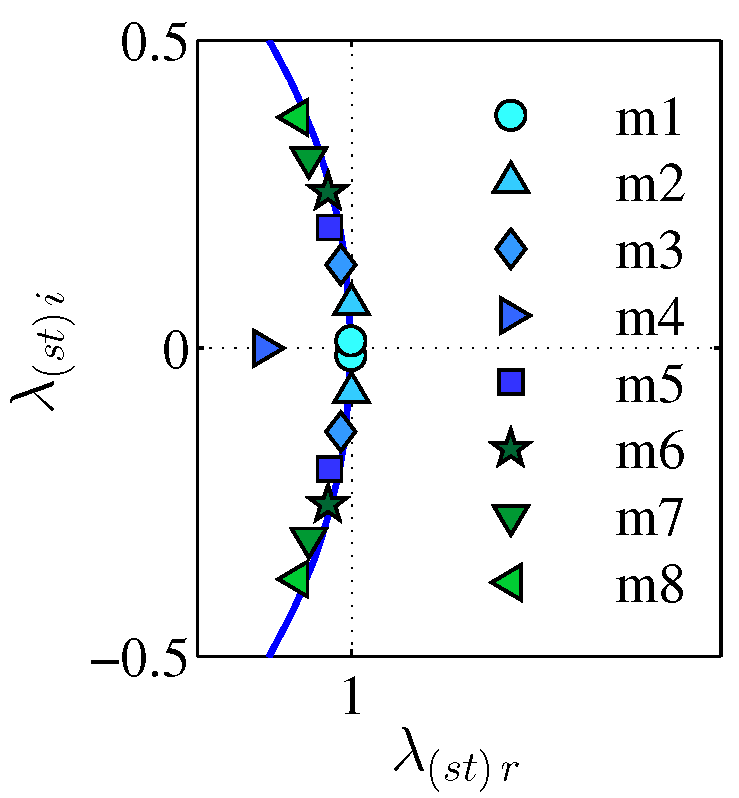}
         \captionsetup{justification=centering}
         \caption{First 8 $\cal{C}$DMD eigenvalues\\
         in spatiotemporal space}
         \label{fig:EarlyVort_EV_ST_Cut}         
    \end{subfigure} 
    
    \vspace{0.4 cm} 
    
    \begin{subfigure}[b]{0.5\linewidth}
         \centering
         \includegraphics[scale=0.4]{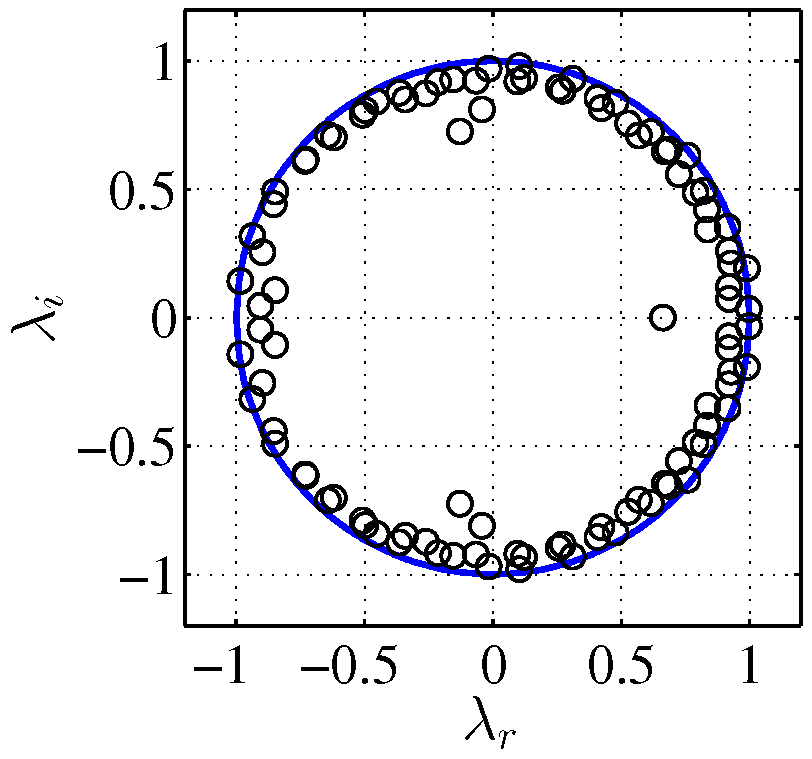}
         \captionsetup{justification=centering}
         \caption{$\cal{C}$DMD spectrum \\
	  in physical space}
         \label{fig:EarlyVort_EV_Ph}         
    \end{subfigure} 
    \begin{subfigure}[b]{0.5\linewidth}
         \centering
         \includegraphics[scale=0.4]{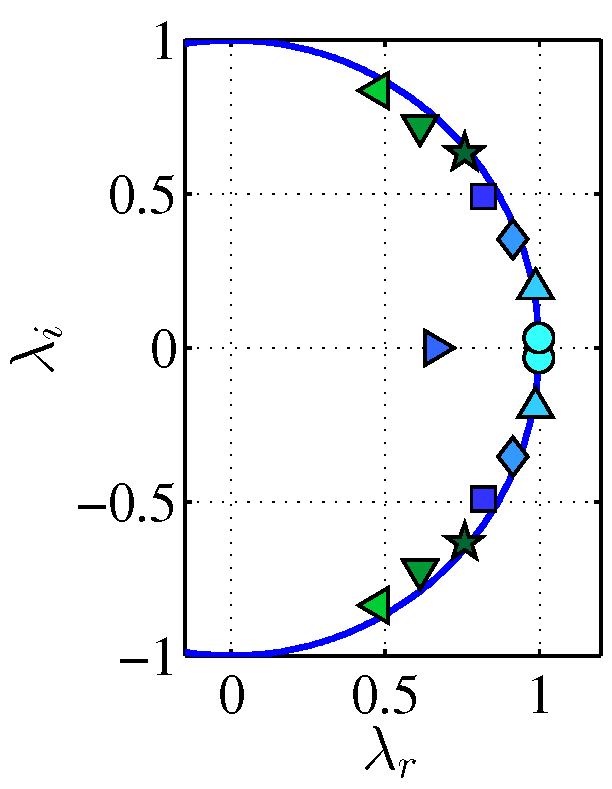}
         \captionsetup{justification=centering}
         \caption{First 8 $\cal{C}$DMD eigenvalues\\
         in physical space}
         \label{fig:EarlyVort_EV_Ph_Cut}         
    \end{subfigure} 
    
    \vspace{0.4 cm} 
    
    \begin{subfigure}[b]{0.5\linewidth}
         \centering
         \includegraphics[scale=0.4]{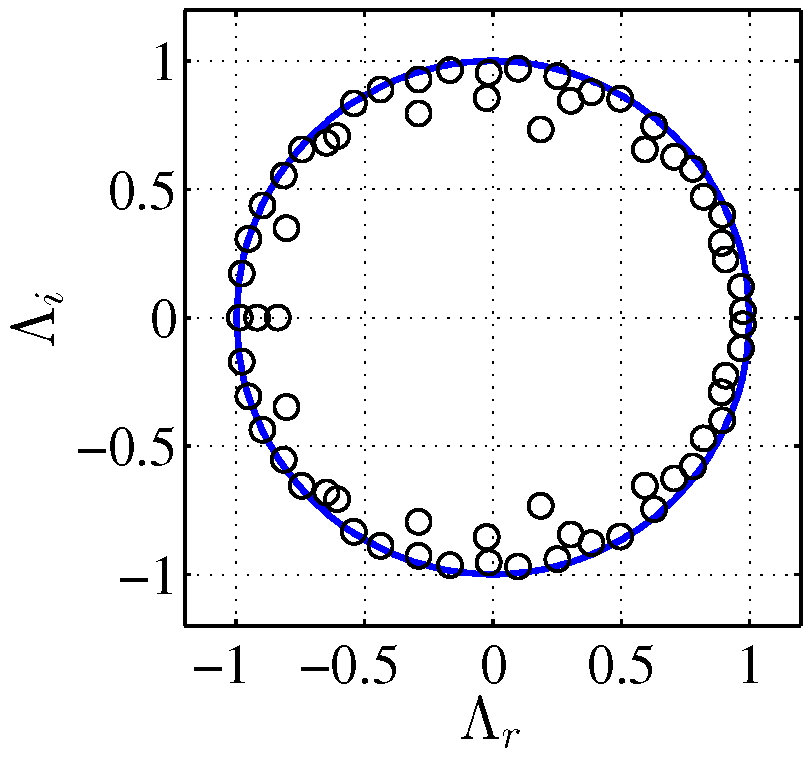} 
         \captionsetup{justification=centering}
         \caption{DMD spectrum}
         \label{fig:EarlyVort_DMD_EV}         
    \end{subfigure} 
    \begin{subfigure}[b]{0.5\linewidth}
         \centering
         \includegraphics[scale=0.4]{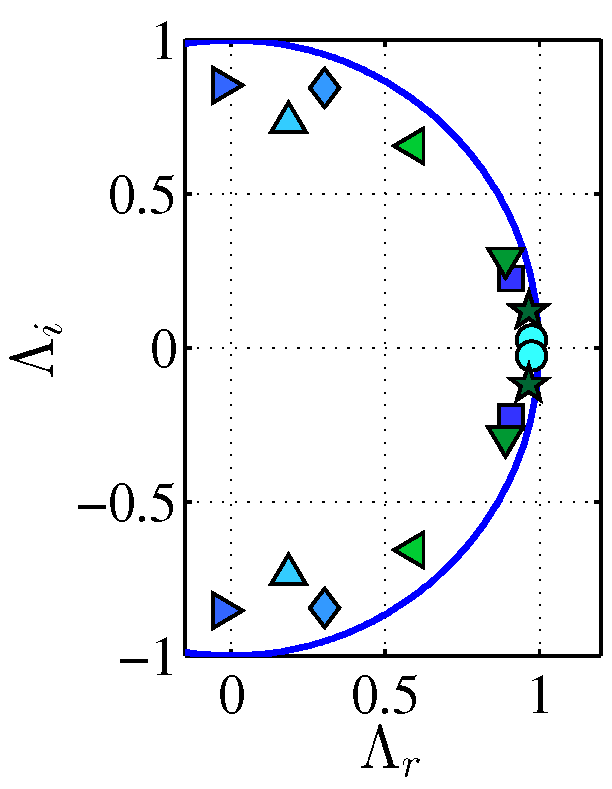}
         \captionsetup{justification=centering}
         \caption{First 8 DMD eigenvalues}
         \label{fig:EarlyVort_DMD_Cut}         
    \end{subfigure} 
   
   \caption{Figures on the left column represent respectively $\cal L$DMD spectrum in spatiotemporal space $\lambda_{st}$ (a),
   $\cal L$DMD spectrum in physical space $\lambda$ (c) and DMD spectrum $\Lambda$ (e). The figures on the right column
   depict the first 8 eigenvalues in the corresponding figure on the left.}
   \label{fig:EarlyVortHead_EV_unity}
 \end{figure}

The eigenvalues which are resulted from the decomposition along $\tau$ are transformed using 
equation \ref{eq:EV_transform} to the physical space. It can be seen in figures \ref{fig:EarlyVort_EV_ST} and \ref{fig:EarlyVort_EV_Ph}
that the eigenvalues resulted in the rotated frame are lying mostly on the unity circle. 
As expected, by being transformed back, they tend towards inside the circle,
signaling a faster decay rate along physical time. \medskip 

\begin{figure}
     \begin{subfigure}[b]{0.5\linewidth}
         \centering
         \includegraphics[scale=0.4]{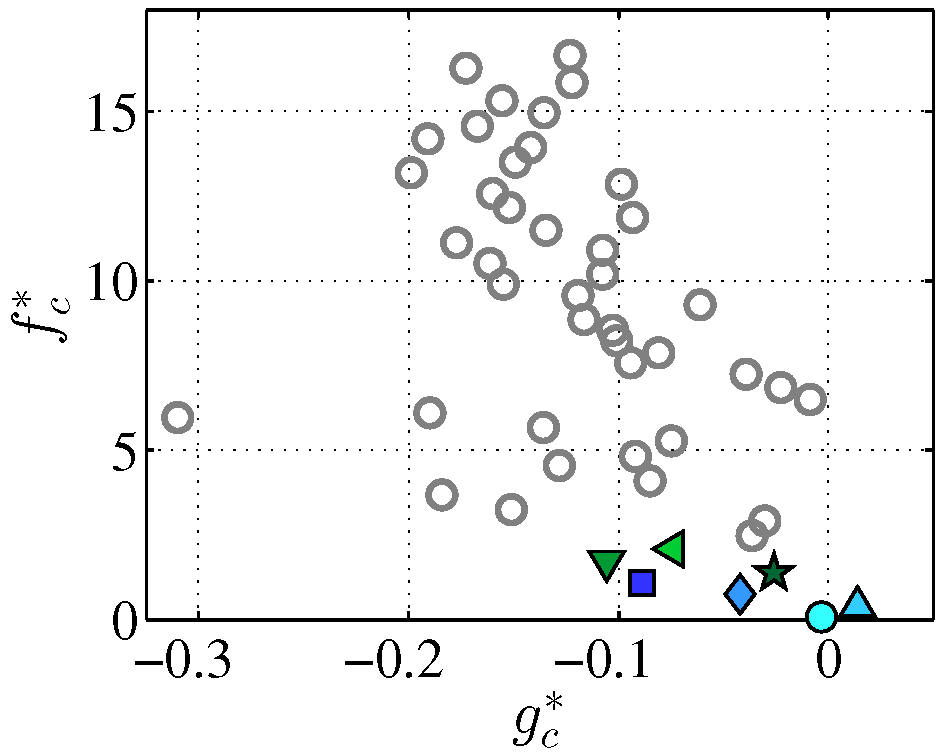} 
         \captionsetup{justification=centering}
         \caption{Frequencies and growth rates\\
         of $\cal{C}$DMD modes in physical space}
         \label{fig:EarlyVort_EV_Ph_dg}         
    \end{subfigure} 
    \begin{subfigure}[b]{0.5\linewidth}
         \centering
         \includegraphics[scale=0.4]{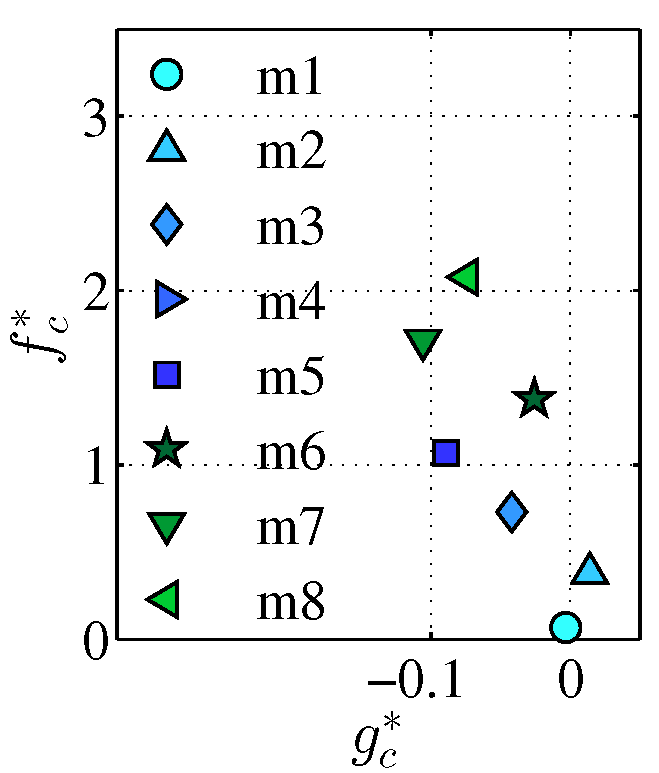}
         \captionsetup{justification=centering}
         \caption{Frequencies and growth rates of \\
         the first 8 $\cal{C}$DMD modes in physical space}
         \label{fig:EarlyVort_EV_Ph_dg_Cut}         
    \end{subfigure} 
    
    \vspace{0.4 cm} 
    
    \begin{subfigure}[b]{0.5\linewidth}
         \centering
         \includegraphics[scale=0.4]{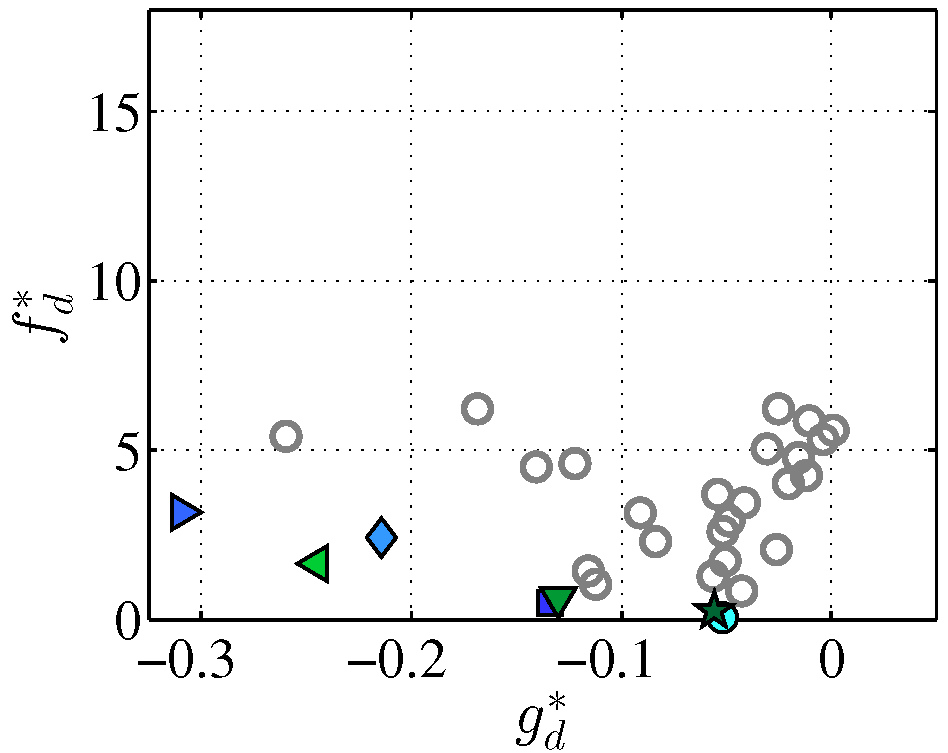}
         \captionsetup{justification=centering}
         \caption{Frequencies and growth rates\\
         of DMD modes}
         \label{fig:EarlyVort_DMD_EV_dg}         
    \end{subfigure} 
    \begin{subfigure}[b]{0.5\linewidth}
         \centering
         \includegraphics[scale=0.4]{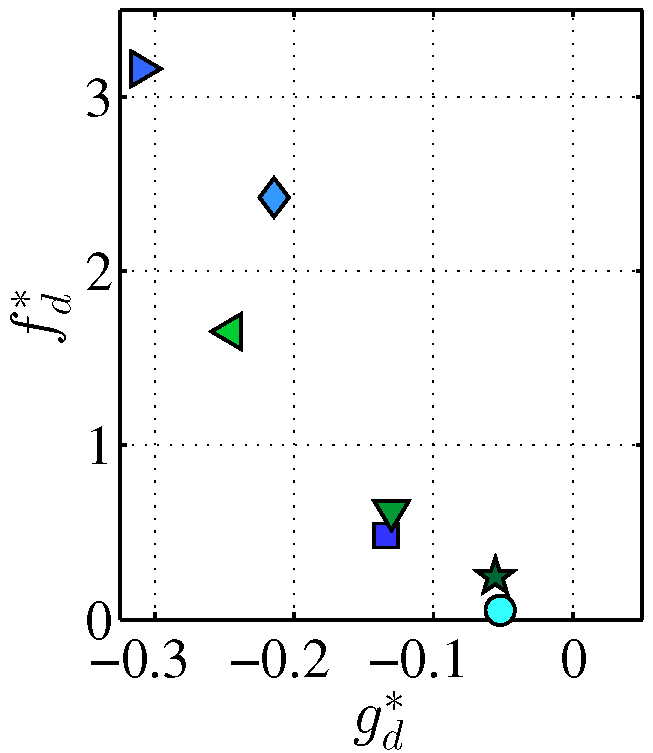}
         \captionsetup{justification=centering}
         \caption{Frequencies and growth rates\\
         of the first 8 DMD modes}
         \label{fig:EarlyVort_DMD_EV_dg_cut}         
    \end{subfigure} 
   \caption{Dimensionless frequencies and growth rates for $\cal{C}$DMD (a,b) and DMD modes (c,d). 
   (indices $c$ and $d$ refer to $\cal{C}$DMD and DMD respectively).}
 \label{fig:EarlyVortHead_EV_gdf}
 \end{figure}

The second effect of this transformation can be noted on the frequency of the modes.
The first 8 eigenvalues which are plotted separately in figures \ref{fig:EarlyVort_EV_ST_Cut} and \ref{fig:EarlyVort_EV_Ph_Cut}, 
cover a wider frequency range in the physical space.
On the other hand, the DMD on the stationary frame, has captured modes with larger decay rates (figures \ref{fig:EarlyVort_DMD_EV} 
and \ref{fig:EarlyVort_DMD_Cut}). The frequency and decay rate of the modes can be studied more clearly in figure
\ref{fig:EarlyVortHead_EV_gdf}.\medskip 

\begin{figure}
     \begin{subfigure}[b]{0.5\linewidth}
         \centering
         \includegraphics[scale=0.4]{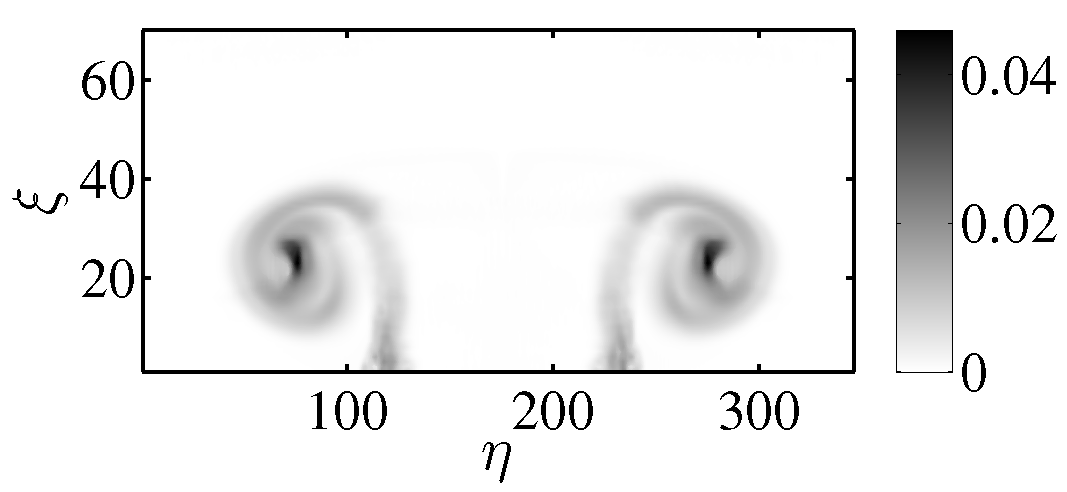} 
         \captionsetup{justification=centering}
         \caption{$\cal{C}$DMD, m=1\\
         $\lambda_{(st)}=0.99\pm0.012i$\\ 
         $c^{*}=0.93$}
         \label{fig:EV_LDMD_st_Mode_01}         
    \end{subfigure} 
    \begin{subfigure}[b]{0.5\linewidth}
         \centering
         \includegraphics[scale=0.4]{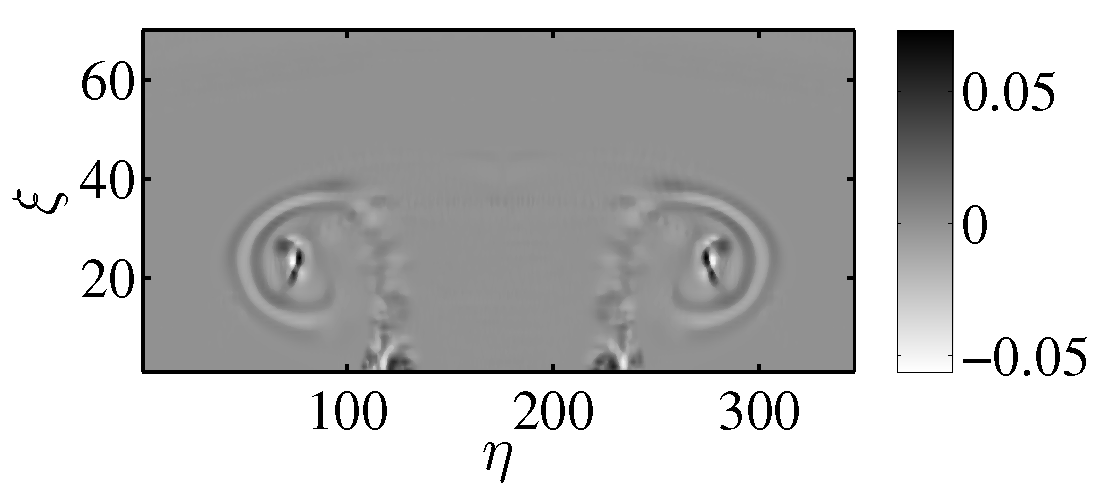}
         \captionsetup{justification=centering}
         \caption{$\cal{C}$DMD, m=2\\
         $\lambda_{(st)}=1\pm0.07i$\\
         $c^{*}=0.27$}
         \label{fig:EV_LDMD_st_Mode_02}
     \end{subfigure}
      
      \vspace{0.4 cm}
      
     \begin{subfigure}[b]{0.5\linewidth}
         \centering
         \includegraphics[scale=0.4]{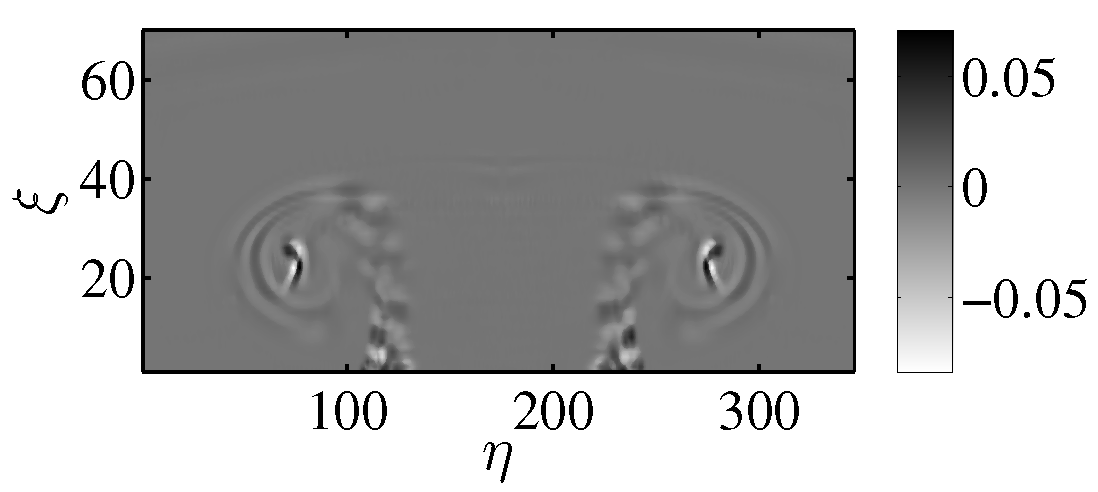} 
         \captionsetup{justification=centering}
         \caption{$\cal{C}$DMD, m=3\\
         $\lambda_{(st)}=0.98\pm0.013i$\\
         $c^{*}=0.17$}
         \label{fig:EV_LDMD_st_Mode_03}         
    \end{subfigure} 
    \begin{subfigure}[b]{0.5\linewidth}
         \centering
         \includegraphics[scale=0.4]{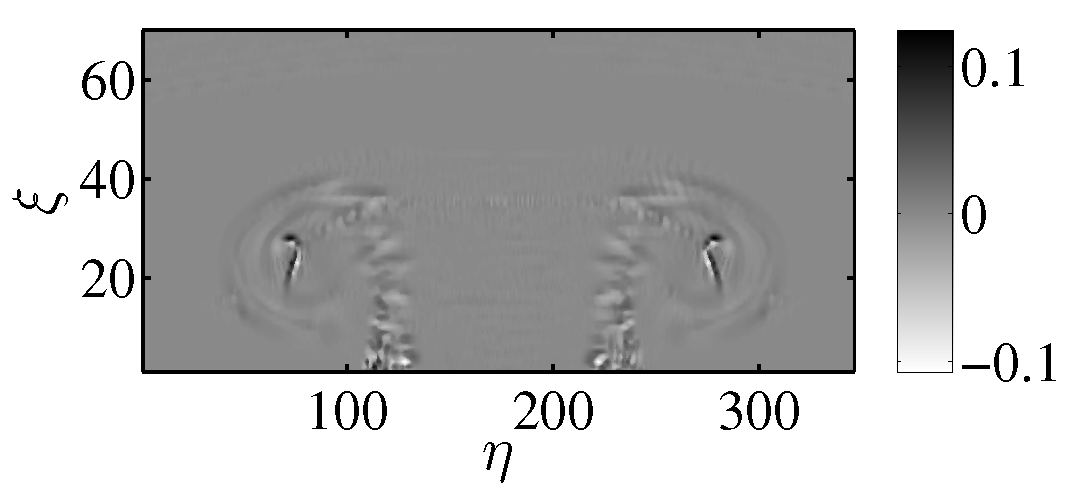}
         \captionsetup{justification=centering}
         \caption{$\cal{C}$DMD, m=4\\
         $\lambda_{(st)}=0.85$\\
         $c^{*}=0.1$}
         \label{fig:EV_LDMD_st_Mode_04}   
    \end{subfigure} 
   \caption{First $4$ $\cal{C}$DMD modes in spatiotemporal space ($c^{*}=<|c_{i}|/\|C\|>_{\tau}$) (the real parts of the modes are shown).}
 \label{fig:EarlyVortHead_LDMD_Modes}
 \end{figure}

The first $\cal{C}$DMD mode is captured with a very small decay rate and frequency of $d^{*}_{c}=0.003$ and $f^{*}_{c}=0.06$ implying a rather
invariant development in space and time. This mode, having the highest amplitude and lowest decay rate, is shown in 
figure \ref{fig:EV_LDMD_st_Mode_01}
in spatiotemporal space, and is regarded as one of the suitable candidates for reconstruction of the vortex head.
The bounds of the vortex are clearly detected without being smeared or bearing traces of preceding or 
following timesteps. \medskip

The second mode, contrary to all the other ones, has a small growth rate. 
This mode is also shown in figure \ref{fig:EV_LDMD_st_Mode_02} in spatiotemporal space. While the third (figure \ref{fig:EV_LDMD_st_Mode_03}) and the 
fifth modes have small decay rates, the fourth mode which is lying far inside the unity circle (figure \ref{fig:EarlyVort_EV_Ph_Cut}) with
a real eigenvalue, possesses a very large decay rate of $d_{c}^{*}=0.82$ in physical space. Since one of the aims of this chapter 
is to use a few modes to describe parts of the flow that is neither decaying nor growing too fast, therefore this mode (figure \ref{fig:EV_LDMD_st_Mode_04}) 
will not be considered as a candidate for reconstruction of the vortex head. It is rather a short lived secondary phenomenon living on that structure.\medskip

\begin{figure}[H] 
     \begin{subfigure}[b]{0.5\linewidth}
         \centering
         \includegraphics[scale=0.4]{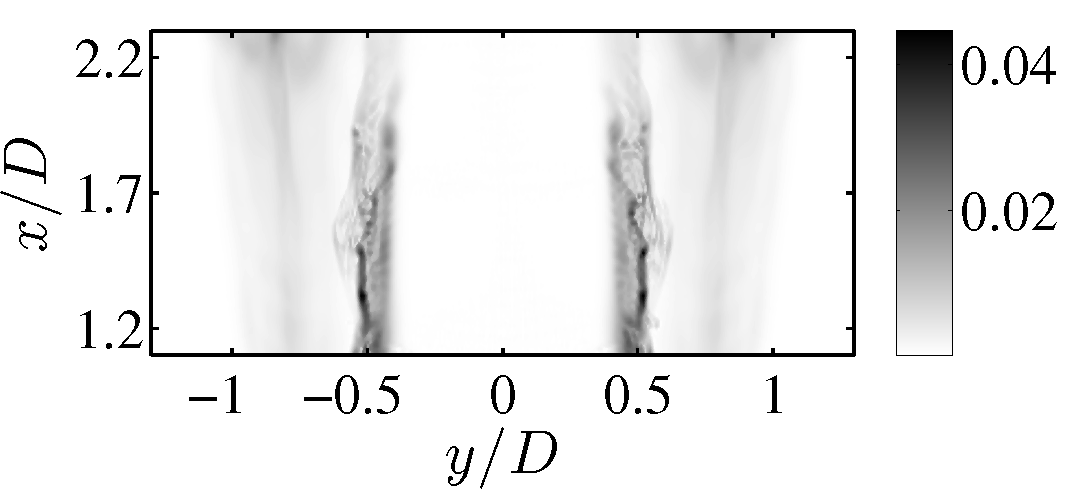} 
         \captionsetup{justification=centering}
         \caption{DMD, m=1\\
         $\Lambda=0.97\pm0.025i$\\
         $c^{*}=0.51$}
         \label{fig:EV_DMD_st_Mode_01}         
    \end{subfigure} 
    \begin{subfigure}[b]{0.5\linewidth}
         \centering
         \includegraphics[scale=0.4]{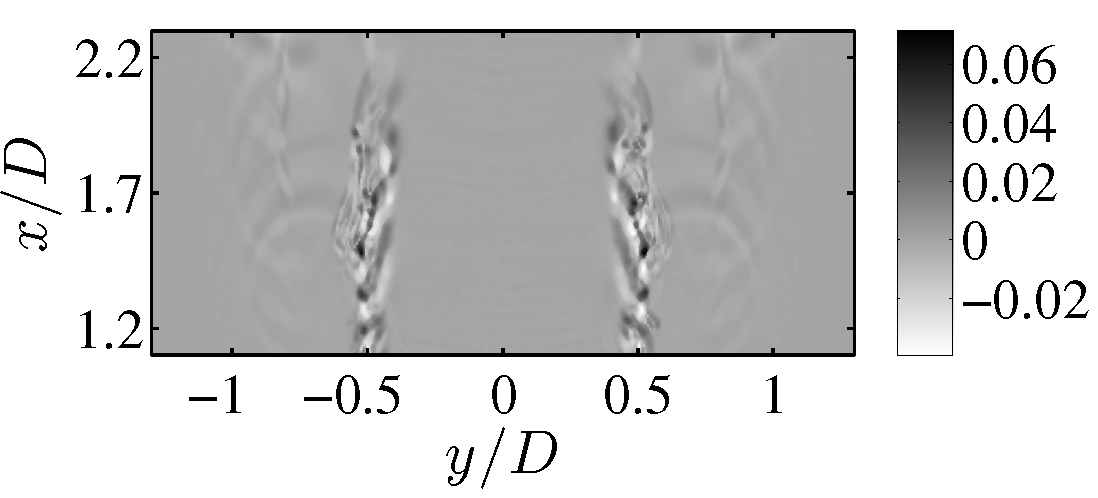}
         \captionsetup{justification=centering}
         \caption{DMD, m=2\\
         $\Lambda=0.18\pm0.73i$\\
         $c^{*}=0.42$}
         \label{fig:EV_DMD_st_Mode_02}
     \end{subfigure}
   \caption{First $2$ DMD modes ($c^{*}=<|c_{i}|/\|C\|>_{t}$) (The real parts of the modes are shown).}
 \label{fig:EarlyVortHead_DMD_Modes}
 \end{figure}

For the decomposition in the stationary frame, the first mode appears with a strongly smeared picture of the vortex head and with the
decay rate of $d^{*}_{d}=0.52$ which is rather large for the first mode compared with that of the first $\cal{C}$DMD mode. This mode has 
captured dominant traces of the shear layer as it will be also demonstrated later for the its reconstruction.
The second, third and fourth modes, have relatively higher decay rates of $d_{d}^{*}=0.5, \,0.2$ and $0.3$ respectively. 
Therefore, given the discussion above, these modes are not selected for the vortex head reconstruction either.
The first two DMD modes are shown in figure \ref{fig:EarlyVortHead_DMD_Modes}.\medskip

\begin{figure}[H]
        \begin{subfigure}[b]{0.5\linewidth}
                \centering
                \includegraphics[scale=0.4]{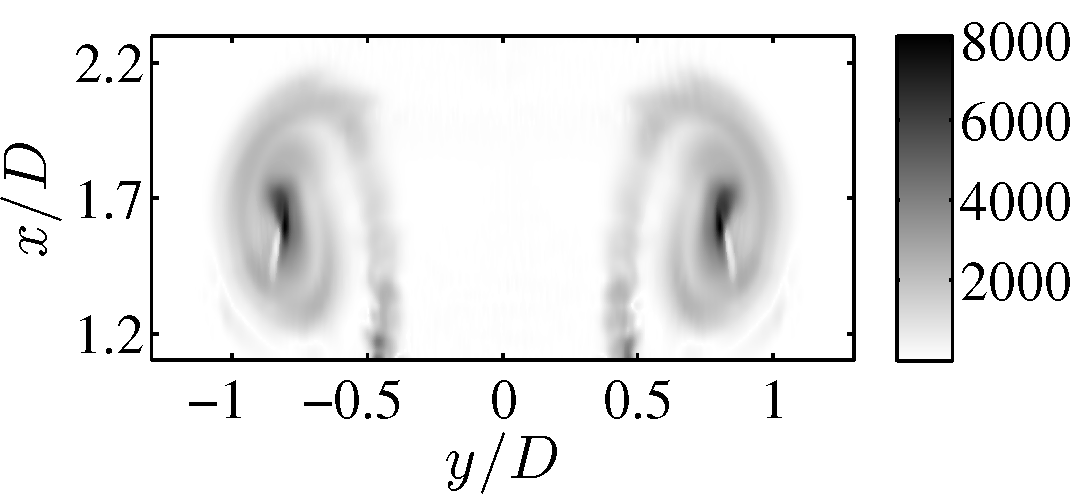} 
                \captionsetup{justification=centering}
                \caption{$\cal{C}$DMD, m=1\\
                $d^{*}_{c}=0.003,\;f^{*}_{c}=0.06$}
                \label{fig:EV_LDMD_recon_01} 
        \end{subfigure}
        \begin{subfigure}[b]{0.5\linewidth}
                \centering
                \includegraphics[scale=0.4]{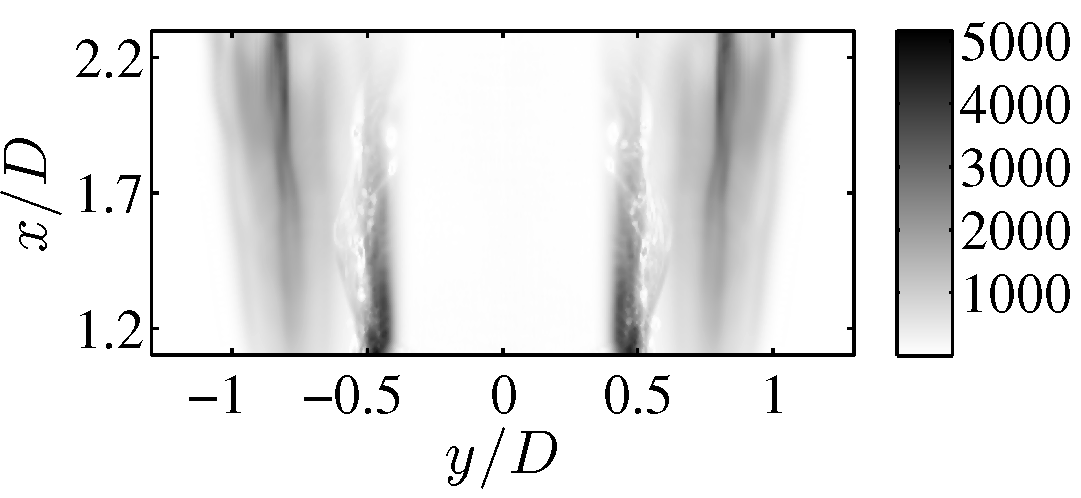} 
                \caption{DMD, m=1\\
                $d^{*}_{d}=0.05,\;f^{*}_{d}=0.05$}
                \label{fig:EV_DMD_recon_01} 
        \end{subfigure}
        
        \vspace{0.4 cm}    
        
        \begin{subfigure}[b]{0.5\linewidth}
                \centering
                \includegraphics[scale=0.4]{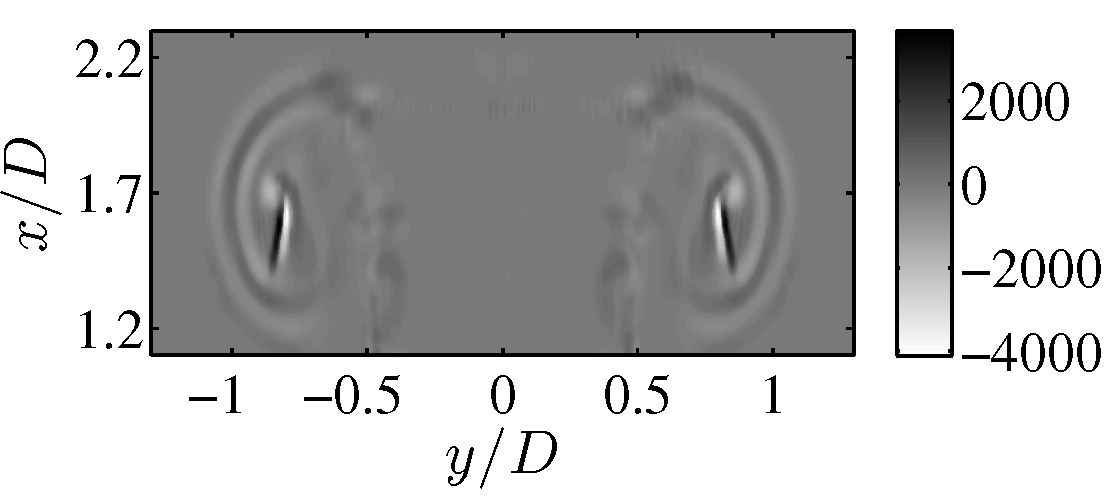} 
                \caption{$\cal{C}$DMD, m=2\\
                $d^{*}_{c}=0.01,\;f^{*}_{c}=0.3$}
                \label{fig:EV_LDMD_recon_02} 
        \end{subfigure}
        \begin{subfigure}[b]{0.5\linewidth}
                \centering
                \includegraphics[scale=0.4]{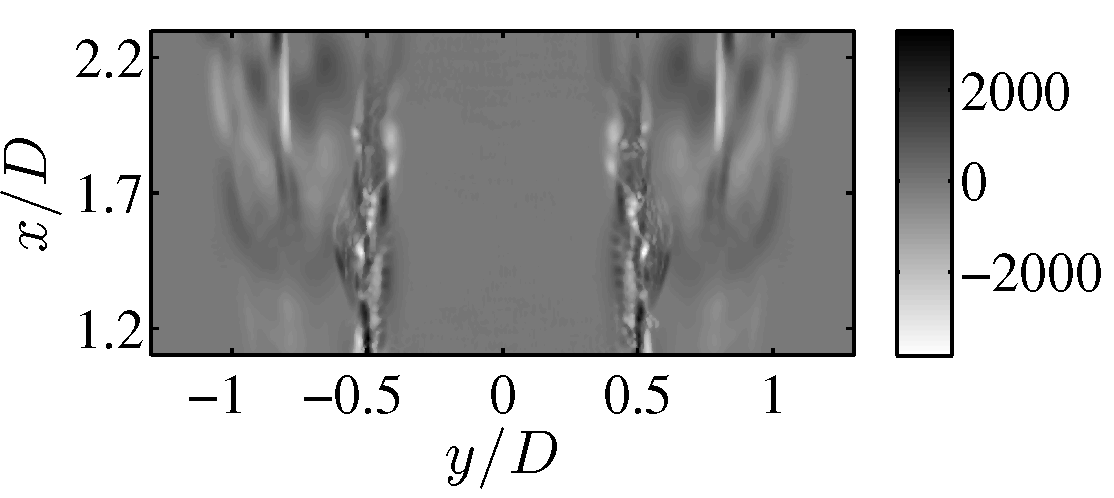} 
                \caption{DMD, m=5\\
                $d^{*}_{d}=0.1,\;f^{*}_{d}=0.4$}
                \label{fig:EV_DMD_recon_05} 
        \end{subfigure}
        
        \vspace{0.4 cm}
        
        \begin{subfigure}[b]{0.5\linewidth}
                \centering
                \includegraphics[scale=0.4]{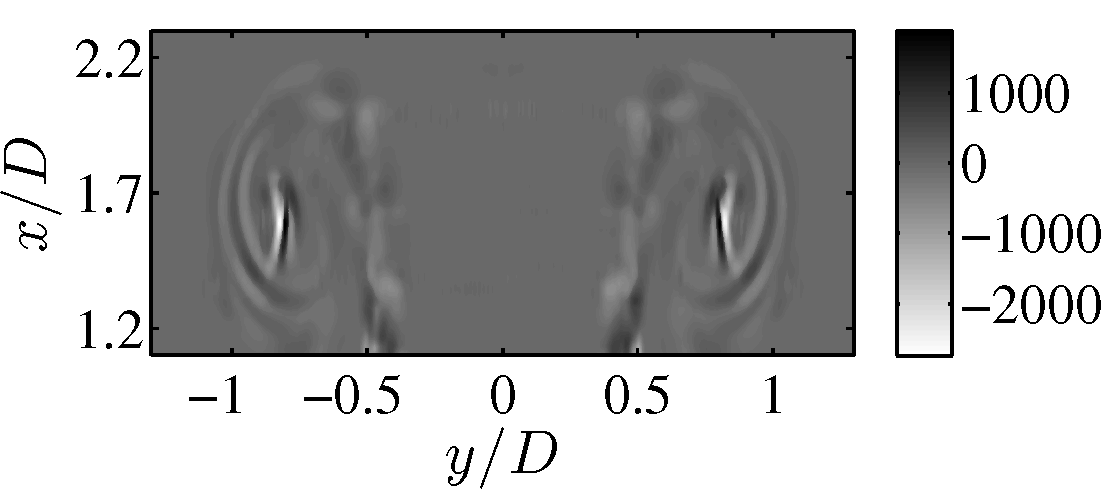} 
                \caption{$\cal{C}$DMD, m= 3\\
                $d^{*}_{c}=0.04,\;f^{*}_{c}=0.7$}
                \label{fig:EV_LDMD_recon_03} 
        \end{subfigure}
        \begin{subfigure}[b]{0.5\linewidth}
                \centering
                \includegraphics[scale=0.4]{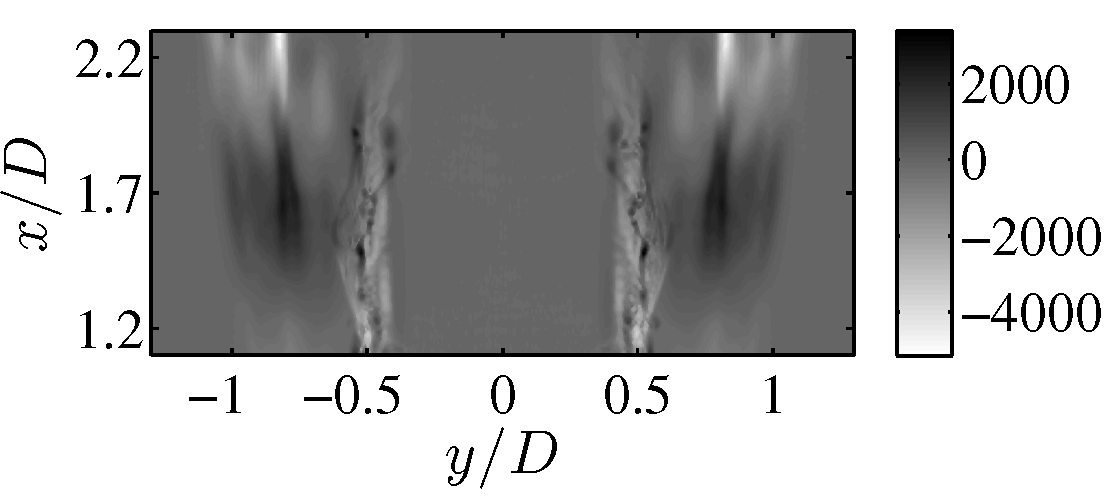} 
                \caption{DMD, m=6\\
                $d^{*}_{d}=0.05,\;f^{*}_{d}=0.2$}
                \label{fig:EV_DMD_recon_06} 
        \end{subfigure}
        
                \vspace{0.4 cm}
        
        \begin{subfigure}[b]{0.5\linewidth}
                \centering
                \includegraphics[scale=0.4]{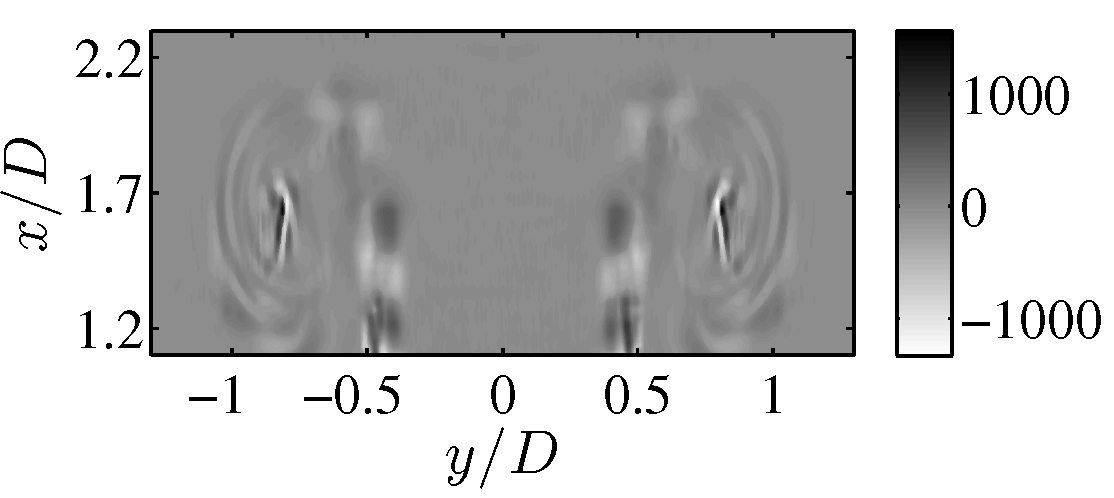} 
                \caption{$\cal{C}$DMD, m= 5\\
                $d^{*}_{c}=0.08,\;f^{*}_{c}=1$}
                \label{fig:EV_LDMD_recon_05} 
        \end{subfigure}
        \begin{subfigure}[b]{0.5\linewidth}
                \centering
                \includegraphics[scale=0.4]{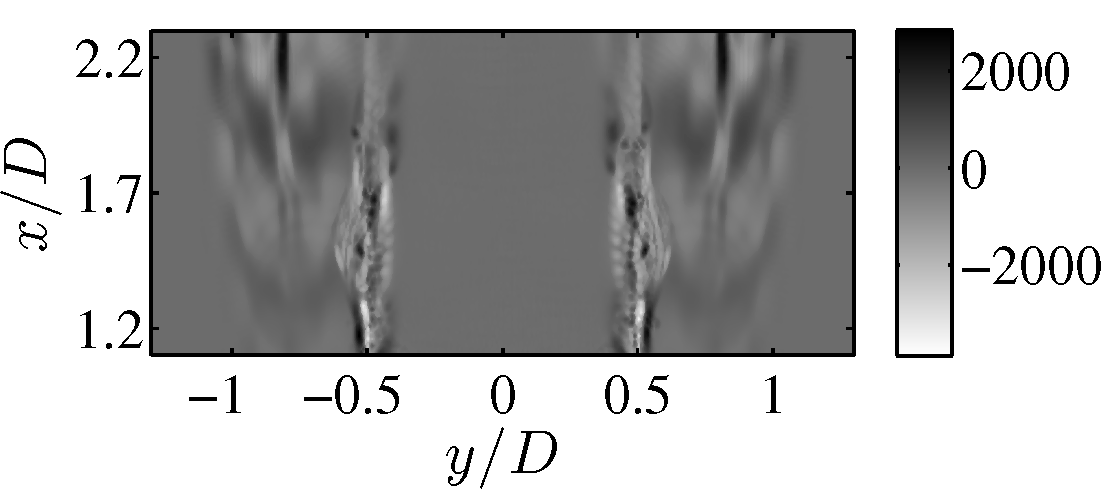} 
                \caption{DMD, m=7\\
                $d^{*}_{d}=0.1,\;f^{*}_{d}=0.6$}
                \label{fig:EV_DMD_recon_07} 
        \end{subfigure}
        
    \caption{Reconstruction of $4$ single $\cal{C}$DMD (left column) and DMD (right column) modes at time $t^{*}=5.6$.}
    \label{fig:EarlyVortHead_SingleModeComp}
  \end{figure}

In the next step, the four $\cal{C}$DMD modes that are selected to represent the vortex head, have been reconstructed along $\tau$ and then transformed 
back to physical space as illustrated in figure \ref{fig:EarlyVortHead_SingleModeComp} at time $t^{*}=5.6$. The DMD modes on the other hand,
are reconstructed along $t$ and confronted against the $\cal{C}$DMD modes in the same figure for the same timestep.\medskip


As the vortex ring propagates downstream at the dominant group velocity, the 
traditional DMD detects instances of the same structure at different spatial locations. As a result, spurious 
shadows are introduced in each mode behind and ahead of the vortex ring (figures \ref{fig:EV_DMD_recon_01}, 
\ref{fig:EV_DMD_recon_05}, \ref{fig:EV_DMD_recon_06}\& \ref{fig:EV_DMD_recon_07}). Therefore, many modes would be required in order to cancel out the shadows 
and to reconstruct the expected form of the structure.\medskip

On the contrary, the Characteristic DMD follows 
the vortex ring in space-time adjusted to its group velocity and results in a clear representation of the 
vortex head with the first mode (figure \ref{fig:EV_LDMD_recon_01}). Each of the next modes in figures \ref{fig:EV_LDMD_recon_02},
\ref{fig:EV_LDMD_recon_03} \& \ref{fig:EV_LDMD_recon_05} subsequently 
accommodate parts of the vortex head, propagating at different frequencies and decay rates, without being smeared or adversely affected
by the vortex being transported.\medskip

\begin{figure}[H]
        \begin{subfigure}[b]{0.5\linewidth}
                \centering
                \includegraphics[scale=0.4]{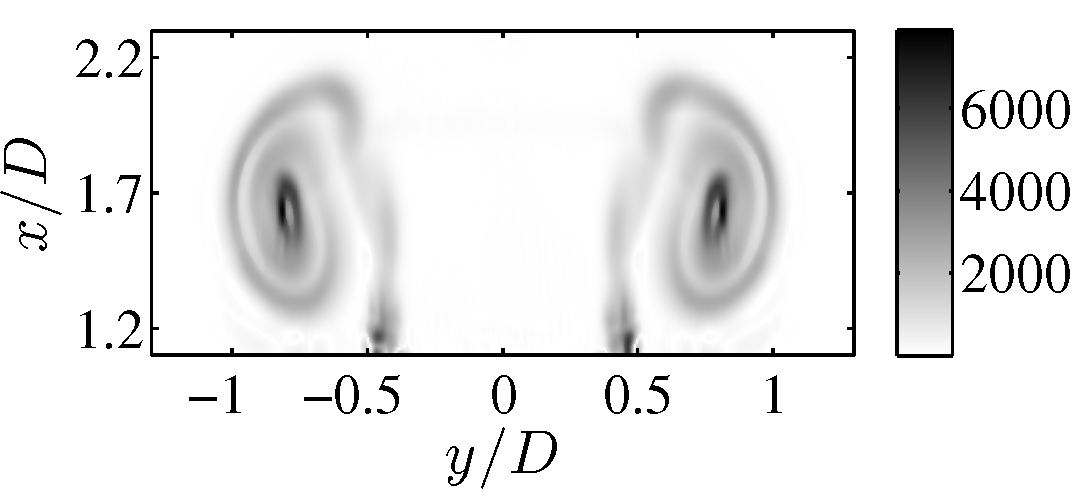} 
                \caption{$\cal{C}$DMD,\\
                modes (1, 2, 3, 5)}
                \label{fig:EV_LDMD_recon_sum_1235} 
        \end{subfigure}
        \begin{subfigure}[b]{0.5\linewidth}
                \centering
                \includegraphics[scale=0.4]{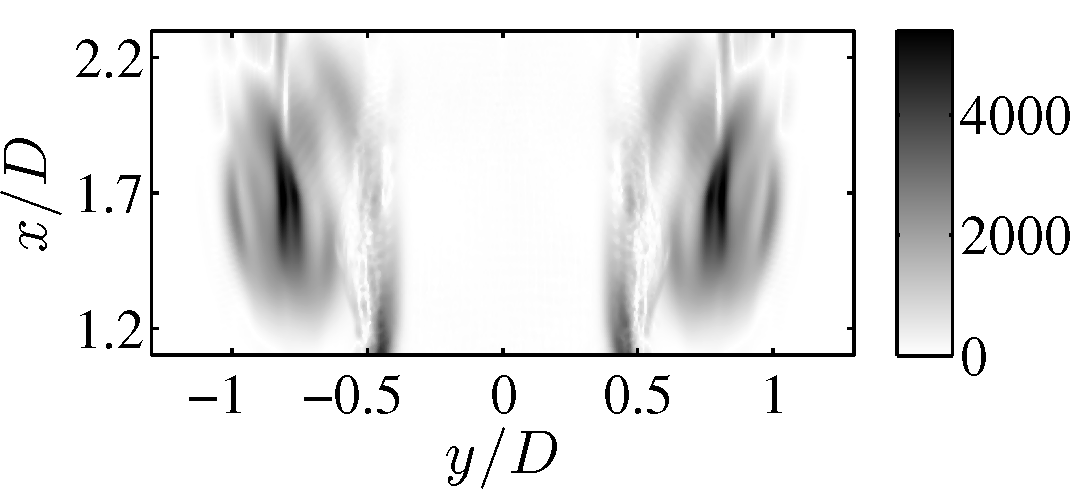} 
                \caption{DMD,\\
                modes (1, 5, 6, 7)}
                \label{fig:EV_DMD_recon_sum_1567} 
        \end{subfigure}
        
                \vspace{0.4 cm}
        
        \begin{subfigure}[b]{0.5\linewidth}
                \centering
                \includegraphics[scale=0.4]{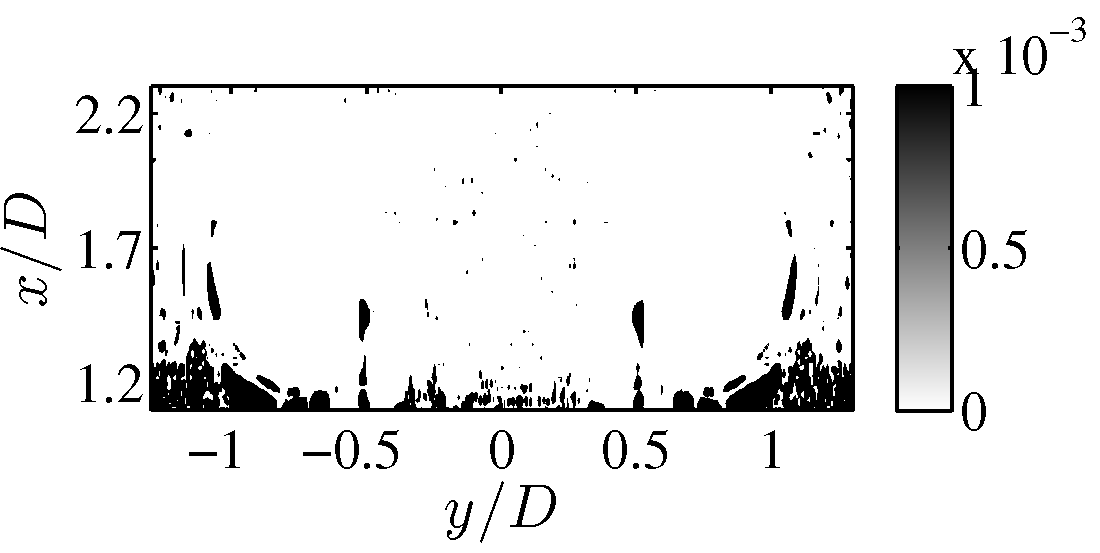} 
                \captionsetup{justification=centering}
                \caption{Relative Error for\\ 
                $\cal{C}$DMD modes (1, 2, 3, 5)}
                \label{fig:EV_LDMD_relErr} 
        \end{subfigure}
        \begin{subfigure}[b]{0.5\linewidth}
                \centering
                \includegraphics[scale=0.4]{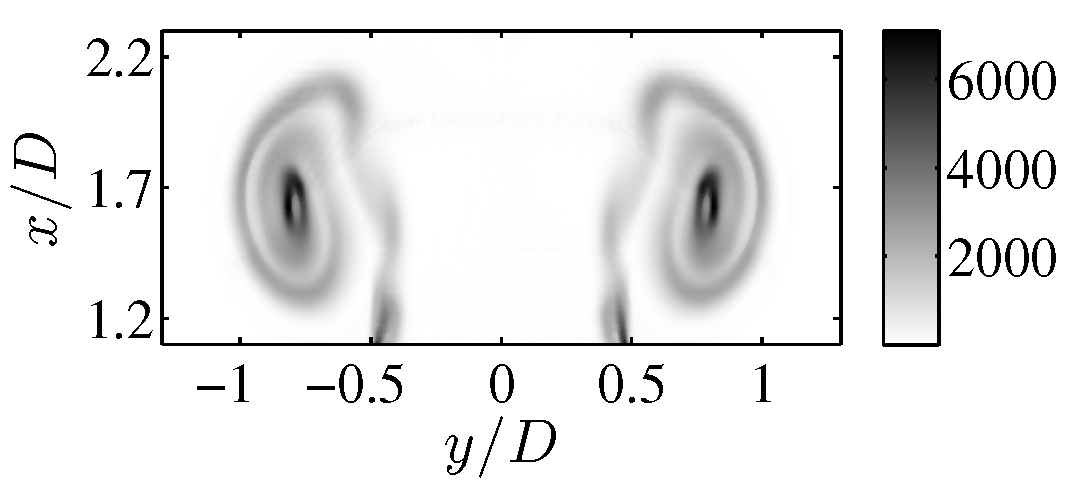} 
                \captionsetup{justification=centering}
                \caption{Full-field vortex head\\
                at time $t^{*}=5.6$}
                \label{fig:EV_fullfield_02} 
        \end{subfigure}
    \caption{Reconstruction of $4$ $\cal{C}$DMD (a) and DMD (b) summed up modes at time $t^{*}=5.6$, relative error for the $\cal{C}$DMD modes (c) in comparison with the fullfield (d).}
    \label{fig:EarlyVortHead_Recon_sum}
  \end{figure}

Summation of the four elected $\cal{C}$DMD modes are shown in 
figure \ref{fig:EV_LDMD_recon_sum_1235} at time $t^{*}=5.6$ and confronted against the summed up DMD modes
in figure \ref{fig:EV_DMD_recon_sum_1567}. The vortex head reconstructed by $\cal{C}$DMD is almost indistinguishable 
from the fullfield depicted in figure \ref{fig:EV_fullfield_02}. The vortex head itself, was extracted nicely in a single mode as shown before.
But the structures with finer scales inside the vortex head move with a slower group velocity as they travel backwards with
respect to the main head motion. Therefore, more than one mode was required to capture them.
That means, the internal structures of
the jet head suffers again from the same deficiencies as did the whole
structure before.\medskip

Nevertheless, already four $\cal{C}$DMD modes give a
clear resemblance of the full structure while the same number of DMD modes fail to 
represent even the boundaries of the vortex head. The same holds true for the structures of the shear layer on the trailing jet, as it moves
with a faster velocity than the vortex head and to capture these structures the same practice should be applied using the velocity of the trailing edge.\medskip

\begin{figure}
          \begin{subfigure}[b]{0.32\linewidth}
	    \centering
	    \includegraphics[scale = 0.4]{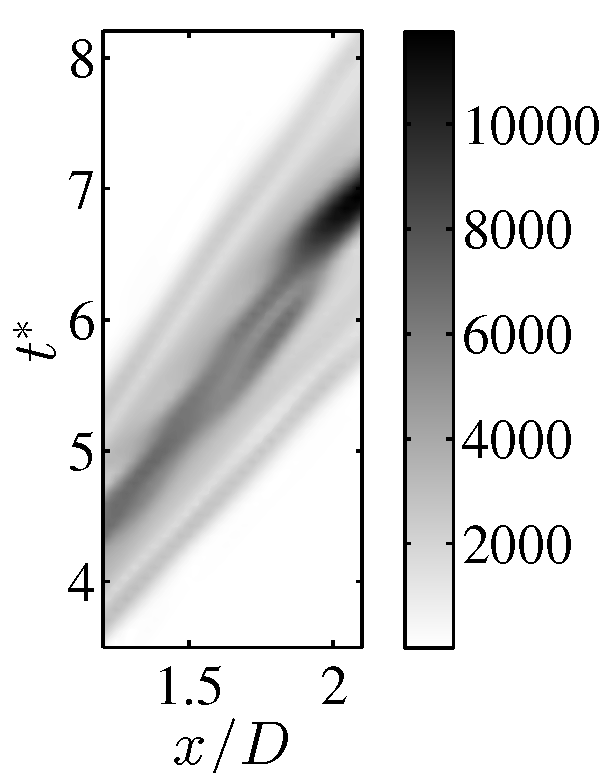}
	    \caption{}
	    \label{fig:EarlyVortHead_CharDiagRecon_a}  
      \end{subfigure}
          \begin{subfigure}[b]{0.32\linewidth}
	    \centering
	    \includegraphics[scale = 0.4]{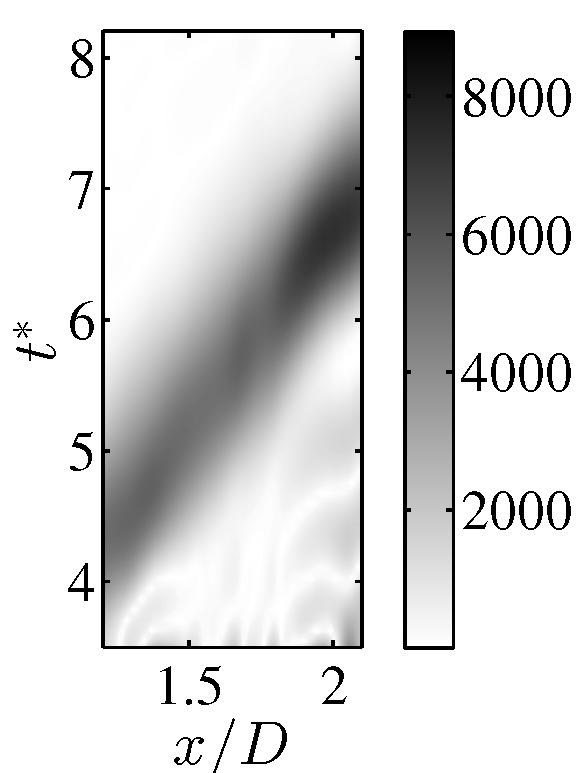}
	    \caption{}
	    \label{fig:EarlyVortHead_CharDiagRecon_b}  
      \end{subfigure}
          \begin{subfigure}[b]{0.32\linewidth}
	    \centering
	    \includegraphics[scale = 0.4]{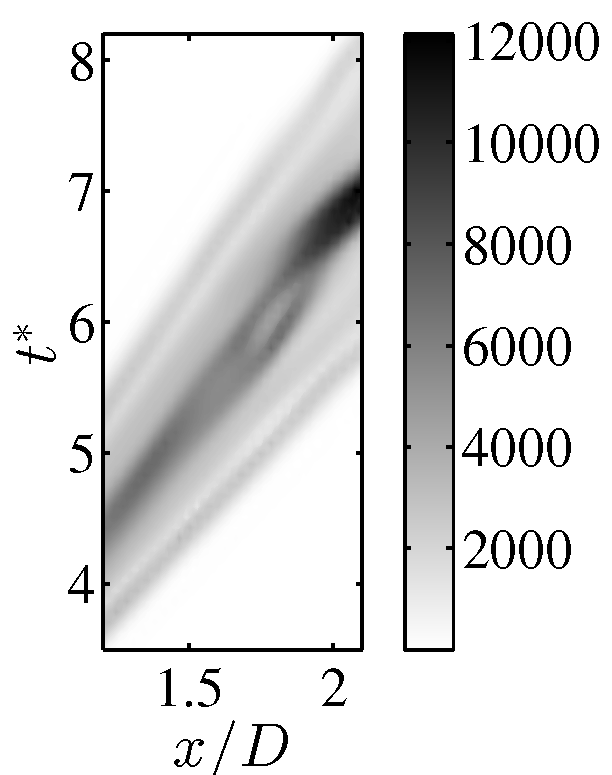}
	    \caption{}
	    \label{fig:EarlyVortHead_CharDiagReconn_c}  
      \end{subfigure}

    \caption{Characteristic diagram reconstructed using $3$ $\cal{C}$DMD (a) and DMD (b) 
    modes in comparison with the fullfield (c).}
    \label{fig:EarlyVortHead_CharDiagRecon}
 \end{figure}

To quantify the accuracy by which the structures have been captured,
the relative error is calculated for the resulted reconstruction in figure \ref{fig:EV_LDMD_relErr}. 
The area on the vortex head shows clearly less than $0.1 \%$ relative error. By looking into the reconstructed part of characteristic diagram, it is observable that the same agreement between the fullfield and
$\cal{C}$DMD modes, holds also valid for the rest of the timesteps (figure \ref{fig:EarlyVortHead_CharDiagRecon}).
The space time diagram reconstructed using $\cal{C}$DMD modes, has captured 
many of the details originally existing in the fullfield. 
As expected of course, the DMD reconstruction has led to a vague reproduction of the field, 
missing out many of the details which can be only captured using many modes.\medskip

\section{ Spatiotemporal vs. spatial decomposition}
\label{sec:StVsSp}  
  
\begin{figure}
  \centering  
  \includegraphics[scale = 0.4]{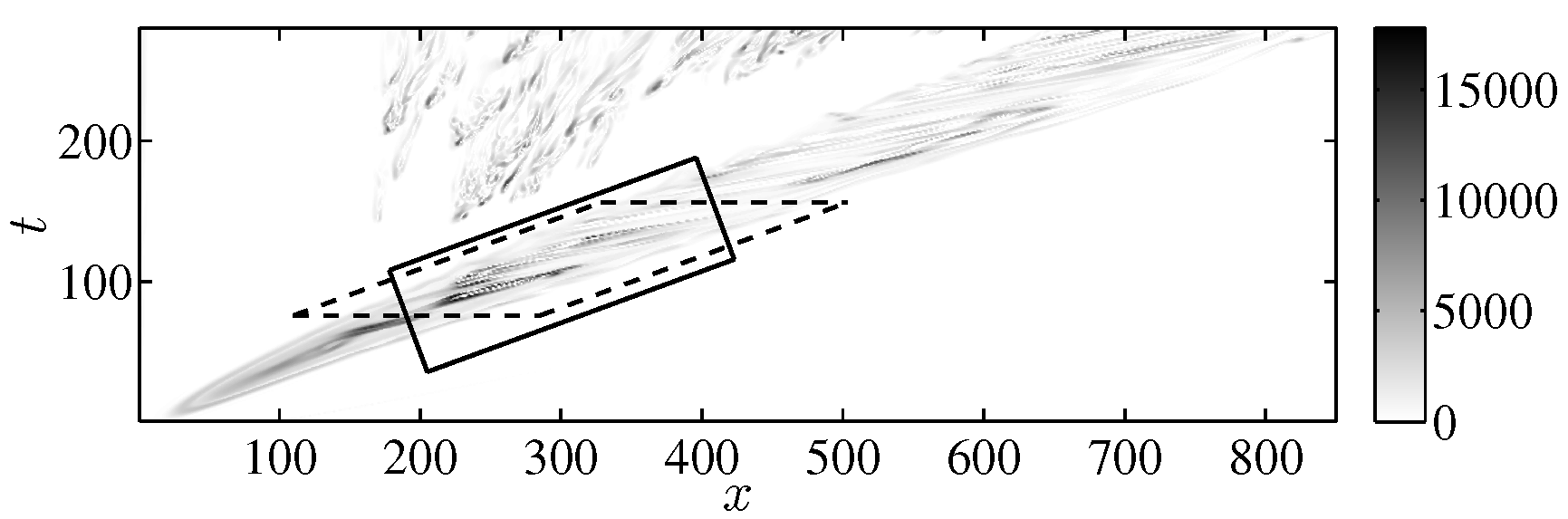}
  \caption{Characteristic diagram at the center of the vortex head at $y/D=0.8$. Solid and dashed windows correspond respectively 
  to a decomposition window in the spatiotemporal and shifted space.}
  \label{fig:CharDiag_turb}
\end{figure}

What sets our proposed method apart from similar studies in the past, is that via a Characteristic DMD, moving structures are 
detected in spatiotemporal space in planes normal to the characteristics, while in other data-driven methods, they are sought in a shifted space. In order to clarify the differences
between the two approaches, a comparative analysis is presented in this chapter between the Characteristic DMD ($\cal{C}$DMD), Shifted DMD ($\cal{SH}$DMD) and a traditional DMD. 
For this purpose, and to verify the dependence of the decompositions on the number of snapshots, the turbulent stage of the vortex head has been selected. This is due to the fact, 
that as shown in figure \ref{fig:CharDiag_turb}, the time span during which the vortex head remains turbulent, is longer than the laminar development of the vortex head. \medskip

\begin{figure}
      \begin{subfigure}[b]{0.5\linewidth}
	    \centering
	    \includegraphics[scale = 0.4]{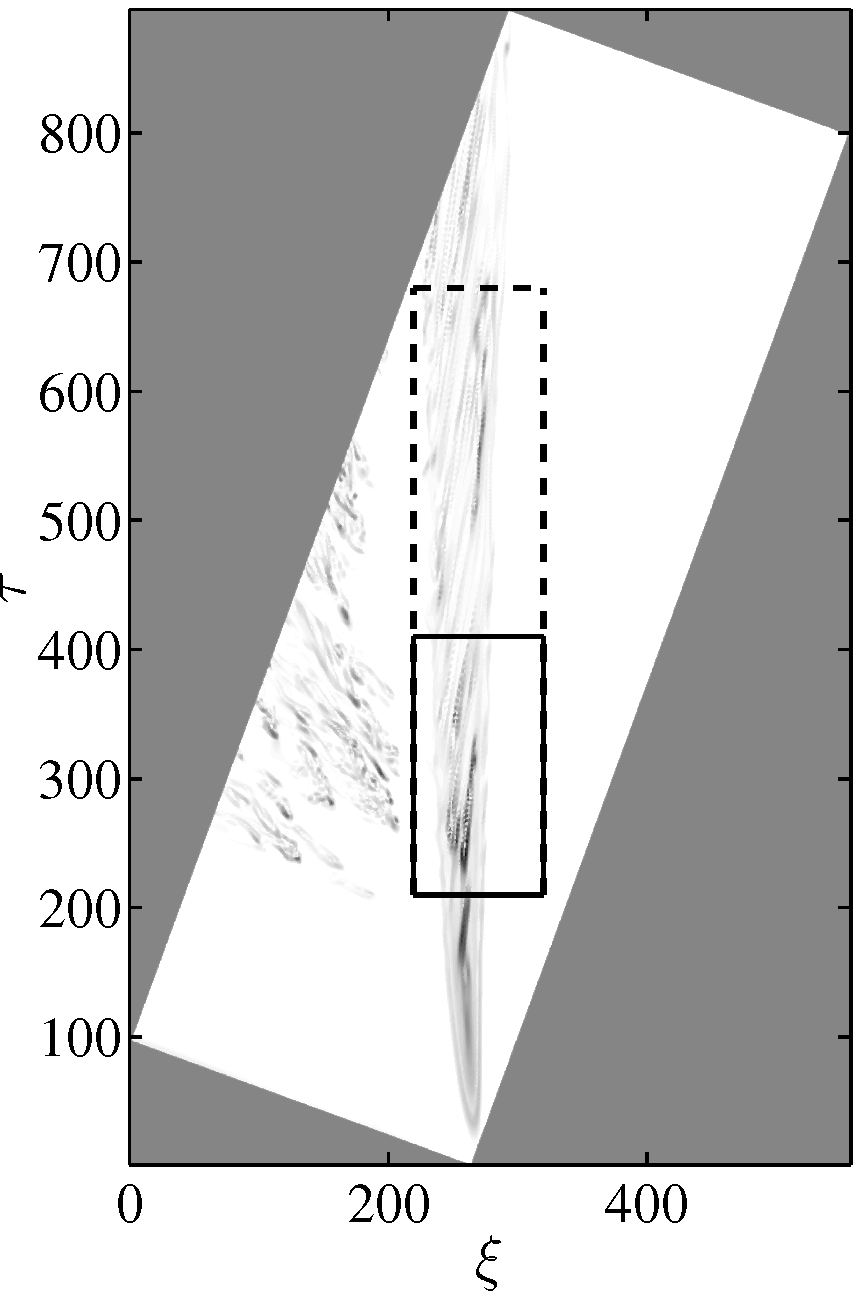}
	    \caption{}
	    \label{fig:CharDiag_rot}  
      \end{subfigure}
      \begin{subfigure}[b]{0.5\linewidth}
	    \centering
	    \includegraphics[scale = 0.4]{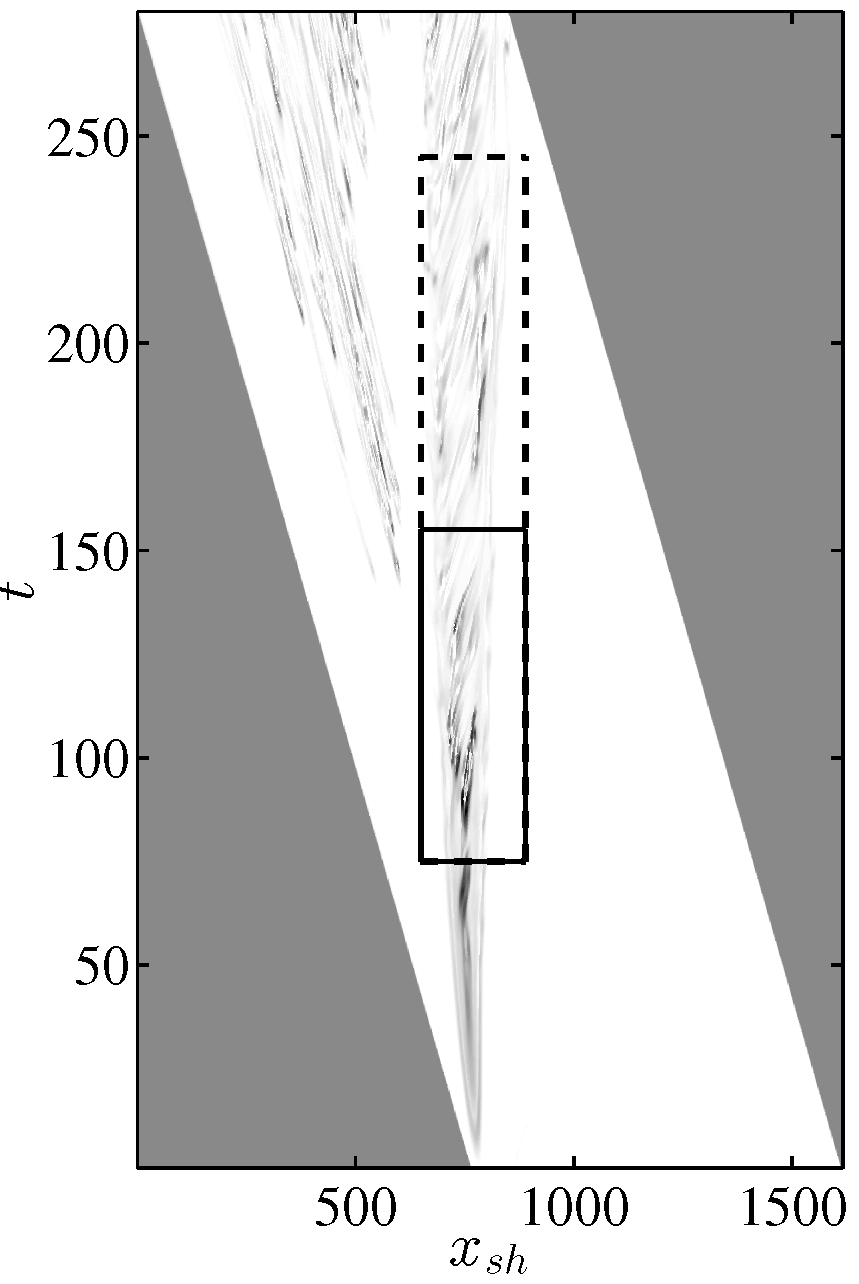}
	    \caption{}
	    \label{fig:CharDiag_shift}  
      \end{subfigure}
    \caption{Transformed snapshots matrix in the spatiotemporal (a) and shifted (b) reference frames.}
    \label{fig:CharDiag_rot_shift}
 \end{figure}

As illustrated in figure \ref{fig:CharDiag_rot_shift}  two different transformations have been applied to the snapshots matrix, both corresponding to the same 
group velocity $u_{g}^{*} = 0.4$. Applying a rotation in space and time, the snapshots matrix with columns and rows along $x$ and $t$, has been transformed to a spatiotemporal space, resulting 
in a matrix with columns and rows along $\xi$ and $\tau$ (figure \ref{fig:CharDiag_rot}). Via another transformation in form of a shift in space, the snapshots matrix has been transformed 
to a shifted space, aligning the columns and rows of the resulted matrix along $x_{sh}$ and $t$ (\ref{fig:CharDiag_shift}).
The space time diagrams in this figure, are both taken along the stream wise direction at the center of the vortex head.
Figures \ref{fig:CharDiag_turb} and \ref{fig:CharDiag_rot_shift} are both plotted in computational space with the axes corresponding to the relevant matrix elements. \medskip

To examine the the dependence of the dynamics of the resulted modes on the number of snapshots, 
ten decomposition windows have been selected for each set of transformed data. Each of the ten windows in the rotated frame are comparable with the corresponding 
one in the shifted one. On each frame of reference, all ten windows have the same spatial extents accommodating the vortex heads and excluding the 
trailing jet. This choice has been made due to the fact that the two mentioned parts, have different group velocities and with this analysis we are aiming at 
treating only the vortex head. Extra space has been allowed downstream of the vortex head to have a measure of how well each decomposition has captured the bounds of the vortex ring. \medskip

The temporal extents have been chosen to ensure maximum overlap between each two corresponding windows in the rotated and shifted frames. To demonstrate the latter, 
the smallest window in each frame, is plotted in physical space in figure \ref{fig:CharDiag_turb} as solid and dashed lines representing the rotated and shifted decomposition windows respectively. 
As it can be seen in this figure, the lower and upper bounds of the two illustrated windows, coincide exactly at the center of the vortex ring. The same holds true for all the decompositions.\medskip

While keeping the lower temporal extent fixed, the upper temporal bounds have been incrementally increased for each window. The increments along $\tau$ in spatiotemporal space $i_{\tau}$ and 
along $t$ in the shifted space $i_{t}$ are related as $i_{\tau} = \alpha~ i_{t} $, with $\alpha$ defined by equation \ref{eq:alpha}. Solid and dashed black lines in figure \ref{fig:CharDiag_rot_shift} correspond 
respectively to the bounds of the smallest and largest windows.\medskip

The gray areas in both figures, show the zeros which have been added to the snapshots matrix via the transformations. The first observation at this point, is that the full dataset can be used neither in the rotated nor in the shifted space. In other words, while treating non-periodic data, 
both methods will essentially lead to a certain level of data loss. This limitation will not exist in the shifted decomposition for periodic datasets.\medskip

\begin{figure}
  \centering  
  \includegraphics[scale = 0.4]{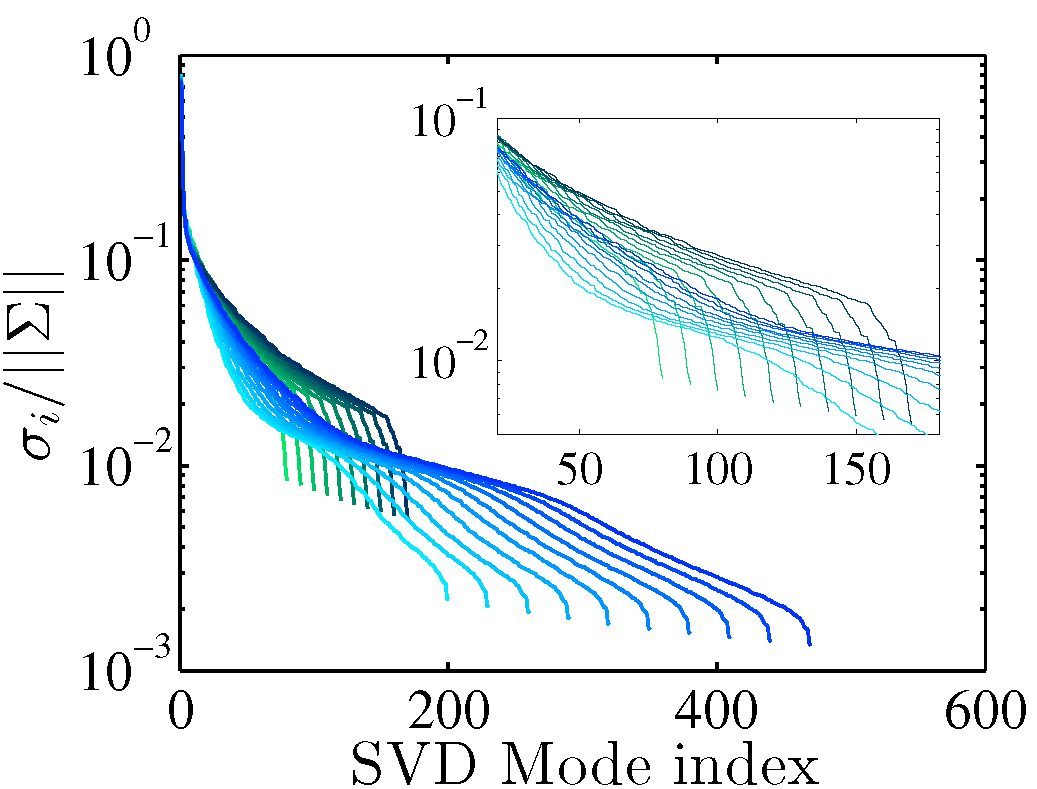}
  \caption{Singular values in spatiotemporal and shifted reference frames presented respectively in blue and green. Darker colors correspond to larger windows.}
  \label{fig:SingVals_mult_rot_shift}
\end{figure}

In the next step, a singular value decomposition is carried out on each frame of reference and for each window. In figure \ref{fig:SingVals_mult_rot_shift}, singular values resulted 
in the rotated and shifted frames are plotted in blue and green respectively with the darker colors depicting the larger decomposition windows. As expected, as the number of 
snapshots increases, a slower drop of singular values is resulted in both frames. But It can be observed 
that for all decompositions, there is a faster drop of singular values in the spatiotemporal space. This means that independent of the number of snapshots, the vortex ring 
can be reduced using fewer modes in the spatiotemporal space in comparison with the shifted space.\medskip

\begin{figure}
      \begin{subfigure}[b]{0.5\linewidth}
	    \centering
	    \includegraphics[scale = 0.4]{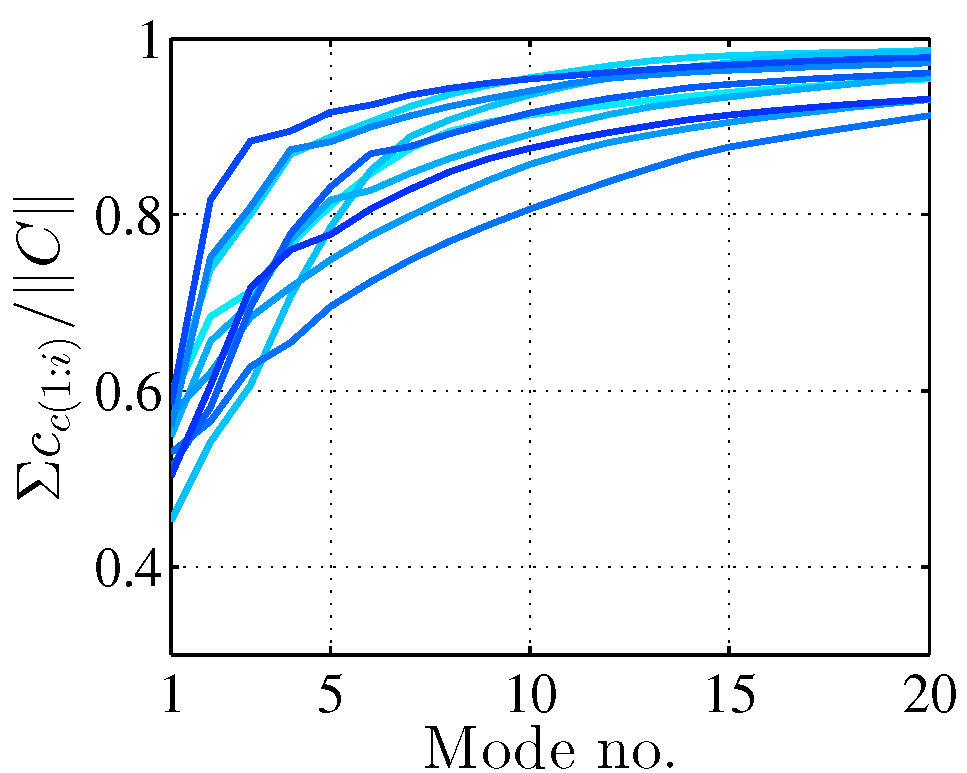}
	    \caption{}
	    \label{fig:CumulRot}  
      \end{subfigure}
      \begin{subfigure}[b]{0.5\linewidth}
	    \centering
	    \includegraphics[scale = 0.4]{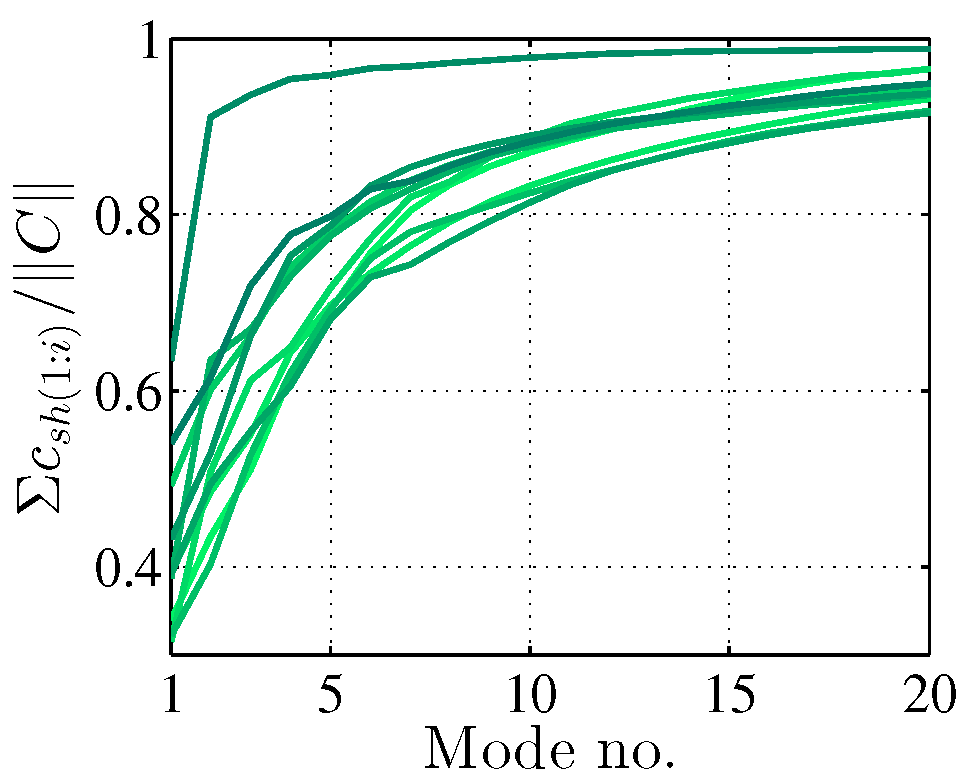}
	    \caption{}
	    \label{fig:CumulShift}  
      \end{subfigure}
    \caption{Cumulative mode contents in the spatiotemporal (a) and shifted (b) reference frames.}
    \label{fig:CumulModeContent}
 \end{figure}

Next, a standard DMD is carried out for each of the windows and the resulted modes are sorted based on their average contribution in all the timesteps. 
The cumulative mode amplitudes are plotted for $\cal{C}$DMD and $\cal{SH}$DMD in figure \ref{fig:CumulModeContent}. It can be observed that the first 2 or 3 modes in the spatiotemporal space, 
have a higher cumulative content than the same number of modes in the shifted space. \medskip

To reach an elaborate assessment of the modes and their evolutions in space and time, one of the ten groups of decompositions have been selected with 320 and 120 number of snapshots along $\tau$ and $t$ respectively. To compare the results with a traditional DMD in physical space, another decomposition is also carried out on a stationary frame frame of reference. The temporal bounds for the latter DMD, are similar to those selected for the $\cal{SH}$DMD.\medskip

\begin{figure}
      \begin{subfigure}[b]{0.5\linewidth}
	    \centering
	    \includegraphics[scale = 0.4]{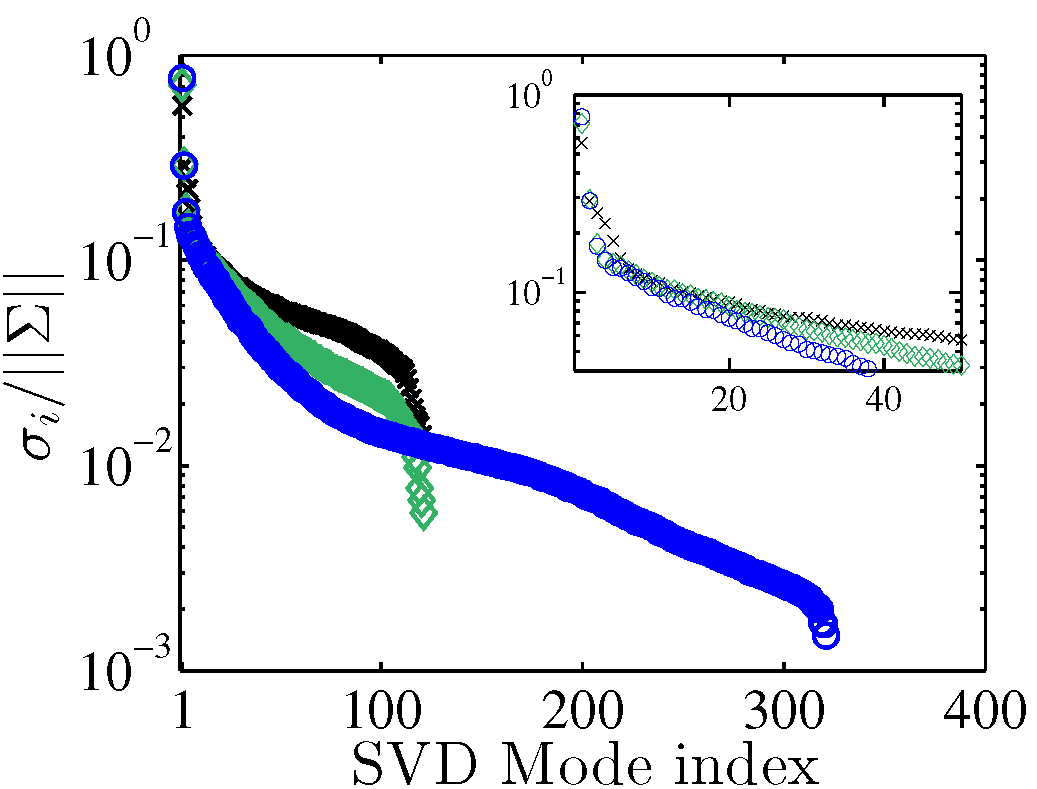}
	    \caption{}
	    \label{fig:SingVals_rotShift}  
      \end{subfigure}
      \begin{subfigure}[b]{0.5\linewidth}
	    \centering
	    \includegraphics[scale = 0.4]{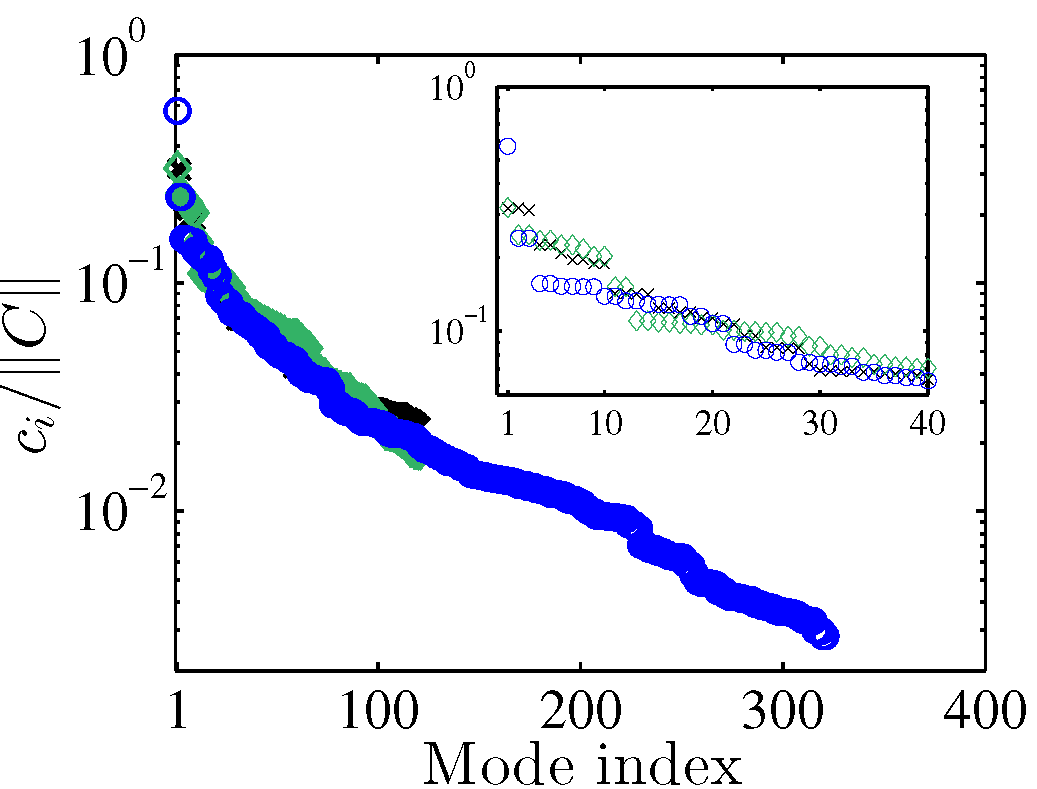}
	    \caption{}
	    \label{fig:ModeAmp_rot_shift}  
      \end{subfigure}
    \caption{Singular values (a) and mode contents (b) in the spatiotemporal (\textcolor{blue}{$\circ$}), shifted (\textcolor{ForestGreen}{$\diamond$}) and stationary ($\times$) reference frames.}
    \label{fig:SingVal_ModeAmp_rot_Shift}
 \end{figure}

The resulted singular values are shown in figure \ref{fig:SingVals_rotShift} with blue circles, green squares and black crosses representing the rotated, shifted and physical spaces respectively. 
Since via the spatiotemporal and the shifted transformations, the vortex head is being followed on a moving frame of reference, an SVD leads to a faster drop of singular values compared to the singular values in physical space. Nevertheless, as mentioned earlier, the fastest drop happens along $\tau$. \medskip

For the three mentioned dynamic mode decompositions in spatiotemporal, shifted and physical spaces, mode amplitudes are plotted in figure \ref{fig:ModeAmp_rot_shift} with the symbol 
colors similar to what was defined earlier. It can be seen in this figure and in its magnified part, that except for the first 10 modes, the overall trend of the modal content is 
not hugely different for the three decompositions. This similarity can be explained by the fact that the vortex ring and the events inside it are traveling at different velocities as 
it can be seen in figure \ref{fig:CharDiag_turb}. To capture these events efficiently, a transformation to a different frame of reference would be required. \medskip

The result is that on a frame of reference which is chosen by the group velocity of the vortex head, many modes will be still needed to capture the events inside, which are traveling at a different velocity. 
Therefore except for the first few modes, the modal decay will follow the same trend as that of the DMD modes on physical space. In other words, all three decompositions treat the mentioned  instabilities similarly 
and the main difference is noticeable only for the part of the flow, whose group velocity is closest to the velocity of the moving frame. \medskip

Similar to the analysis in the previous section, it is the vortex head which fits most closely the definition of a coherent structure, and therefore the frame of reference is chosen 
based on this coherent part of the flow. The first 10 mode amplitudes show that the decomposition on each frame of reference, treats the vortex head differently. While a 
$\cal{C}$DMD along the characteristics leads to the first mode with average amplitude of $c_{c}=0.57$, the first $\cal{SH}$DMD and DMD modes have amplitudes of $c_{sh} = 0.33$ 
and $c_{d} = 0.31$ respectively. To reach a cumulative mode content of 0.57, 3 modes have to be taken on each of the other frames. Taking 12 modes will amount to cumulative mode contents 
of $c^{*}_{c}=0.88$ along $\tau$, $c^{*}_{sh}=0.85$ along $t$ in the shifted space and $c^{*}_{d}=0.87$ along $t$ in the physical space.\medskip

\begin{figure}
  \centering  
  \includegraphics[scale = 0.4]{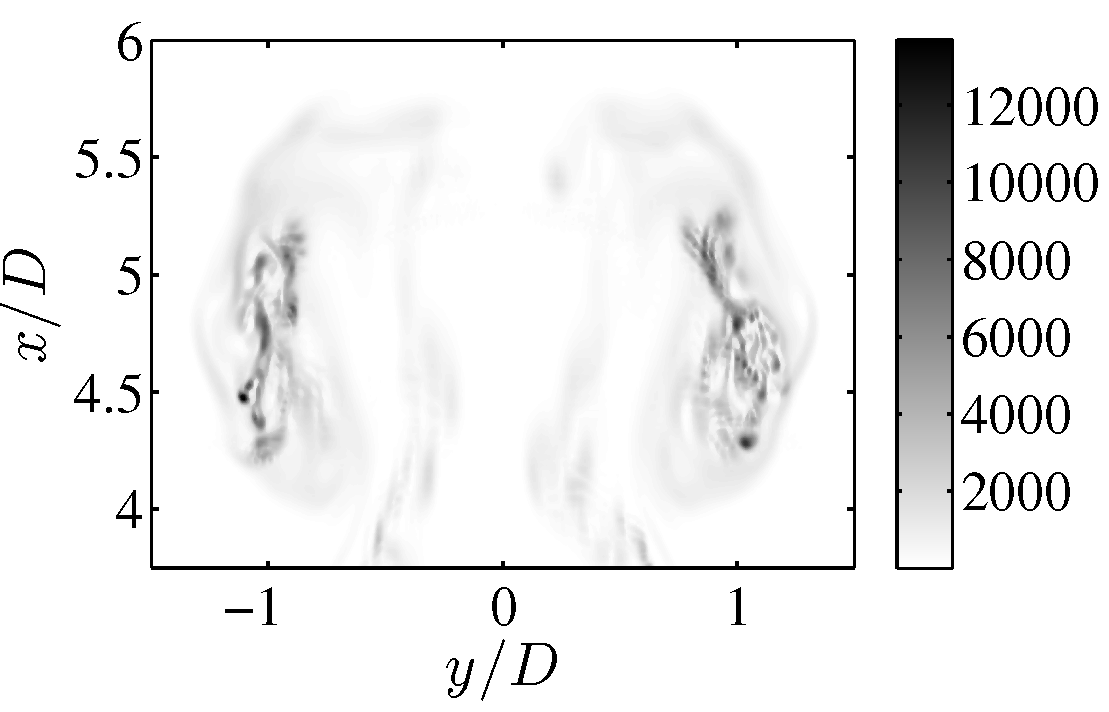}
  \caption{Representation of the fullfield vortex head at time $t^{*}=13.7$.}
  \label{fig:vortHead_turb_full}
\end{figure}

\begin{figure}
      \begin{subfigure}[b]{0.5\linewidth}
	    \centering
	    \captionsetup{justification=centering}
	    \includegraphics[scale = 0.4]{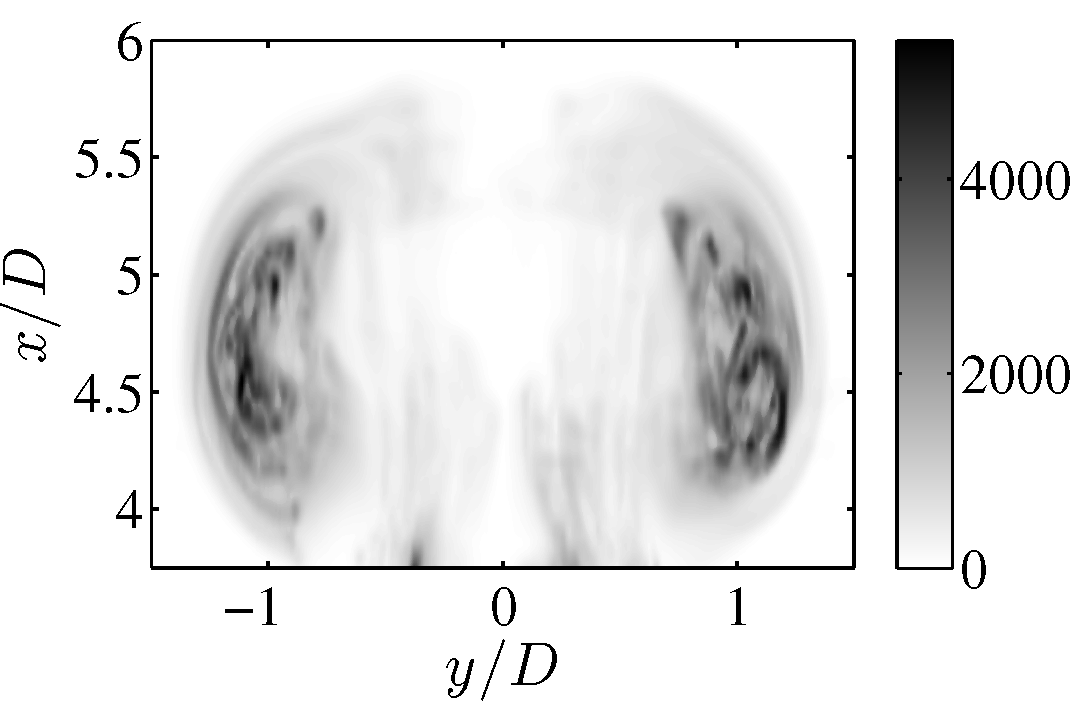}
	    \caption{$\cal{C}$DMD, m=1\\
	        $c^{*}_{c}=0.57$\\
                $d^{*}_{c}=0.02,\;f^{*}_{c}=0$}
	    \label{fig:CDMD_1}  
      \end{subfigure}
      \begin{subfigure}[b]{0.5\linewidth}
	    \centering
	    \includegraphics[scale = 0.4]{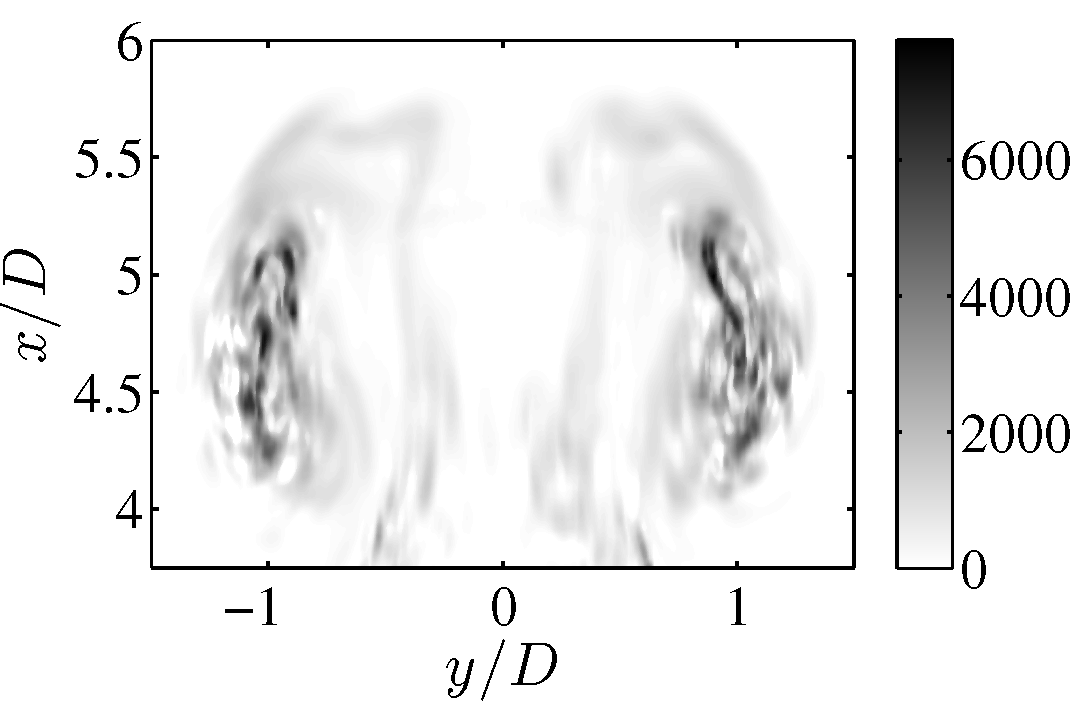}
	    \caption{$\cal{C}$DMD\\ 
	        modes: 1-12\\
                $c^{*}_{c}=0.88$}
	    \label{fig:CDMD_12}      
      \end{subfigure}
      
      \vspace{0.4 cm}
      
      \begin{subfigure}[b]{0.5\linewidth}
	    \centering
	    \captionsetup{justification=centering}
	    \includegraphics[scale = 0.4]{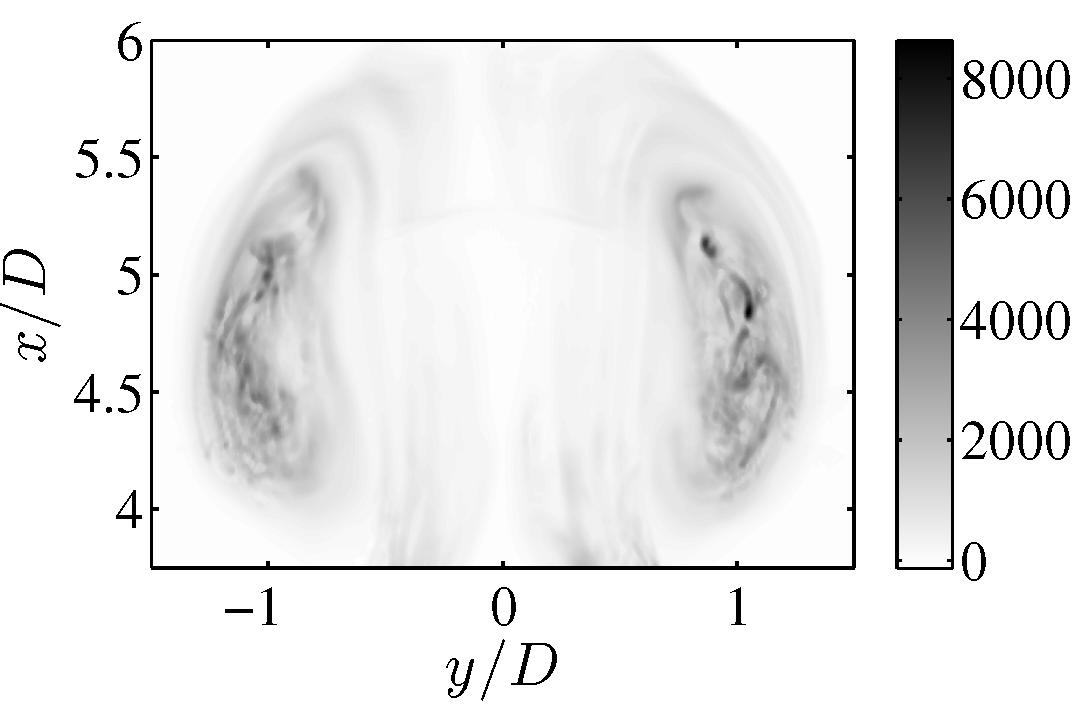}
	    \caption{$\cal{SH}$DMD, m=1\\
	    $c^{*}_{sh}=0.33$\\
	    $d^{*}_{sh}=0.006,\;f^{*}_{sh}=0$}
	    \label{fig:ShDMD_1}  
      \end{subfigure}
      \begin{subfigure}[b]{0.5\linewidth}
	    \centering
	    \includegraphics[scale = 0.4]{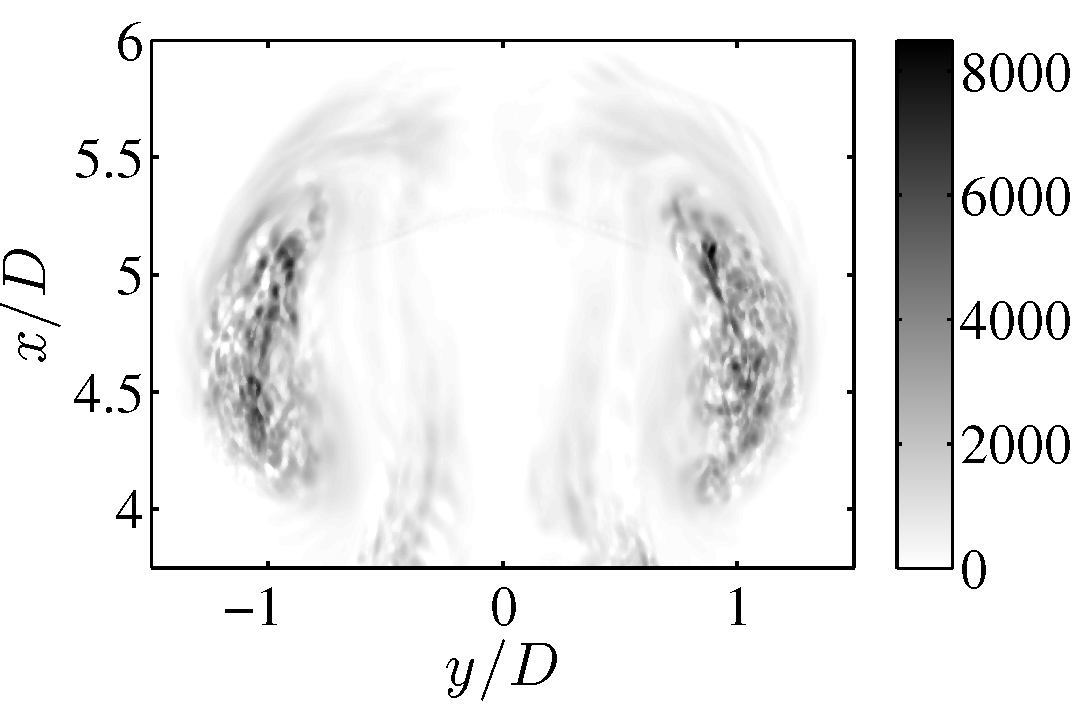}
	    \caption{$\cal{SH}$DMD\\
	    modes: 1-12\\
	        $c^{*}_{sh}=0.85$}
	    \label{fig:ShDMD_12}      
      \end{subfigure}
      
      \vspace{0.4 cm}
      
      \begin{subfigure}[b]{0.5\linewidth}
	    \centering
	    \captionsetup{justification=centering}
	    \includegraphics[scale = 0.4]{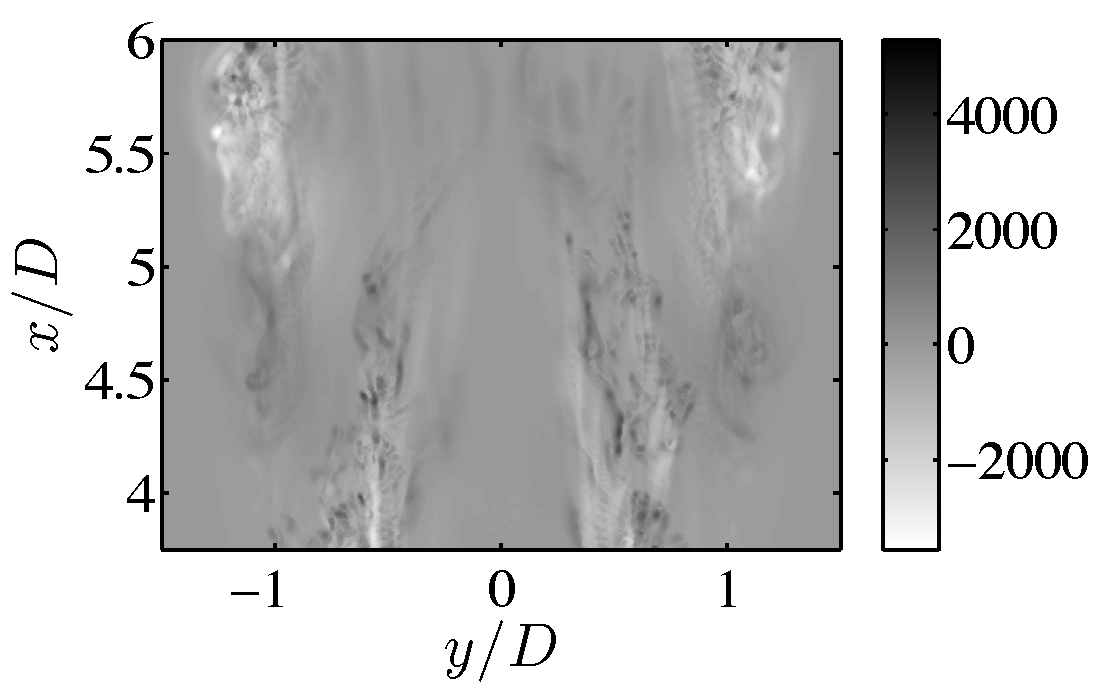}
	    \caption{DMD, m=1\\
	        $c^{*}_{d}=0.31$\\
                $d^{*}_{d}=0.07,\;f^{*}_{d}=0.2$}
	    \label{fig:DMD_1}  
      \end{subfigure}
      \begin{subfigure}[b]{0.5\linewidth}
	    \centering
	    \includegraphics[scale = 0.4]{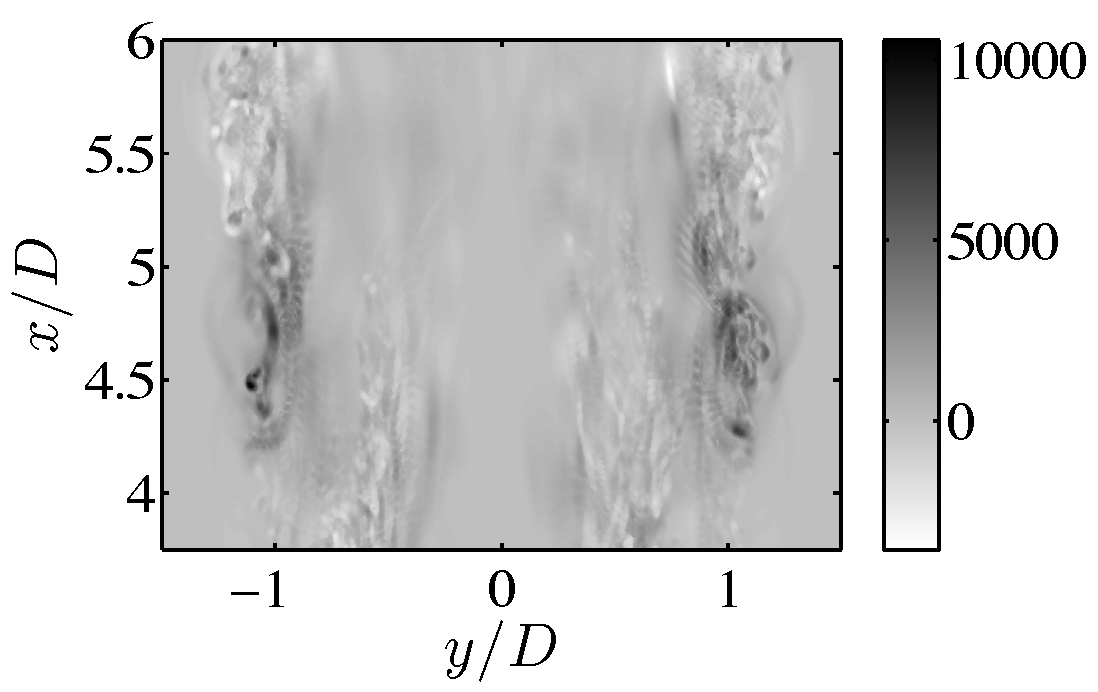}
	    \caption{DMD\\
	    modes: 1-12\\
	        $c^{*}_{d}=0.87$}
	    \label{fig:DMD_12}      
      \end{subfigure}
    \caption{Reconstruction of the first mode (left column) and the first 12 modes (right column) for $\cal{C}$DMD (a,b), $\cal{SH}$DMD (c,d) and DMD (e,f) at timestep $t^{*}=13.7$.}
    \label{fig:CDMD_ShDMD_DMD}
 \end{figure}

\begin{figure}
      \begin{subfigure}[b]{0.5\linewidth}
	    \centering
	    \includegraphics[scale = 0.4]{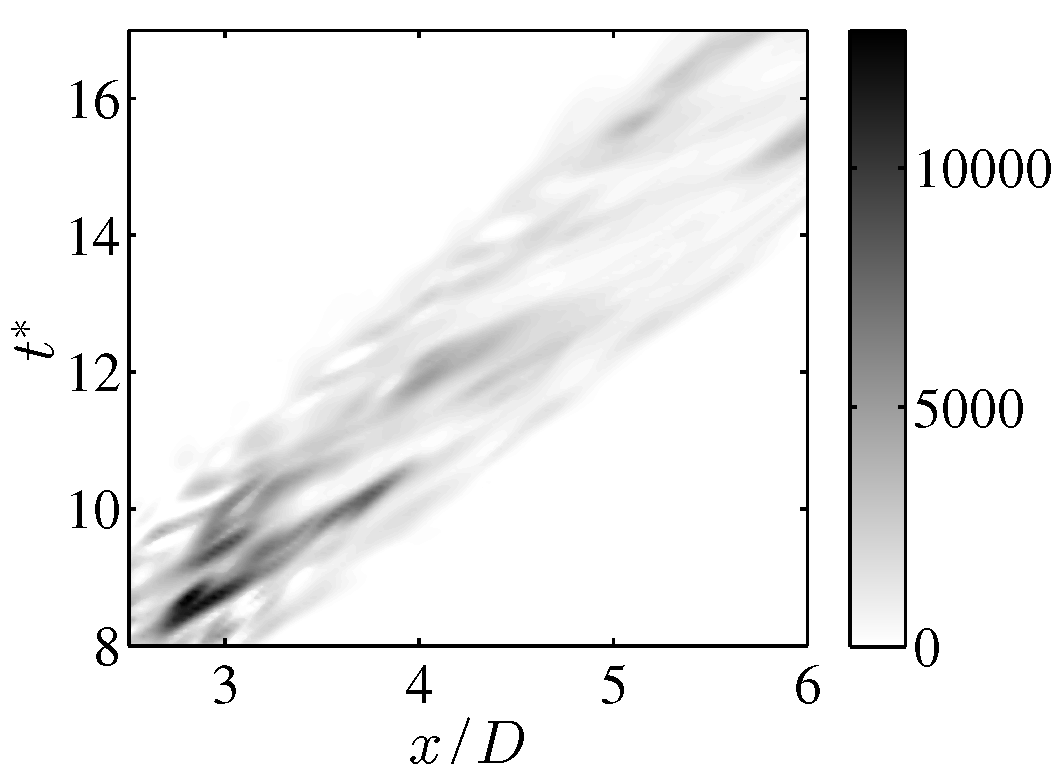}
	    \caption{$\cal{C}$DMD, modes: 1-12}
	    \label{fig:CharDiag_CDMD_12m}  
      \end{subfigure}
      \begin{subfigure}[b]{0.5\linewidth}
	    \centering
	    \includegraphics[scale = 0.4]{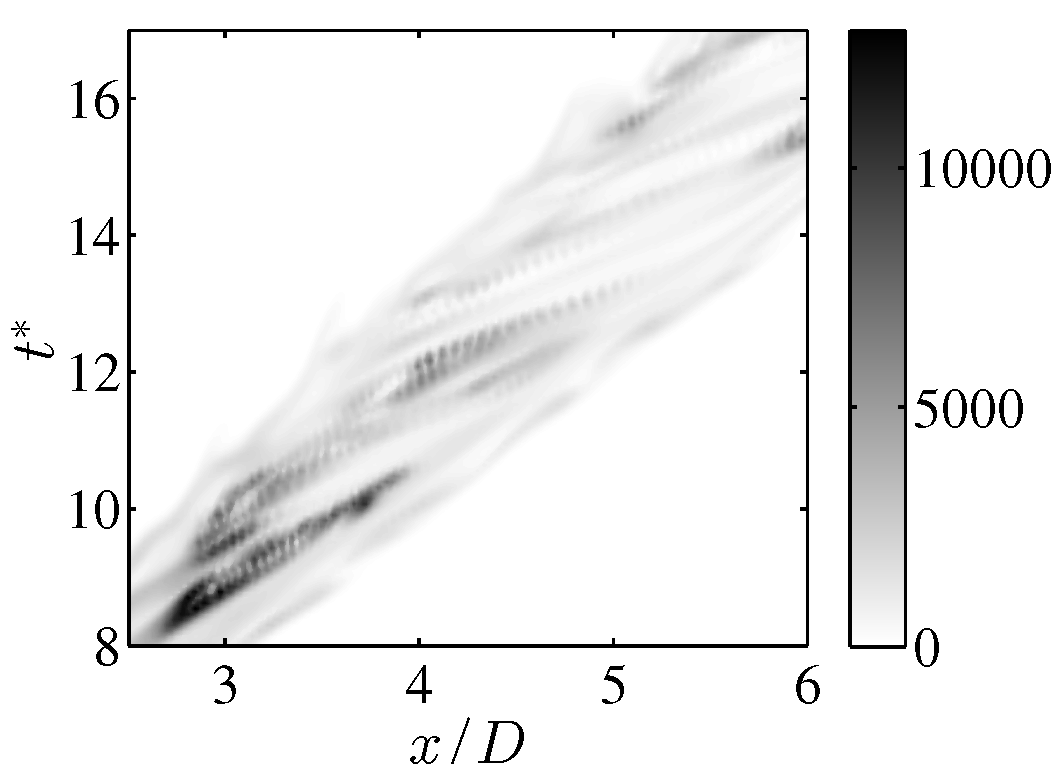}
	    \caption{fullfield}
	    \label{fig:CharDiag_fullfield_turb}  
      \end{subfigure}
      
      \vspace{0.4 cm}
      
      \begin{subfigure}[b]{0.5\linewidth}
	    \centering
	    \includegraphics[scale = 0.4]{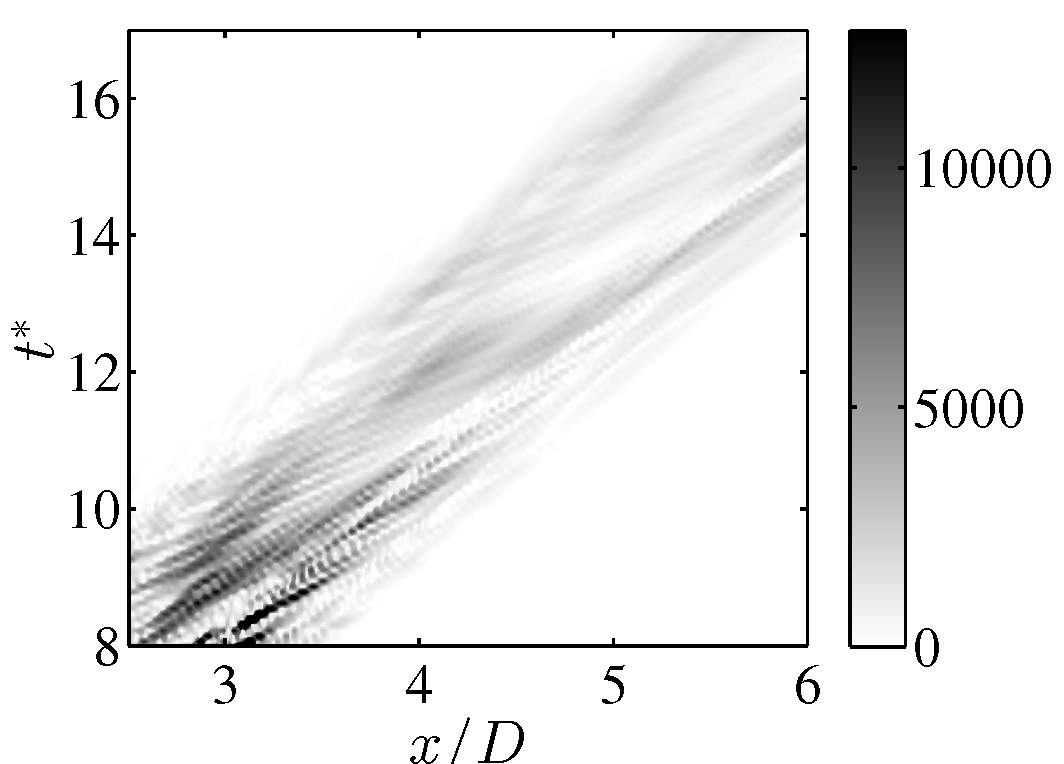}
	    \caption{$\cal{SH}$DMD, modes: 1-12}
	    \label{fig:CharDiag_ShDMD_12m}  
      \end{subfigure}
      \begin{subfigure}[b]{0.5\linewidth}
	    \centering
	    \includegraphics[scale = 0.4]{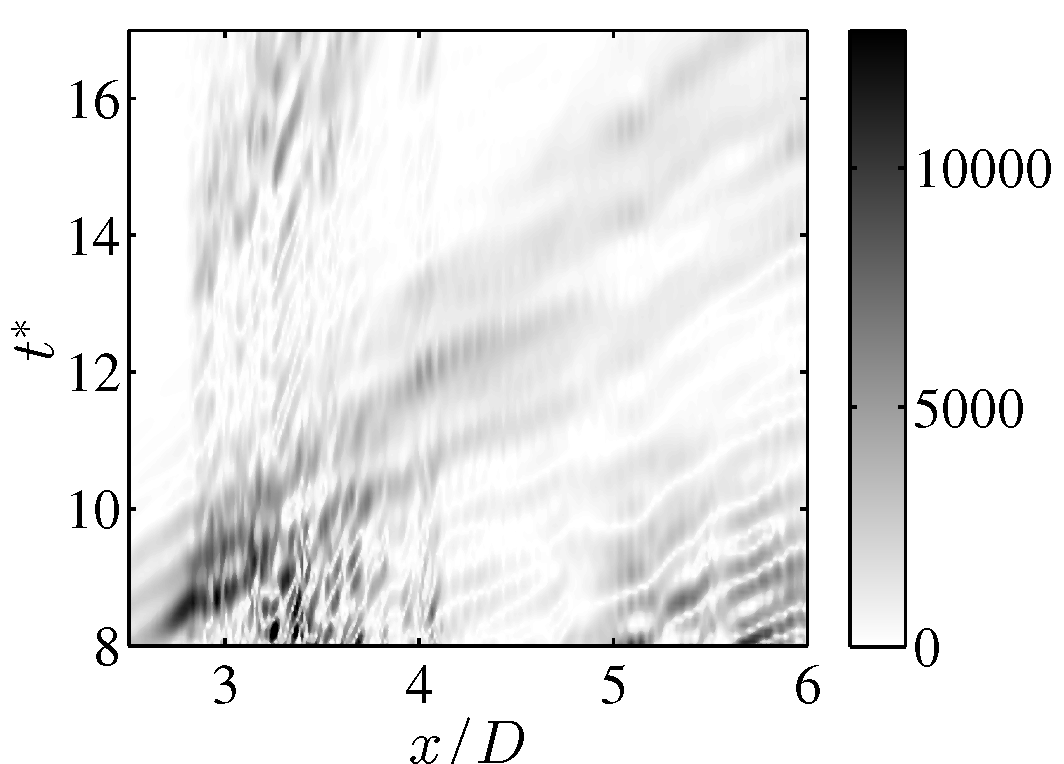}
	    \caption{DMD, modes: 1-12}
	    \label{fig:CharDiag_DMD_12m}  
      \end{subfigure}
      
    \caption{Space time diagram plotted at the center of the vortex head, reconstructed using 12 $\cal{C}$DMD, $\cal{SH}$DMD and DMD modes, compared against the fullfield.}
    \label{fig:CharDiag_CDMD_ShDMD}
 \end{figure}

\begin{figure}
      \begin{subfigure}[b]{1\linewidth}
	    \centering
	    \includegraphics[scale = 0.32]{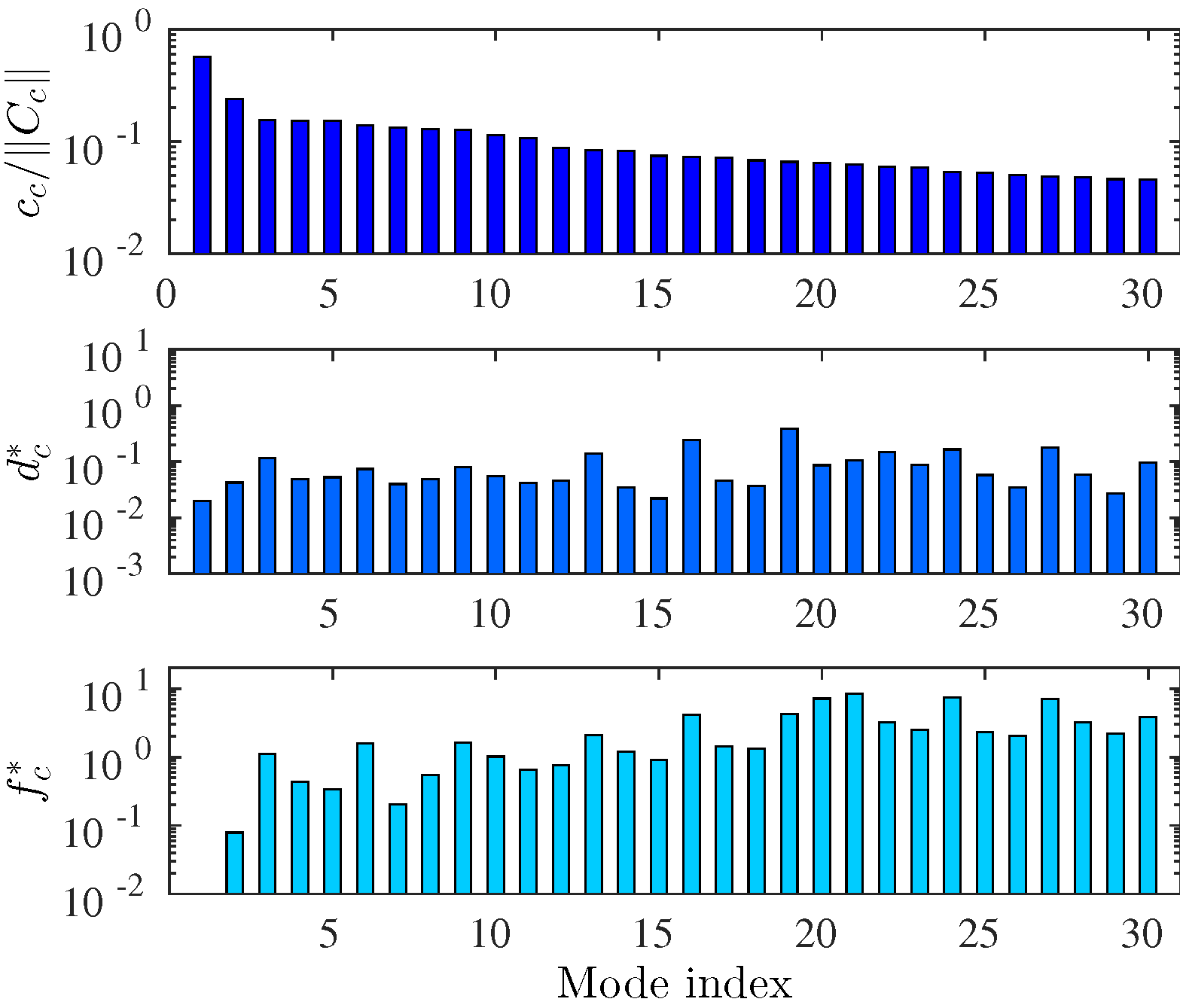}
	    \caption{}
	    \label{fig:Spectrum_CDMD}  
      \end{subfigure}
      \begin{subfigure}[b]{1\linewidth}
	    \centering
	    \includegraphics[scale = 0.32]{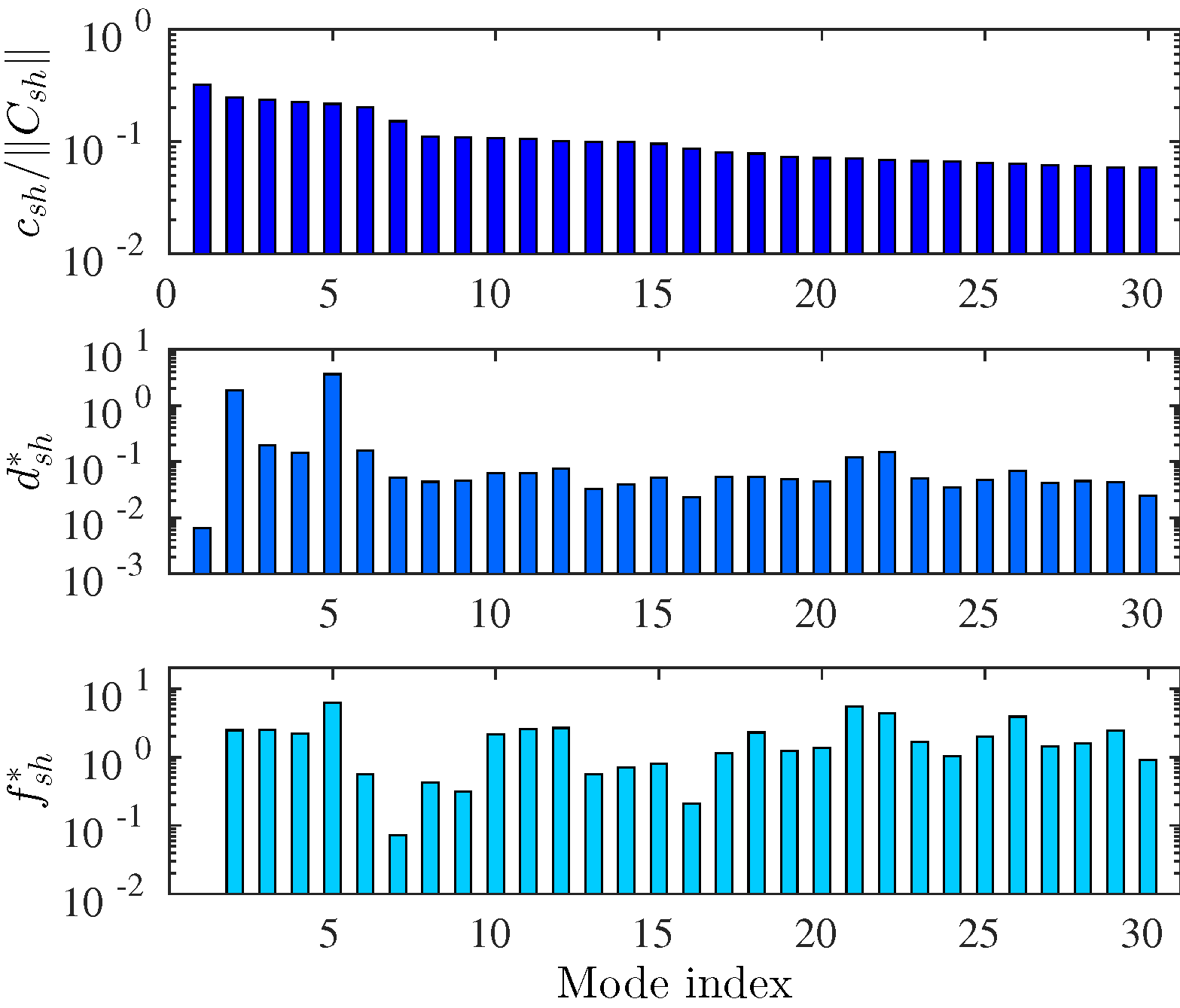}
	    \caption{}
	    \label{fig:Spectrum_SHDMD}  
      \end{subfigure}
      \begin{subfigure}[b]{1\linewidth}
	    \centering
	    \includegraphics[scale = 0.32]{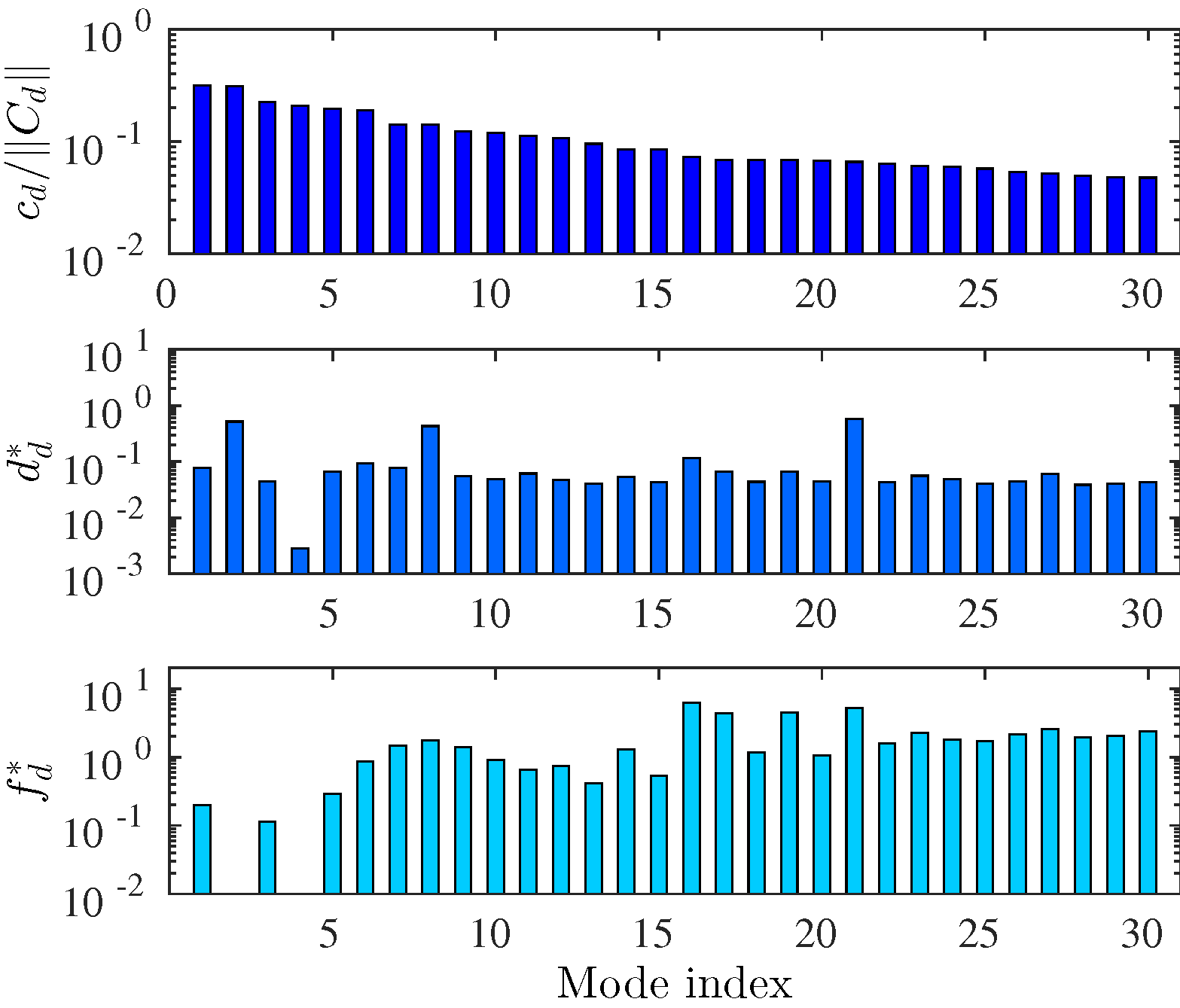}
	    \caption{}
	    \label{fig:Spectrum_DMD} 
      \end{subfigure}
    \caption{Mode amplitudes, dimensionless decay rates and frequencies for the first 30 $\cal{C}$DMD (a), $\cal{SH}$DMD (b) and DMD (c) modes.}
    \label{fig:Spectrum_CDMD_ShDMD}
 \end{figure}

The first modes and summations of the first 12 modes are reconstructed in the rotated and shifted frames and transformed back to physical space, in order to compare the reduced vortex head in the physical
space (figure \ref{fig:vortHead_turb_full}) between the three methods. The mentioned reconstructed modes are plotted in figure \ref{fig:CDMD_ShDMD_DMD} at the same physical time $t^{*}=13.7$. 
To present an overview of all the timesteps, space-time diagram reconstructed using 12 modes in each frame, is depicted in figure \ref{fig:CharDiag_CDMD_ShDMD} and compared against the fullfield. For each decomposition, 
the spectrum is presented in form of dimensionless decay rates and frequencies for the first 30 modes in figure \ref{fig:Spectrum_CDMD_ShDMD} along with the time averaged mode contents in each frame. 
In this figure, complex conjugate mode amplitudes and eigenvalues are presented only once for each mode. The spectrum presented for the decomposition in spatiotemporal space, 
is resulted by transformation from spatiotemporal to physical space by equations \ref{eq:freq} and \ref{eq:decay}.\medskip

As expected, DMD modes captured on the stationary frame of reference, fail to describe a clear picture of the vortex head (figures \ref{fig:DMD_1} and  \ref{fig:DMD_12} ). The first 
mode represents a smeared trace of the vortex ring distorted by the trailing jet which requires many modes to be added to reach a more clear 
reconstruction of the large-scale features of the flow. But even after adding 12 modes, the result is still far beyond sufficient. 
The same can be observed in all other timesteps as demonstrated in the reconstructed space-time diagram in figure \ref{fig:CharDiag_DMD_12m} using 12 
DMD modes. Unlike the first modes on the moving reference frames, which both possess a zero frequency, the eigenvalue of the first DMD mode 
shows frequency of $f^{*}_{d} = 0.2$. The first DMD mode also has a larger decay rate of $d^{*}_{d} = 0.07$ compared to the first modes 
in the rotated and shifted spaces. \medskip
 
Both decompositions on the shifted and rotated frames, deliver much better descriptions of the vortex head with the first mode, 
while the borders of the vortex head appear to be more distinct in the first $\cal{C}$DMD mode. The first $\cal{C}$DMD and $\cal{SH}$DMD modes have respectively 
decay rates of $d^{*}_{c} = 0.02$ and $d^{*}_{sh} = 0.006$. Adding up 12 modes for each decomposition, will capture most of the large 
scale features of the vortex head. The reconstructed space-time diagram using 12 modes for each analysis in figure \ref{fig:CharDiag_CDMD_ShDMD}, provides a 
good measure of how well all timesteps have been described using 12 modes. While both methods have captured the overall shape 
of the vortex ring in all timesteps, closer resemblance is observable between the fullfield and the $\cal{C}$DMD reconstruction.\medskip

The observed differences between the $\cal{C}$DMD and $\cal{SH}$DMD results, can be explained by the fact that, the modes and the structures in spatiotemporal space, 
do not contain only spatial information about the flow, as they do not belong to only one timestep. Rather, each spatiotemporal mode or structure, carries information
about its history, its present state and its future.\medskip

\section{ Summary and Conclusion}
\label{sec:conclusion}

In this study a modal decomposition was carried out along the characteristics direction
given by the group velocity of a structure allowing the extraction
of the moving features. Using this approach, we seek to define the structures in planes normal
to the direction of the characteristics. This means they are defined as
coherent events in space and time, as opposed to a snapshot in time or
a time series at a fixed location. To fulfill this purpose, one rotation 
in four dimensional space would be necessary. In general, this can be done by two rotations, but in
the studied case of a jet, the main direction coincides with one of the axis and 
therefore one rotation was sufficient. In the rotated frame, a modal
decomposition was performed and the vortex head of a starting jet was
extracted with a few modes only.\medskip 

The physical structure was recovered after
rotation back into the physical frame, where the results were compared against a traditional DMD carried out on a stationary frame of reference. 
This comparison revealed a much faster drop of singular values along the characteristics. The vortex head, that was treated as a coherent structure 
with a small decay rate, was reconstructed much more accurately using a $\cal{C}$DMD. The borders were distinctly captured with only one mode, 
and adding 4 spatiotemporal modes provided a very good description of the instabilities inside the vortex head.\medskip

In the final chapter, a comparison was carried out between a characteristic DMD and a shifted DMD, 
extracting the modes in a rotated and a shifted frame respectively. The dependence of the decompositions on the number of snapshots was also verified by 
employing 10 different windows in each frame of reference. It was shown that a singular value decomposition, led to a faster drop of singular values in
all the windows along the characteristics in the rotated reference frame. Both methods resulted in considerable improvements in 
describing the vortex head in comparison with a traditional DMD. Nevertheless, reconstructing the first mode and the summed up first 12 modes for each method, 
the $\cal{C}$DMD modes appeared to capture the large-scale feature of the flow more efficiently with fewer modes.\medskip 

Using the introduced approach, transport-dominated 
structures as well as their development along
their paths can be described. The method can be used for the definition
of empirical coherent structures with discreet translational symmetry, given as a reduced order model described by
a few modes.\\

\textbf{Acknowledgements}:
The authors would like to gratefully acknowledge Juan Jos\'e Pe\~na Fern\'andez for providing simulated data for this study.
Major parts of this work were performed during the sabbatical leave of the first author in 2015. He wishes to thank, both his own university
as well as La Sapienza, Rome, for this opportunity. The second author would like to appreciate Prof. Christoph Egbers (Brandenburg University of Technology - BTU Cottbus - Senftenberg),
for the invaluable support during this research.

\bibliographystyle{spbasic}    
\bibliography{CDMD.bib}   

%
%

\end{document}